# The 2019 Motile Active Matter Roadmap

Gerhard Gompper[1], Roland G. Winkler[1], Thomas Speck[2], Alexandre Solon[3], Cesare Nardini[4], Fernando Peruani[5], Hartmut Löwen[6], Ramin Golestanian[7], U. Benjamin Kaupp[8], Luis Alvarez[8], Thomas Kiørboe[9], Eric Lauga[10], Wilson Poon[11], Antonio De Simone[12], Frank Cichos[13], Alexander Fischer[13], Santiago Muiños Landin[13], Nicola Söker[13], Raymond Kapral[14], Pierre Gaspard[15], Marisol Ripoll[1], Francesc Sagues[16], Julia Yeomans[17], Amin Doostmohammadi[17], Igor Aronson[18], Clemens Bechinger[19], Holger Stark[20], Charlotte Hemelrijk[21], Francois Nedelec[22], Trinish Sarkar[23], Thibault Aryaksama[23], Mathilde Lacroix[23], Guillaume Duclos[23], Victor Yashunsky[23], Pascal Silberzan[23], Marino Arroyo[24], Sohan Kale[24]

1. Theoretical Soft Matter and Biophysics, Institute of Complex Systems and Institute for Advanced Simulation, Forschungszentrum Jülich
2. Institute of Physics, University of Mainz
3. Sorbonne Université, CNRS
4. Service de Physique de l'État Condensé, CEA-Saclay
5. Université Côte d'Azur
6. Institut für Theoretische Physik II: Weiche Materie, Heinrich-Heine-Universität Düsseldorf
7. MPI-DS Göttingen and Oxford University
8. Molecular Sensory Systems, Center of Advanced European Studies and Research, Bonn
9. Centre for Ocean Life, DTU-Aqua, Technical University of Denmark
10. DAMTP, University of Cambridge
11. SUPA and School of Physics & Astronomy, The University of Edinburgh
12. The BioRobotics Institute, Scuola Superiore Sant'Anna, Pisa
13. Molecular Nanophotonics Group, Leipzig University
14. Chemical Physics Theory Group, Department of Chemistry, University of Toronto
15. Center for Nonlinear Phenomena and Complex Systems, Université Libre de Bruxelles
16. Departament de Ciència de Materials I Química Física, Universitat de Barcelona
17. The Rudolf Peierls Centre for Theoretical Physics, Clarendon Laboratory, Oxford
18. Department of Biomedical Engineering, Pennsylvania State University
19. Physics Department, University of Konstanz
20. Institute of Theoretical Physics, Technische Universität Berlin
21. University of Groningen, The Netherlands
22. Sainsbury Laboratory, Cambridge University
23. Laboratoire PhysicoChimie Curie, Institut Curie, PSL Research University - Sorbonne Université - CNRS
24. Universitat Politècnica de Catalunya – BarcelonaTech, Barcelona

## Abstract

Activity and autonomous motion are fundamental in living and engineering systems. This has stimulated the new field of "active matter" in recent years, which focuses on the physical aspects of propulsion mechanisms, and on motility-induced emergent collective behavior of a larger number of identical agents. The scale of agents ranges from nanomotors and microswimmers, to cells, fish, birds, and people. Inspired by biological microswimmers, various designs of autonomous synthetic nano- and micromachines have been proposed. Such machines provide the basis for multifunctional, highly responsive, intelligent (artificial) active materials, which exhibit emergent behavior and the ability to perform tasks in response to external stimuli. A major challenge for understanding and designing active matter is their inherent nonequilibrium nature due to persistent energy consumption, which invalidates equilibrium concepts such as free energy, detailed balance, and time-reversal symmetry. Unraveling, predicting, and controlling the behavior of active matter is a truly interdisciplinary endeavor at the interface of biology, chemistry, ecology, engineering, mathematics, and physics.

The vast complexity of phenomena and mechanisms involved in the self-organization and dynamics of motile active matter comprises a major challenge. Hence, to advance, and eventually reach a comprehensive understanding, this important research area requires a concerted, synergetic approach of the various disciplines. The 2019 *Motile Active Matter* Roadmap of *Journal of Physics: Condensed Matter* addresses the current state of the art of the field and provides guidance for both students as well as established scientists in their efforts to advance this fascinating area.

# Introduction

*Gerhard Gompper and Roland G. Winkler*
*Theoretical Soft Matter and Biophysics, Institute of Complex Systems and Institute for Advanced Simulation, Forschungszentrum Jülich, 52425 Jülich, Germany*

Active matter is a novel class of nonequilibrium systems composed of a large number of autonomous agents. The scale of agents ranges from nanomotors, microswimmers, and cells, to crowds of fish, birds, and humans. Unraveling, predicting, and controlling the behavior of active matter is a truly interdisciplinary endeavor at the interface of biology, chemistry, ecology, engineering, mathematics, and physics. Recent progress in experimental and simulation methods, and theoretical advances, now allow for new insights into this behavior, which should ultimately lead to the design of novel synthetic active agents and materials. This roadmap provides an overview of the state of the art, and discussed future research directions on natural and artificial active agents, and their collective behavior.

## General Principles and Methods

Active systems are persistently out of equilibrium, due to the continuous energy consumption of its constituent agents. This implies the absence of equilibrium concepts like detailed balance, Gibbs ensemble and free energy, as well as time-reversal symmetry. Therefore, theories of active matter have to be constructed on the basis of symmetries – like polar or nematic shape and interactions of the agents – as well as conservation laws and dynamic rules. Agent-based standard models, such as active Brownian particles and squirmers, have emerged to account for the underlying physical mechanisms in dry and wet active matter. They are complemented by continuum field theory. Methods and techniques to analyze these models and theories range from simulations, including mesoscale hydrodynamics approaches, to field-theoretical methods and dynamic density-functional theory.

## Biological Nano- and Microswimmers

Evolution has provided a large diversity of biological swimmers on the microscale, such as sperm cells, bacteria, and algae. Their propulsion mechanisms and navigation strategies are tailored to their function and natural environment, ranging from living organisms to soil and the open seas. Microswimmers prototypically employ cilia or flagella for propulsion, which beat or rotate, but also periodic changes of their body shape. Understanding of the underlying principles and search strategies, e.g., chemotaxis and phototaxis, allows for their targeted manipulation and control in medicine, ecology, and multiple technical applications. The motion of these swimmers on the microscale naturally raises the question how small a swimmer can be to still display directed motion, maybe even on the scale of a single macromolecule.

## Synthetic Nano- and Micromachines

Various strategies for the design of autonomous synthetic nano- and micromachines have been proposed. This includes phoresis – inhomogeneous catalysis of chemical reactions (diffusiophoresis), thermal gradients (thermophoresis) –, planktonic body deformations, and biology-inspired concepts. Such machines provide the basis for multifunctional and highly responsive (artificial) materials, which exhibit emergent behavior and the ability to perform specific tasks in response to signals from each other and the environment. The development of novel techniques facilitates control of the locomotion of individual nano- and micromachines as well as their interactions, and the design of intelligent active materials.

An important model system to study collective motion far from equilibrium is a mixture of semiflexible polar filaments and motor proteins, which form highly dynamic swirl patterns. Here, the motion of topological (nematic) defects and their creation and annihilation plays a central role. Active nematic theory successfully captures many observed phenomena.

In potential applications, the external control of nano-and micromachines is essential. Here, light has become a prime candidate, because it can be switched on and off at will, affects the particles without delay, and can be modified in strength individually for each particle.

## Swarming

Active agents are able to spontaneously self-organize when present in large numbers, resulting in emergent coordinated and collective motion on various length scales. Examples range from the cytoskeleton of cells, swarming bacteria and plankton, to flocks of birds and schools of fish. The mechanisms determining the emergence and dynamics of a swarm include the shape of the agents, steric interactions, sensing, fluctuations, and environmentally mediated interactions. Novel phenomena range from motility-induced phase separation to active turbulence.

In biological systems, the reaction of swarms to external signals is crucial, and is often the reason for the formation of swarms in the first place. Examples range from the reaction of bird flocks to predators to the collective motion of bottom-heavy swimmer like algae in gravitational fields.

In synthetic systems, light control allows for a completely new approach, in which the interactions between particles are partially direct, partially calculated numerically from measured particle conformations and then imposed externally.

## Cell and Tissue Dynamics

Fundamental biological processes, such as morphogenesis and tissue repair, require collective cell motions. Diverse inter- and intracellular processes are involved in migration, ranging from cytoskeleton-generated forces to deform the cell body to intercellular and substrate adhesion. This gives rise to specific phenomena, for particular cell types, such as fingering-like instabilities and spreading, or glass-like arrest as the cellular adhesions mature.

Tissues are nature's active materials, and are therefore very interesting as blueprints for synthetic active materials. Here, diverse aspects are combined synergistically. For example, tissues – in particular epithelial monolayers – respond elastically when stretched, but can also be "super-elastic" (supporting enormous areal strain) due to active cytoskeletal polymerization dynamics. In contrast, fluidization of tissues on long time scales is induced by apoptosis and cell division, but also by mechanical stimuli, implying rheological properties controlled by activity.

## Active Brownian particles: From collective phenomena to fundamental physics

Thomas Speck, Institute of Physics, University of Mainz

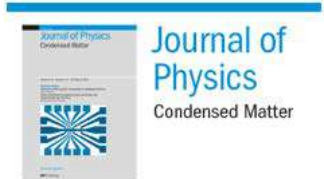

**Status**

Persistence of motion is the hallmark of a range of systems that fall under the umbrella of "active matter". Instead of a global preferred direction, e.g., due to an external field, symmetry is broken locally. Active transport in aqueous environments, e.g., within cells due to motor proteins moving along microtubules, plays a major role in biological systems, which also comprise motile organisms such as bacteria as well as sperm and other types of motile cells. In contrast, synthetic active matter exploits different physical phenomena to achieve locomotion such as particle shape and ultrasound, but the arguably prevalent experimental strategy is Janus particles with two hemispheres that have different surface properties (for a detailed review and references, see Ref. 1). A popular and instructive example are colloidal particles, where one hemisphere is coated with a catalyst for the decomposition of hydrogen peroxide. Suspended into a solvent containing hydrogen peroxide, a local concentration gradient ensues, and the particles are propelled individually along their symmetry axis. This axis is not fixed in space but undergoes rotational diffusion due to fluctuations, which leads to single-particle trajectories that are characterized by a persistence length in analogy with polymers. All propulsion mechanisms have in common that they break detailed balance, which requires the constant input of (free) energy that is dissipated into the environment.

Minimal many-body model systems stripped off microscopic details are pivotal in statistical physics. The arguably most famous example is the Ising model of spins, which describes the minimal ingredients for phase transitions and phase coexistence. In the same spirit, a class of models combing persistent motion with short-range–typically purely repulsive–interactions called Active Brownian Particles (ABPs) has become a playground to study novel collective behavior that would not be possible in thermal equilibrium. While constantly driven away from thermal equilibrium, the observed collective behavior of active particles nevertheless is reminiscent of phases delineated by abrupt changes when tuning external parameters such as particle density. Most notably, self-propelled particles aggregate into clusters even in the absence of cohesive forces through a dynamic feedback between speed and density (Figure 1), which has been coined motility-induced phase separation (MIPS) [3]. Much work has been devoted to establishing a general statistical physics framework that allows to predict the large-scale behavior and phase boundaries, e.g., through dynamical mean-field theories in the spirit of liquid-state theory [4].

Synthetic active particles also have possible applications as micro-engines. Exploiting their directed motion, they can exert forces on their environment and thus convert chemical free energy (e.g., as released by the reduction of hydrogen peroxide) into mechanical work. In particular how propulsion modifies the pressure, a fundamental thermodynamic concept, has sparked quite some interest [6]. It has direct consequences for predicting forces on objects that have been immersed into an active medium, which again might strongly diverge from our expectations as shaped by equilibrium concepts. For example, bacterial suspensions have been used to drive the forward rotation of micrometer-sized gears [5] and asymmetric obstacles can generate a density gradient. Maybe even more fundamentally, the role of entropy production and efficiency of such active engines has recently moved into the focus.

Some propulsion mechanisms require light, e.g., hematite acting as a catalyst for the mentioned decomposition of hydrogen peroxide or to heat one hemisphere through absorption, which triggers the local demixing of a binary solvent [2]. Such light-driven active colloidal particles offer additional control through different illumination patterns that, e.g., guide particles towards target regions, which are thus reached much faster than would be possible through passive diffusion alone. Moreover, judicious choices of the confining geometry can be exploited for this purpose. The reversible coupling of active particles with passive "cargo" is being studied as means to transport selected materials on the microscale, quite in analogy with molecular motors transporting material within cells. Moreover, mixtures of chemically active *uniform* particles with passive particles open the route to design synthetic modular microswimmers [7]. These are clusters composed of a few particles that break symmetry in a way that leads to linear and/or angular propulsion of the whole aggregate. These aggregates can spontaneously assemble from their building blocks, which has raised interest in the control and prediction of the self-assembly pathways of such dissipative structures.

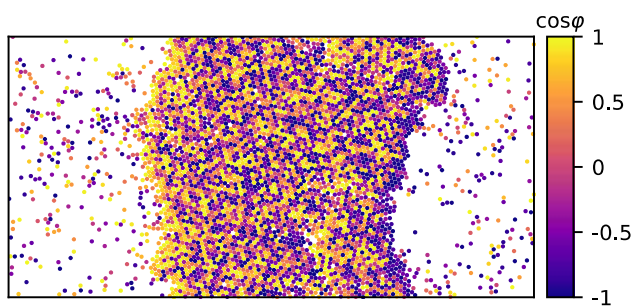

**Figure 1.** *Simulation snapshot of active Brownian particles in an elongated box employing periodic boundary conditions. Clearly visible is the separation into a dense domain spanning the box surrounded by a dilute active gas. Particles are colored according to their orientation, note that in the interfacial region particles predominantly point toward the dense region. Image courtesy of A. Fischer.*

**Current and Future Challenges**

As with any minimal model, a central issue is the range of its applicability. It is worth emphasizing that the goal is not necessarily quantitative agreement, but to identify a dominant physical mechanism underlying a particular observation with the understanding that other mechanisms are at play. For example, for interacting synthetic swimmers, long-range hydrodynamic and phoretic interactions are present (and covered by others in this Roadmap article). Even though neglecting these, active Brownian particles are able to reproduce experimental data on phase separation [2], supporting the notion that directed motion is the crucial ingredient. It remains to be explored more systematically how the MIPS scenario is modified through phoretic and hydrodynamic interactions. At what point does new, qualitatively different behavior emerge?

Even for plain ABPs, an open question is the exact nature of the critical point terminating coexistence [8]. In passive systems, critical fluctuations due to a diverging correlation length show universality, i.e., they are determined by symmetries and conservation laws but independent of microscopic details. Can such a classification scheme of universality classes be extended to active particles? This would be helpful in predicting the large-scale macroscopic properties of active materials. Away from the critical point, new tools will have to be developed to deal with multi-component mixtures of active particles and the associated phase behavior.

A major theoretical undertaking is to develop a better understanding for the statistical foundations of active matter, ideally comparable to equilibrium statistical mechanics. While this goal seems to be futile for non-equilibrium in general, quite some progress has already been made for idealized active particles. The benefit of a more abstract, model-independent understanding is two-fold: first, it would allow to derive general bounds against which models can be validated, and second, it would allow to develop and adapt advanced numerical methods. As one example, one could then study rare events such as nucleation. At the moment, often models are constructed around, and in order to reproduce, experimental observations; with limited use for predicting novel phenomena.

But beyond challenges, there are opportunities. Synthetic active particles allow to disentangle physical mechanisms from more complex, regulatory feedbacks in biological systems [9]. Models in the spirit of ABPs are now being employed for this purpose, and simple variations can reproduce surprisingly complex behavior. For example, some microorganisms such as bacteria and sperm cells can "sense" chemical gradients and adapt their behavior. This can be used for communication purposes through a mechanism called quorum sensing, in which members of a population excrete signaling molecules that diffuse in the surrounding solvent and create a non-uniform concentration profile. The local concentration (more precisely, its relative change) of these molecules is sensed by other members and determines changes in their phenotype, e.g., bioluminescence and motility changes. Implementing such quorum sensing for synthetic active particles has be exploited to achieve aggregation at low densities [10]. While most studies explore the "forward" direction, i.e., they study the emergent collective behavior for a given model, it will also be interesting to consider the reverse direction and to determine rules that lead to a desired target state. In passive materials, this task is intimately related to coarse-graining and the determination of a pair potential from a radial pair distribution. Comparable tools for non-equilibrium active systems are missing, but it seems safe to predict that we will see the deployment of machine learning concepts in this area.

**Concluding Remarks**

Active Brownian particles and its growing number of variants will continue to be valuable models to study basic questions of non-equilibrium many-body statistical physics. One vision is that these models will eventually lead to a comprehensive theoretical framework on a par with equilibrium statistical mechanics. Leaning towards the application side, these models can provide valuable design guidelines for tasks such as directed transport at the microscale and controlling assembly pathways of dissipative materials and structures. Such an understanding opens the route to soft functional materials that could implement responses not possible in passive materials.

**Acknowledgements**

I acknowledge funding from the DFG through the priority program SPP 1726. I thank Clemens Bechinger, Hartmut Löwen, Friederike Schmid, and Peter Virnau for long-standing and fruitful collaborations on the topics discussed here.

## Perspectives on motility-induced phase separation

Alexandre Solon[1] and Cesare Nardini[2]

[1]Sorbonne Université, CNRS, Paris, France

[2]Service de Physique de l'État Condensé, CEA-Saclay, France

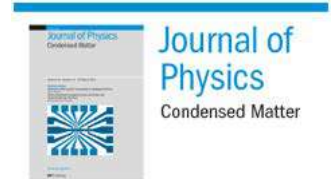

**Status**

Acting on the propagation speed of self-propelled particles can lead to dramatic collective behaviors. This is the case when interactions between particles tend to slow them down, for example because of crowding or a gregarious interaction. Homogeneous collection of active particles can then become unstable and phase separate between a dilute phase of fast movers and a dense phase of slow movers, a phenomenon known as motility-induced phase separation (MIPS). This was observed in the two active systems most commonly encountered in a laboratory: suspensions of motile bacteria and self-propelled colloids. Understanding MIPS in these (relatively) simple systems has both a fundamental interest as a nonequillibrium phase transition and a more practical one toward manipulating active systems through motility and deciphering complex active environments such as the intra-cellular medium.

Two microscopic models are commonly used to study MIPS numerically and analytically. It was first predicted for active particles with quorum-sensing interactions (QSAPs) [1], that adapt their propulsion speed according to the mean density $\rho$ surrounding them through a prescribed velocity function $v(\rho)$. The second model features active particles interacting through pairwise forces (PFAPs) [2], more precisely through a short-ranged repulsion like hard or soft-core interactions. PFAPs model more closely self-propelled colloids while QSAPs can be seen as a microscopic model for some bacterial species which sense the local density through a chemical that they release in the external medium. Notably, in both models phase separation occurs in absence of any attractive interaction and it is thus a purely non-equilibrium phenomenon.

In the following, we consider only "dry" active matter which neglects hydrodynamic interactions. Their effects are not fully elucidated, although they have been shown in some cases to limit bulk phase separation to a micro-phase separation. Additionally, in the literature, one often distinguishes different types of angular dynamics for active particles such as active Brownian particles or run-and-tumble particles; this distinction will not be important here [3].

The mechanism destabilizing an homogeneous system is relatively well understood: When interactions in the system slow down the active particles efficiently enough, an homogeneous suspension becomes linearly unstable [1] and the system phase-separates. This note summarizes recent and ongoing attempts at describing the (nonlinear) phase-separated state.

**Mapping to equilibrium**. At first sight, MIPS is very similar to an equilibrium liquid-gas phase separation. More particularly, it was shown for QSAPs [1] that as a first approximation in a gradient expansion, the dynamics of the density field of the active suspension can be mapped on an equilibrium system with effective free-energy density $f(\rho)=\rho(\log \rho-1)+ \int_0^\rho \log v(s)\, ds$. However, the coexistence densities do not obey the common tangent construction on the free energy density (see Fig. 1, left), meaning that $f(\rho)$ is not minimized.

Somewhat similarly, for PFAPs one can derive an equation of state $P(\rho)$ giving the mechanical pressure as a function of density [4]. However, the phase equilibria are not given by the Maxwell equal-area construction (see Fig. 1, right) as they would in a passive system.

These two results indicate that it is necessary to go further and take into account the nonequilibrium character of active systems.

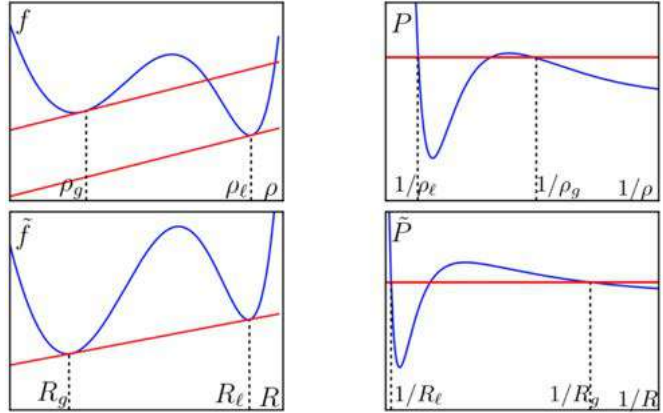

**Figure 1.** Failure of the common tangent construction (upper left) and Maxwell equal-area construction (upper right): The coexistence densities $\rho_g$ and $\rho_\ell$ are not given by the usual thermodynamic construction. Using the effective density $R(\rho)$ defined in the text, the values at coexistence $R_g$ and $R_l$ are obtained through a generalized common tangent construction (lower left) or Maxwell equal-area construction (lower right) albeit on generalized thermodynamic functions.

**Nonequilibrium interfaces.** To describe MIPS, it was proposed to generalize the Cahn-Hilliard equation, the classical model for phase separation in equilibrium. In the passive version, one starts from a free energy functional $F[\rho]$ such that the weight of a coarse-grained density field reads $P[\rho] \propto e^{-\beta F}$ with $\beta$ the inverse temperature. Expanding the free energy in gradients, at first order we have $F[\rho] = \int f(\rho) + c(\rho)|\nabla\rho|^2 dr$, where $f$ is the (bulk) free energy density and the gradient term gives a surface tension. The Cahn-Hilliard equation then gives the evolution of the density field

$$\dot{\rho}(r,t) = -\nabla \cdot J; J = -M(\rho)\nabla\frac{\delta F}{\delta\rho} \qquad (1)$$

where $M(\rho)$ is a mobility term, unimportant for the phase equilibrium, that is only assumed to be positive. One identifies the chemical potential as the quantity driving the dynamics in Eq. (1)

$$\mu(r,t) \equiv \frac{\delta F}{\delta\rho} = f'(\rho) - \frac{c'}{2}|\nabla\rho|^2 - c\Delta\rho$$

where the primes denote the derivative with respect to $\rho$. A first non-equilibrium extension was proposed in Ref. [5] which considered a non-equilibrium chemical potential

$$\mu = g_0(\rho) + \lambda|\nabla\rho|^2 - \kappa\Delta\rho, \qquad (2)$$

In the special case $\lambda = const$ and $\kappa = 1$. When $\lambda \neq 0$, Eq. (2) cannot be written as the derivative with respect to $\rho$ of a free energy functional. Consequently, the extra $\lambda$ term was found to break detailed balance and the common tangent construction.

Going one step further, Ref. [6,7] considered arbitrary coefficients $\lambda(\rho)$ and $\kappa(\rho)$ (i.e. the most general expression for $\mu$ at this order in gradient) and showed how to compute analytically the phase equilibria of this generalized Cahn-Hilliard (GCH) equation. This relies on introducing an "effective density" $R(\rho)$ satisfying the differential equation $\kappa R'' = -(2\lambda + \kappa')R'$. Up to this change of variable, one recovers a thermodynamic structure: The coexistence densities are set by the equality of chemical potential $g_0$ and a generalized pressure $\mathcal{P}$, which defines the common-tangent construction on a generalized free energy density $\mathcal{f}$ (see Fig. 1). The generalized quantities are all defined with respect to $R$: $\mathcal{f}$ is such that $d\mathcal{f}/dR = g_0$ and $\mathcal{P} = Rg_0 - \mathcal{f}$. Note that the equilibrium case corresponds to $2\lambda+\kappa' = 0$ so that $R = \rho$ and the usual thermodynamic construction is recovered. In the active case, $R \neq \rho$ and the interfacial terms thus affect the phase equilibria because they enter in the definition of $R$. This is a major difference with the equilibrium situation where the phase diagram depends only on bulk properties. On the other hand, the non-equilibrium character of the system stems only from gradient terms and, consistently, the entropy production, when defined locally, is concentrated at the interfaces [8].

For QSAPs, one can explicitly coarse-grain the microscopic dynamics to write a dynamical equation for the density field [3]. At high density, when fluctuations can be neglected, it takes exactly the form of the GCH equation; the phase diagram and the coexisting densities quantitatively agree with those found in numerical simulations [6,7].

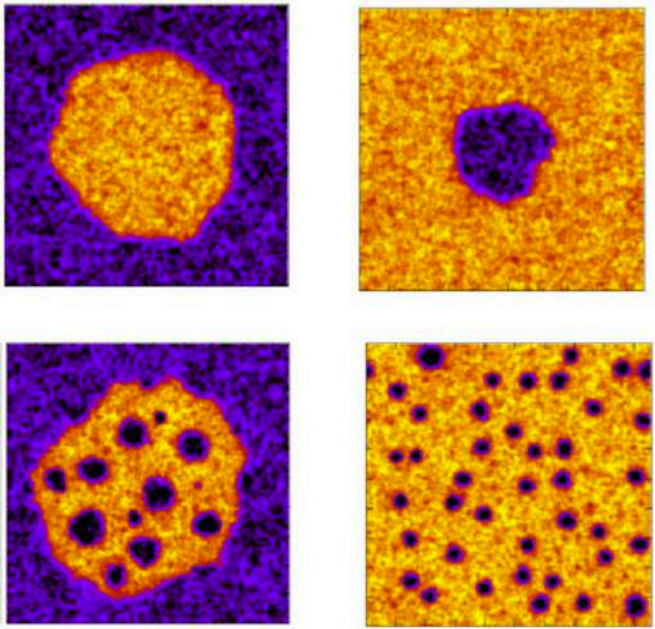

Figure 2. Simulations of Eq.(3), with liquid regions in yellow and vapor in purple for small (upper row) and large (lower row) $\zeta$. Upon increasing $\zeta$, a phase transition is crossed between full phase separation and a microphase separation showing a population of vapor bubbles forming in the liquid due to reverse Ostwald ripening [10].

## Current and Future Challenges

It was recently realized that PFAPs actually show more complex nonequilibrium features than QSAPs, which are not accounted for by the GCH. Simulations indeed revealed [9] that, at phase coexistence, the liquid phase hosts a population of mesoscopic vapor bubbles that are continuously nucleated in the bulk, coarsen, and are ejected into the exterior vapor. The inverse process, with the interface buckling to bring a bubble of vapor into the liquid is rarely observed, suggesting that for PFAPs nonequilibrium features are crucial also in the bulk (bubbly phase separation). It is a current challenge to explain this phenomenology.

A first attempt to rationalize such complex steady states consists in a further generalization of the Cahn-Hilliard equation, including a term that cannot be written as the gradient of a chemical potential

$$\dot{\rho} = \nabla \cdot M(\rho)[\nabla\mu + \zeta\nabla^2\rho\nabla\rho], \qquad (3)$$

with Eq. (2) for $\mu$. This is now the most general dynamics that can be written for a conserved quantity at this order in gradients and it was analyzed in [10] for constant coefficients $M, \lambda, \kappa$ and $\zeta$.

In an equilibrium fluid, full phase separation is kinetically achieved through the Ostwald process: as the Laplace pressure is larger in smaller droplets, the bigger ones grow at the expense of smaller ones. Crucially, it was shown in [10] that *the $\zeta$* term can revert the Ostwald process, still keeping interfaces stable. Increasing $\zeta$, one observes a phase transition between full phase separation and, depending on the global density, the "bubbly phase separation" or the microphase separated state represented in Fig. 2 (bottom). It is presently an open question whether such results can be connected to the measurement of negative interfacial tension in PFAPs [7, 11].

Explicit coarse-graining of particle models [10] that treat the effect of two-body forces perturbatively with respect to quorum sensing suggests that the pairwise interactions of PFAPs can generate the type of $\zeta$ term discussed above. Eq. (3) but it remains to show that pairwise forces suffice to generate this term and that it is indeed responsible for the phenomenology of PFAPs. In that regard, an important challenge is to find means to relate the coefficients of the macroscopic equations to microscopic measurements.

The motivation for the extensions of the Cahn-Hilliard equation presented here is to describe phase separation in systems that do not obey detailed balance. As such, it is not limited to active matter and could be relevant to other non-equilibrium systems such as sheared suspensions undergoing shear de-mixing or shear-banding, or to liquid-like droplets in the intracellular medium. Finding whether the ideas of generalized thermodynamics and reverse Ostwald ripening can be applied to other situations will thus be part of future endeavors. Mixtures of active and passive particles are another very active subject that is also of practical interest to describe e.g. mixtures of motile and non-motile bacteria or the intracellular medium where active and passive components coexist. Several types of phase separation have been observed numerically but our theoretical understanding is still rudimentary and it is a future challenge to develop the type of thermodynamic formalism presented in this note for mixtures.

## Self-Organized Collective Patterns in Active Matter: from phase separation to collective motion

Fernando Peruani, Université Côte d'Azur

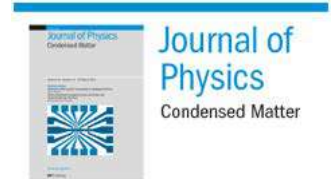

### Status

The emergence of self-organized patterns of actively moving entities, from bacteria to sheep, and including manmade active systems, are typically explained by either invoking the presence of some velocity alignment mechanism that mediates the interactions among the moving individuals [1] or by postulating the existence at the mesoscale of a coupling between density and particle speed [2]. However, there exist alternative mechanisms that lead to similar self-organized active patterns. It is worth noting that velocity alignment and density-speed coupling mechanisms are the results of theoretical speculations formulated before the realization of specifically designed, quantitative, active-matter experiments. Therefore, active-matter experiments have been analyzed assuming, beforehand, the existence of one of these two mechanisms, or a combination of both, overlooking competing, alternative mechanisms. Here, we review several pattern formation mechanisms in active matter, which we classify in two groups: those that lead to self-organized patterns in the absence of (orientational) order and those that involve the presence of it. We pay particular attention to novel, alternative mechanisms such as attractive forces in the absence of Newton's third law, and revisit collective effects induced by particle shape, which are examples where notably the orientational order symmetry is an emergent, dynamical property of the system.

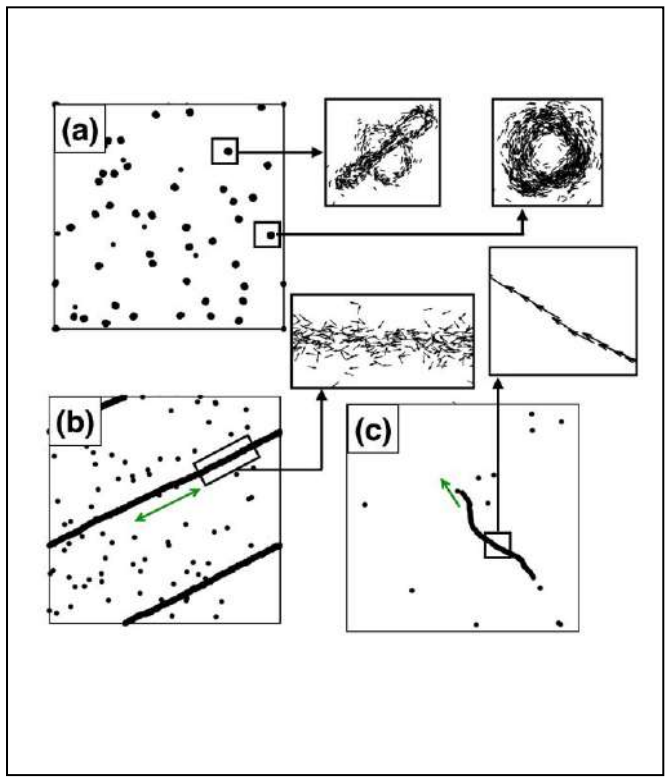

**Figure 1.** Self-organized patterns in a system of active particles with attractive interaction and without velocity alignment [3]. For reciprocal interaction, particles aggregate (a). By breaking the action-reaction symmetry nematic (b) and polar (c) patterns emerge.

### Current and Future Challenges

**Self-organization in the absence of order** -- Spontaneous phase separation is one of the simplest self-organized dynamics and occurs in equilibrium as well as in non-equilibrium systems. In the following, we focus on what we call ``neutral phase separation" (NPS). We define NPS as a phase separation process in which growing (phase) domains exhibit vanishing orientational order -- implying that particle velocities do not display any type of order -- as these regions increase in size as the system phase separates[1]. Active particles that interact via short-range, isotropic, attractive forces, phase separate following a Cahn-Hilliard dynamics [3] (Fig. 1a). A less obvious observation is that purely repulsive interactions can also lead to phase separation (with Cahn-Hilliard dynamics). Here, the mechanism involves an effective decrease in the mobility of the particles in regions with high particle densities [2]. This mechanism is particularly evident in lattice models that possess a maximum occupation number (of particles) per lattice site [4]. In these models, particle movements to occupied sites (i.e. sites occupied by the above-mentioned maximum number) are forbidden. As result of this, particles within high-density areas are virtually trapped and only particles at the boundary of these areas are able to move. The rates at which particles arrive and leave these low-mobility regions determine the fate of these phase-separated domains. It is worth noting that if particles exhibit velocity alignment, the local mobility of particles is not only determined by particle density, but also depends on (orientation) order [4]. In the presence of order, a scalar description of the problem – i.e. a description in terms of a density field – is no longer possible and a zoo of self-organized patterns, with no equilibrium counterparts, emerges [4].

**Self-organization in the presence of order** -- Collective motion patterns with either polar or nematic (local) order are observed in a variety of experimental systems: from swarming bacteria to sheep herds, and including artificial active systems such as Quincke rollers. A large number of models have been proposed and studied with the intention of shedding light on such collective motion patterns. Below, we classify models in two categories: i) models with built-in microscopic interactions whose symmetry sets the symmetry of the emergent order, and ii) models where the order symmetry is an emergent, dynamical property of the system.

        i) When the symmetry of the emergent order is given by the symmetry of microscopic interactions

Many models from equilibrium statistical mechanics fall into this broad category. Let us take as example the Ising model with an interaction potential between neighboring spins $S_i$ and $S_j$ of the form $-JS_iS_j$ with $J>0$ and Glauber dynamics. Such interaction potential favors (local) polar alignment of the spins, while thermal fluctuations tend to randomize the spin orientations. In this model, there is no doubt on the symmetry of the emergent order: if order emerges, it has to be polar. Similarly, in flocking polar models (e.g. Vicsek model [1]) interactions between neighboring particles favor local polar alignment of the particle velocities, while fluctuations tend to randomize them. When fluctuations are weak, the active particles self-organize into large-scale polar patterns [1]. If the symmetry of the interaction is modified to favor nematic alignment of particle velocities -- e.g. by letting particles set their velocity either parallel or antiparallel to the average local velocity -- the emergent collective patterns are nematic [5]. In short, in these models the symmetry of microscopic interactions determines the symmetry of the emergent order. Furthermore, at the onset of (orientational) order, active particles phase separate into high-density, high-order regions (usually called bands) for both interaction symmetries, polar and nematic [1, 5]. However, the spatio-temporal dynamics of these bands are fundamentally different for polar and nematic interactions.

---

[1] In flocking models, the system also phase-separates, but the emerging domains display orientational order. The presence of order dramatically affects the physics of the underlying phase-sepeation process.

### ii) When the symmetry of order is an emergent, dynamical property of the system

In active systems, the symmetry of the emergent order is not necessarily the symmetry of the underlying microscopic interaction rules. Moreover, the symmetry of order can be an emergent, dynamical property of the system. This is illustrated below using two specific active matter models.

**Breaking Newton's third law –** Newton's third law implies that interactions are reciprocal. However, in many biological systems interactions are not related to physical forces and do not need to be reciprocal. As an example, think of a flock of birds. Birds as well as other animal systems flock using visual information. Given that animals possess a field of view, which limits the visual perception of neighbours, it cannot be ensured that if an individual A sees an individual B, B also sees A. This means that interactions do not need to be reciprocal. To understand the consequences of breaking Newton's third law in active systems, let us consider a simple two-dimensional active system where the self-propelled particles interact with those particles inside a field of view via an attractive force [3]. By construction, this model does not possess velocity-velocity interactions and thus, there is no built-in microscopic interaction rule with either polar or nematic symmetry. In [3] it was proved that in this active model both, nematic (Fig. 1b) and polar (Fig. 1c) order spontaneously emerge in different areas of the parameter space, with the emergence of order always being associated with density instabilities and involving a (non-neutral) phase separation process [6]. Notably, the emergence of orientational order requires the absence of the action-reaction symmetry (i.e. Newton's third law), which is what actually drives the system out of equilibrium. The use of a field of view ensures that the action-reaction symmetry is not present. In summary, this model proves that both, polar and nematic order can emerge in the absence of velocity-velocity interactions, and without built-in microscopic rules with a predefined symmetry. The order symmetry is then an emergent, dynamic property of the system.

**Collective effects induced by particle shape --** Collections of self-propelled elongated particles -- e.g. self-propelled rods -- interacting by purely repulsive forces resulting from simple volume exclusion effects form (at the mesoscale) both, polar and nematic patterns [7, 8]. Notably, since particles interact via a potential that penalize particle overlapping and consequently does not distinguish the ``head'' and "tail" of particles, interactions display nematic symmetry. This becomes evident if particles are thermally agitated instead of being self-propelled: above a given density, particles display (local) nematic order. On the other hand, if particles are self-propelled, depending on density and particle aspect ratio, moving polar clusters may emerge in the system, despite the nematic symmetry of interactions [8]. The emergence of polar order results from the kinematics of particles: though the interaction is nematic, only particles moving in the same direction remain together, while particles moving in opposite direction quickly move apart. Thus, the dynamics is such that polar configurations of particles are long-lived. When volume exclusion interactions are strong enough to prevent (local) counter-propagating flows of particles (as in Fig. 1b), the described mechanism leads to polar order (at the mesoscale) as shown in Fig. 2 [8]. Interestingly, nematic patterns can still emerge, but this requires either particle reversal [9] or particle to be able to move over each other [10]. If this occurs, we observe the formation of nematic bands, which for high enough system sizes become transversally unstable [9]. However, note that physical self-propelled rods embedded in two dimensions [8] do not undergo an isotropic-nematic transition and do not form nematic bands, as initially suggested by mean-field approaches that neglect pair-correlations [1]. All this indicates that the role of particle shape in active systems needs to be revisited. These recent reported results show that the symmetry of the emergent order is not given by the symmetry of the interaction potential only, but also for the kinematics of particles. In particular, understanding the emergence of polar order requires to go beyond mean-field assumptions and include pair-correlations. In summary, here again, the order symmetry is an emergent, dynamical property of the system.

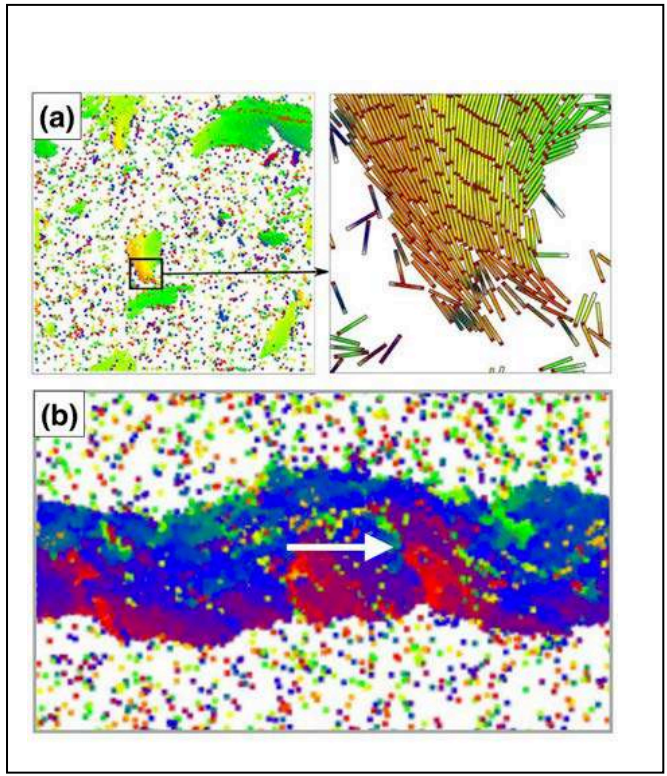

Figure 2. Self-organized patterns in a system of self-propelled rod [8]. Self-propelled rods form moving polar clusters despite the fact that particles interact via a potential that promotes local nematic alignment of the particles. The moving direction is colour-coded.

**Concluding Remarks**

Self-organized patterns of actively moving entities may emerge by alternative mechanisms to velocity alignment and speed-density coupling. Attractive forces in the absence of Newton's third law or the interplay between self-propulsion and particle shape can lead to polar as well as nematic collective patterns. These observations invite to revise the interpretation of existing experimental data. Importantly, the order symmetry for these alternative mechanisms is an emergent, dynamic property of the system and not the mere reflection of the microscopic interaction symmetry. A theoretical understanding on how such order symmetries emerge remains an open question. Arguably, such order symmetries result from a non-trivial interplay between the kinematics of particles and the microscopic interaction rules.

**Acknowledgements**

Financial support from Grant ANR-15-CE30-0002-01, project "BactPhys", is acknowledged.

# Dynamical density functional theory for microswimmers

Hartmut Löwen

Institut für Theoretische Physik II: Weiche Materie, Heinrich-Heine-Universität Düsseldorf, 40225 Düsseldorf, Germany

## 1. Status

Classical density functional theory (DFT) has become a versatile tool to calculate and predict *equilibrium* properties of inhomogeneous fluids [1] including sedimentation, wetting near walls and bulk phase transitions such as melting and freezing [2] for given interparticle interaction. The cornerstone of DFT is the existence of a functional $\Omega$ ($[\rho(\vec{r})]$, T, $\mu$) of the one-particle density $\rho(\vec{r})$ for the grandcanonical free energy which becomes minimal for the equilibrium density such that

$$\frac{\delta \Omega}{\delta \rho(\vec{r})} = 0$$

holds in equilibrium characterized by temperature T and chemical potential $\mu$. Unfortunately, for interacting particles this functional is not known exactly, but there are reliable approximations in particular for hard-core particles. For overdamped Brownian dynamics of passive colloidal particles, the *time-dependent* dynamical extension of classical DFT, so-called dynamical density functional theory (DDFT) was subsequently developed [3, 4]. In its simplest form, DDFT is a generalized diffusion equation (or continuity equation) for the time dependent one-particle density $\rho(\vec{r},t)$ driven by the density functional derivative

$$\frac{\partial \rho(\vec{r},t)}{\partial t} = D\vec{\nabla} \cdot \left( \rho(\vec{r},t)\vec{\nabla}\left(\frac{\delta \Omega}{\delta \rho(\vec{r},t)}\right) \right)$$

where D is the short-time diffusion coefficient of the colloids. For a noninteracting system (an ideal gas of Brownian particles) the standard diffusion equation is recovered. DDFT can be derived from the underlying many-body Smoluchowski equation invoking one additional (so-called "adiabatic") approximation [4]. The DDFT formalism was also generalized towards solvent-mediated hydrodynamic interactions between the passive colloids [5]. As compared to Brownian dynamics computer simulations, in general DDFT provides an accurate 'ab initio' (i.e. microscopic) theory and has been applied to various problems such as crystal growth as well as flow of colloidal fluids and crystals through constrictions  and over barriers. The predictive power of DDFT can also be exploited for microswimmers. First, DDFT was applied to describe dry active matter [6] and used to study cluster formation and swarming of self-propelled particles near a wall. One of the complications is the polar nature of self-propelled particles which requires the inclusion of additional orientational degrees of freedom for the particles in the functional and therefore more sophisticated one-particle densities $\rho(\vec{r}, \vec{u},t)$ that depend on both position $\vec{r}$ and orientations described by a unit vector $\vec{u}$.

## 2. Current challenges and future directions

More recent DDFT studies have also included the hydrodynamic flow fields of microswimmers (pushers or puller) into the formalism thereby constructing a particle-scale statistical theory of Brownian microswimmers which unifies thermal fluctuations, self-propulsion and hydrodynamic solvent flow fields. This was exemplified for linear microswimmers (i.e. for microswimmers which move straight on average) in 2016 by Menzel et al [7]. A closed theory - much more sophisticated than the DDFT for the dry active matter case - was proposed. One application was for microswimmers in a confining potential which due to hydrodynamic effects self-organize into a hydrodynamic fluid pump. An example is shown in Figure 1 for both pushers and pullers demonstrating that the hydrodynamic flow field of the microswimmers explicitly enters into the theoretical description. This behaviour is in agreement with simulations.

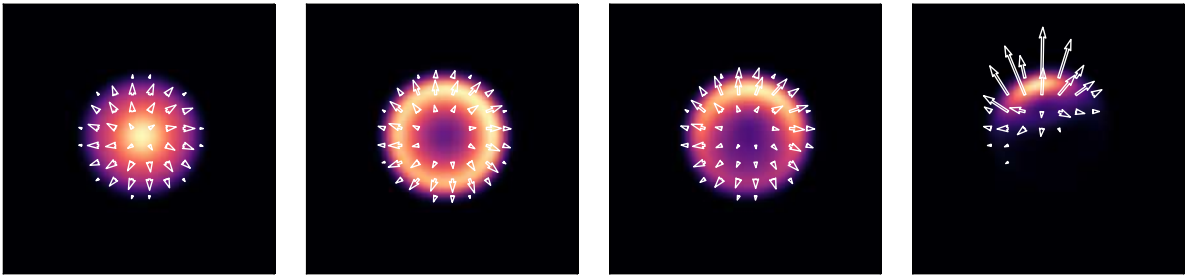

***Figure 1*** Time evolution of the density profiles (color maps) and orientation profiles (white arrows) for an ensemble of pusher microswimmers in an external trapping potential. Starting from an equilibrated non-active system, the active drive is switched on at time t=0. From left to right, snapshots at times t=0.05, 0.5, 1.5, 2.5 are shown. The pushers first form an outwardly-orientated high-density ring, which then collapses towards a single spot due to their mutual hydrodynamic interactions.

The formalism was subsequently applied to swimmers which propel in circles [8]. Under confinement circular swimming increases the localization in traps. Third, the spontaneous polar order in pusher and puller suspensions was also explored theoretically using DDFT [9]. An inherent evolution of polar order in planar systems of puller microswimmers was identified, if mutual alignment due to hydrodynamic interactions overcomes the thermal dealignment by rotational diffusion. Conversely, disordered pusher suspensions remain stable against homogeneous polar orientational ordering. Future problems lying ahead are numerous, we just mention of few of those here. First of all the formalism can be straightforwardly generalized towards multicomponent mixtures of microswimmers such as active-passive mixtures or that of pushers and pullers [10]. Second, as equilibrium DFT and DDFT describes the crystallization transition, DDFT provides also an appropriate framework to describe freezing of microswimmers at high densities. Third, while actual numerical implementations for DDFT were performed in two spatial dimensions (though the solvent flow field was kept three- dimensional as arising from an unbounded surrounding fluid), an extension towards the third dimension should be done which, however, will require much higher numerical efforts. Next, a big challenge is to include inertial effects as relevant for microflyers at higher Reynolds number and for self-propelled granular particles. This is challenging as already the DDFT for passive particles with underdamped dynamics has not been developed much so far since the additional particle momenta make the description much more complex. Finally, swimming in a viscoelastic solvent is relevant for many applications. Again DDFT will provide a framework for the collective properties of microswimmers to include memory effects of the solvent but this is far from straightforward.

## 3. Concluding remarks

The recently developed dynamical density functional theory for microswimmers is a predictive and powerful approach which unifies thermal effects, self-propulsion, interactions and hydrodynamics for pushers and pullers in a solvent of a Newtonian fluid at low Reynolds numbers. No other method so far has achieved this. Apart from further applications concerning microswimmers in confining geometry, this formalism should be extended in the future towards binary mixtures, towards freezing of concentrated suspensions of microswimmers and to viscoelastic solvents and underdamped dynamics.

## 4. Acknowledgements  I thank A. M. Menzel, C. Hoell and G. Pessot for the fruitful collaboration within the SPP 1726. This work was supported by the microswimmer SPP 1726 of the German Research Foundation (DFG) within the project LO418/17-2.

# Enzymes and Molecular Swimmers
Ramin Golestanian, MPIDS and Oxford University

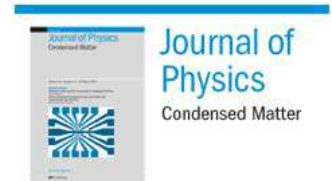

## Status

Enzymes are molecular machines that catalyse chemical reactions. The appropriate description of a chemical reaction is a Kramers (escape) process in the reaction space in which the system aims to go from an initial higher energy state, which corresponds to the reactant (or "substrate" as it is called in the biochemistry literature) to a final lower energy state, corresponding to the product, by overcoming an energy barrier. If the barrier is substantially larger than the thermal energy, the reaction will not take place often enough; however, when an enzyme comes along, it can lower the barrier in its own vicinity and usher the transition. Therefore, enzymes are drivers of non-equilibrium activity at the right time at the right place. In recent years, it has been observed experimentally that the non-equilibrium chemical activity of enzymes leads to intrinsically non-equilibrium behaviour and interactions already at the microscopic scale, and in particular substrate dependent enhanced diffusion [1] and chemotaxis towards [2] and away [3] from the substrate source. Enhanced diffusion and chemotaxis of enzymes are subtle non-equilibrium stochastic processes and need to be well understood (see Fig. 1).

There are several mechanisms that can contribute to enhanced diffusion of enzymes with varying degrees of significance [4]: (1) self-phoresis (due to self-generated chemical gradients), (2) boost in kinetic energy (caused by release of the energy of the reaction to the centre of mass translational degrees of freedom by equipartition), (3) stochastic swimming (due to cyclic stochastic conformational changes associated with the catalytic activity of the enzyme) [5,6,7], (4) collective heating (caused by treating enzymes as mobile sources of heat for exothermic reactions), and (5) modified equilibrium (due to the changes in the hydrodynamic couplings between the different modules of an enzyme while undergoing thermal fluctuations, caused by binding and unbinding of chemicals; see Fig. 2) [8]. The existence of multiple mechanisms highlights that it is perhaps misleading to think about enhanced diffusion of catalytically active enzymes as one phenomenon with universal features. Similarly, it has been argued that several mechanisms contribute to enzyme chemotaxis and that the different tendencies observed in the experiments can be explained as a competition between these contributions [9]. In particular, phoresis tends to direct enzymes towards their substrates, due to the predominantly attractive interaction between them, whereas enhanced diffusion tends to have the opposite effect as the enzymes will evacuate the region with higher substrate concentration more quickly due to enhanced diffusive activity.

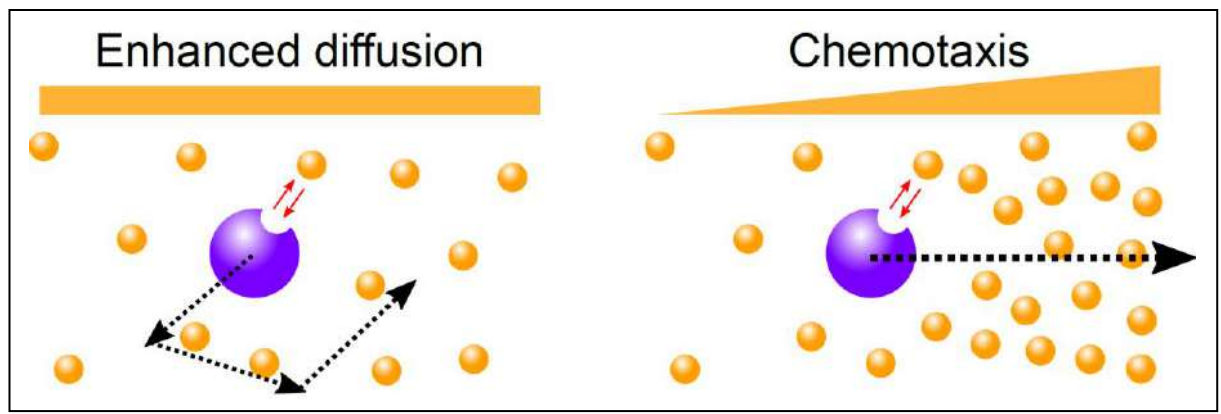

**Figure 1.** Schematic representation of enhanced diffusion and chemotaxis of catalytically active enzymes. Credit: Jaime Agudo-Canalejo.

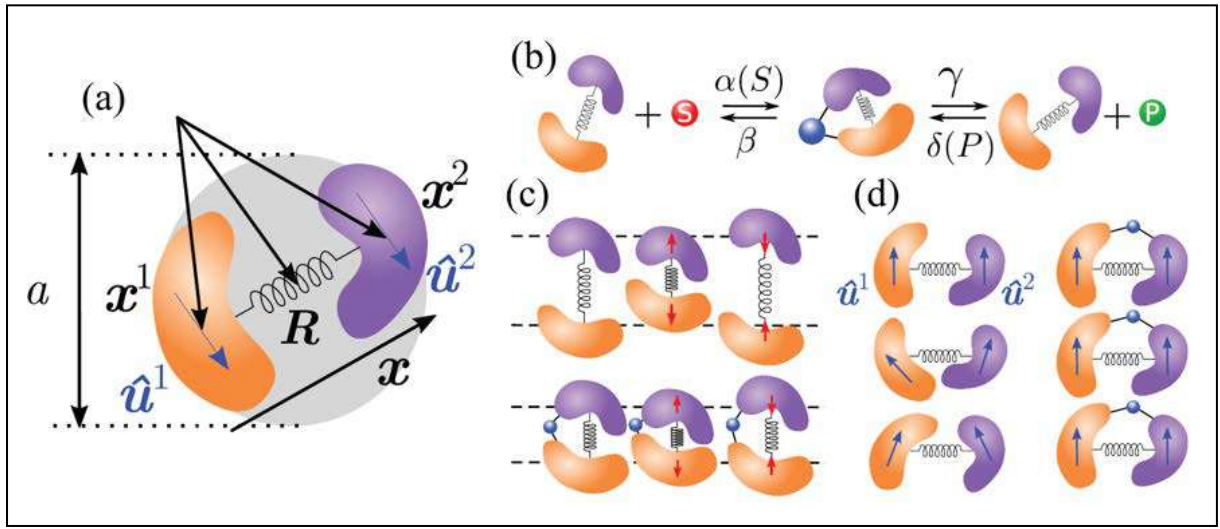

Figure 2. Modified Equilibrium Mechanism for Enhanced Diffusion. (a) Generalized dumbbell model with two subunits that represent the modular structure of an enzyme interacting via hydrodynamic interactions and a harmonic-like potential. (b) The mechano-chemical cycle explored by the enzyme in the presence of substrate and product molecules: when the enzyme is free, it can bind to a substrate molecule and transform it into a product molecule. These transformations are assumed to be reversible. (c) Modification of the extent of elongation fluctuations of the dumbbell due to substrate or product binding. (d) Modification of the orientational fluctuations upon substrate or product binding. Reproduced from [8]. Credit: Pierre Illien.

**Current and Future Challenges**

There are a number of unresolved issues and mechanistic questions regarding the existing reports on enzyme enhanced diffusion, including questions about artefacts that might arise when using fluorescence correlation spectroscopy (FCS) and the multimeric structure of some enzymes [10]. To resolve these issues, it will be important to use complementary experimental techniques. A comprehensive understanding of these mechanisms, albeit complex and technical, will be of utmost importance when studying the collective behaviour of active enzymes since many of the mechanisms involved in single-enzyme activity have the capacity to introduce long-range interactions between enzymes.

A general challenge in such experiments come from the non-equilibrium nature of the phenomena. If the experiments are performed on closed samples with an initial supply of the substrate molecules, the observations will be performed through a transient regime that ultimately relaxes to equilibrium. To create a stationary state for the substrate concentration level a microfluidic setup needs to be used with inlets and outlets. This might introduce additional effects to do with the existence of non-vanishing fluid drift in the experimental sample. One way to deal with this challenge is to perform theoretical analysis that takes into account such effects.

**Concluding Remarks**

Catalytically active enzymes exhibit remarkable non-equilibrium behaviour, including enhanced diffusion in the presence of substrate with uniform concentration profile and net directional movement in the presence of substrate gradient. These studies will naturally be extended to denser suspensions of enzymes and collective behaviour of active enzyme suspensions. Some preliminary work on collective behaviour of enzymatic systems has been done, showing great promise for exciting new directions.

**Acknowledgements**

The theoretical work on molecular swimming and nonequilibrium activity of enzymes has been performed jointly with Tunrayo Adeleke-Larodo, Jaime Agudo-Canalejo, Armand Ajdari, Pierre Illien, and Ali Najafi.

## Modeling flagellated microswimmer

Gerhard Gompper and Roland G. Winkler
Theoretical Soft Matter and Biophysics, Institute of
Complex Systems and Institute for Advanced Simulation,
Forschungszentrum Jülich, 52425 Jülich

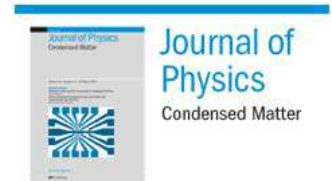

**Status**

**Swimming in Bulk and near Surfaces:** The majority of motile prokaryotes and eukaryotes exploit flagella for propulsion. However, the structure and active dynamics of their flagella differs substantially. Bacteria typically use one or several rotating helical flagella driven by a rotary motor in the cell wall, whereas eukaryotes employ beating flagella (or cilia) with a propagating deformation wave [1]. Since such cells are of micrometer size only, their dynamics is governed by low-Reynolds-number hydrodynamics, and viscous effects dominate over inertia.

Self-propulsion and fluid flow around flagellated microswimmers have been modeled on various levels of coarse-graining. The simplest and most generic approach is a force and rotlet dipole for E. coli and time-dependent force triplets for both Chlamydomonas and sperm [1,2]. Explicit models for flagellated microswimmer cells range from rigid bodies with an attached single filament taking the fluid into account implicitly by a hydrodynamic tensor [1], to cell bodies and filaments composed of point particles embedded in an explicit fluid [3,4]. Such studies emphasize the importance of hydrodynamic interactions for, e.g., the synchronization of bacteria flagella in the bundling process [5], the synchronized beating of Chlamydomonas flagella [6], or the formation of metachronal waves in cilia arrays [1]. Moreover, they illustrate the complexity of the hydrodynamic flow field adjacent to a cell, which is important for cell-cell scattering/interaction processes as well as swimming close to surfaces. Here, the nature of the equivalent force dipole (pusher/puller/neutral swimmer) determines the far-field hydrodynamics, and provides a guide for the type of interaction. However, ultimately the near-field hydrodynamics often dominates.

Bacteria often move in complex environments, which are rather viscoelastic than Newtonian. Such environments break the time-reversible symmetry of Newtonian fluids, and allow self-propulsion even for a time-symmetric internal motion, seemingly violating the scallop theorem. Controversial results on the swim speed in viscoelastic fluids have been found, suggesting a more or less significant increase of swimming speed. In addition, viscoelastic modifications of the run-and-tumble behavior have been observed [7].

Interactions of microswimmers with surfaces are fundamental in many biological processes, e.g., biofilm formation and egg fertilization. In the vicinity of surfaces, hydrodynamic interactions with surfaces and fluid flows generated by locomotion determine the swimmer orientation. Pushers, like E. coli and sperm, orient preferentially parallel to a surface, whereas pullers, like Chlamydomonas, align perpendicular to surfaces. For bacteria, the rotation of the helical bundle and the counter-rotation of the cell body lead to circular trajectories with a handedness and circle radius depending on the surface-slip length—CW trajectories follow for no-slip and CCW for perfect-slip boundary conditions [1,4,8] (Fig. 1, top). Interestingly, simulations show that the cell trajectory is sensitive to nanoscale changes in the surface slip length, being itself significantly larger. This can be exploited to direct bacterial motion on striped surfaces with different slip lengths, which implies a transformation of the circular motion into a snaking motion along the stripe boundaries [4].

Geometrical restrictions play a major role for microswimmers in microchannels and microfluidic devices. A new feature is here the presence of corners (for straight channels of rectangular cross section) or curved or sharp edges (for structured channels) [9]. These structural features are interesting because they facilitate the control of motility and cell sorting. An example is the swimming of sperm cells in periodic zigzag channels (Fig. 1, bottom) [9]. Several properties emerge: (i) In the straight segments, sperm swim along the side walls; (ii) this can happen with the (planar) flagellar beat plane perpendicular or parallel to the side wall; (iii) when the cell encounters a sharp edge, it is scattered away from the side wall—with a scattering angle depending on the beat-plane orientation (nearly straight motion for parallel orientation, deflection toward the originating side wall for perpendicular orientation). This behavior can be well understood from the steric interaction of the flagellar beat envelope with the side wall [9]. Finally, for curved edges, sperm is able to follow the wall if the curvature is sufficiently small.

**Collective Behavior:** Dense bacterial suspension on surfaces show an intriguing chaotic state of collective motion called 'active turbulence' [10]. As far as bacteria are concerned, simulations of the collective behavior for models with explicit flagella are extremely demanding in terms of computation time. This is particularly relevant for ``swarmer'' cells, which are characterized by long cell bodies and a large number (of order 100) rotating flagella [11]. Therefore, to study collective phenomena, simpler coarse-grained models are often applied, in particular a squirmer model. Squirmers are of spheroidal shape (typically spheres), with a prescribed surface velocity, which determines the swimming velocity and the flow field [12]. Hydrodynamic simulations of spherical squirmers show cluster formation in two dimensions (2D), in thin films with no-slip boundary conditions (quasi-2D), and in three dimensions. Figure 2 shows examples of motility-induced phase separation (MIPS) of non-hydrodynamic active Brownian particles (ABPs), cluster formation of spherical squirmers, and MIPS of spheroidal squirmers. For spherical squirmers, the formation of small clusters, rather than the appearance of MIPS, is attribute to changes in the orientational dynamics by interference of the flow fields of the individual squirmers [12]. In contrast, spheroidal squirmers exhibit MIPS already for rather small activities [12]. Hence, shape and hydrodynamics govern structure formation of active matter. Squirmer simulations have not shown any significant turbulent collective motion so far.

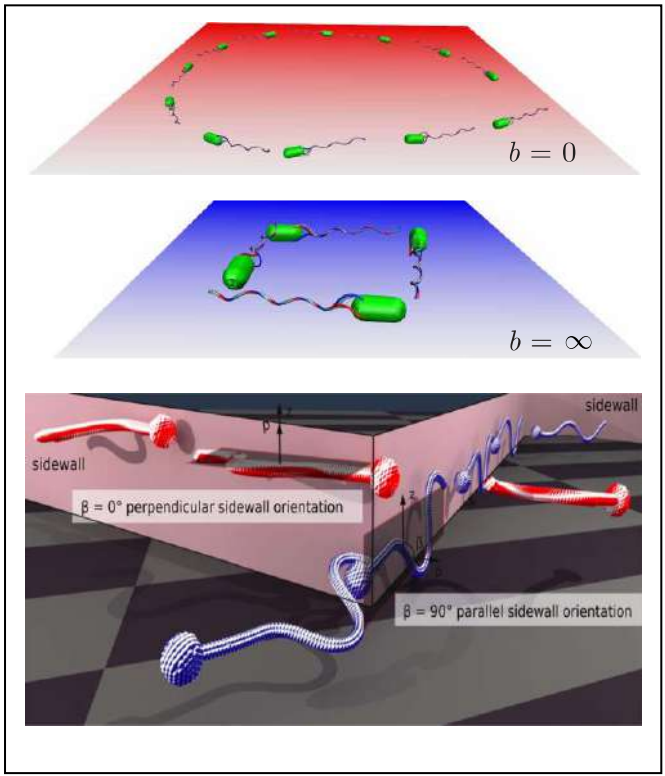

**Figure 1.** (Top) Clockwise and counterclockwise trajectories from hydrodynamic simulations of a bacterium swimming near homogeneous surfaces with no-slip (b = 0) and slip ( b = ∞) boundaries [4]. (Bottom) Overlay of a sperm cell swimming along the sidewall of a zigzag channel with parallel (blue) and perpendicular (red) beat-plane orientation to the sidewall [9].

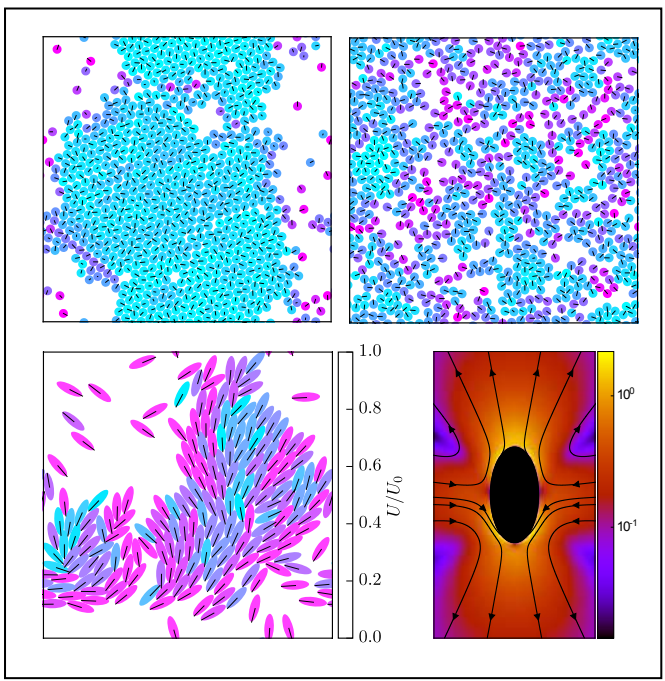

**Figure 2.**  Cluster formation in systems of self-propelled colloidal particles in quasi-2D (slit) [12]. The color code indicates the microswimmer velocity with respect to the propulsion velocity $U_0$. (Top left) Mobility-induced phase separation (MIPS) of active Brownian particles (ABPs) (no hydrodynamics). (Top right) Spherical squirmer show cluster formation, but no MIPS at the same density and Péclet number as for the ABPs (left) (Pe = 115, 2D packing fraction $\phi = 0.6$). (Bottom left) Elongated squirmers (spheroids, aspect ratio three) exhibit MIPS even at very low Péclet numbers, here, Pe=12, which is enhanced by hydrodynamic interactions in contrast to spheres. (Bottom right) Flow field of a spheroidal squirmer (pusher). The magnitude of the velocity field is color coded logarithmically.

**Current and Future Challenges**

In recent years, significant progress has been achieved in the understanding of the swimming mechanisms of flagellated microswimmers, and, more generally, the collective motion of active matter. Nevertheless, there remain many unresolved issues, and new questions have emerged.

- The explicit modeling of microswimmers, with cell body and flagella, is demanding in terms of simulation time, specifically in studies of their collective behavior. Here, questions on the appropriate level of modeling arise. The dipole description of microswimmers is obviously too simple in many cases, but what level of detail is required to capture the essential mechanisms? Does flagellar motion have to be modeled explicitly? Is it important to consider also flagellar elasticity and feedback on the motor forces?

- To what extent does the run-and-tumble motion of bacteria depend on the properties of the flagella? How is a complex environment, e.g., a viscoelastic fluid or a fluid with large objects (obstacles), affecting the run-and-tumble motion?

- Every microswimmer creates its specific, shape-dependent flow field. Simulation studies point toward an essential influence of near-field hydrodynamics on the collective motion of microswimmers. To which extent are collective swimming patterns universal or specific to particular microswimmers?

- Swarming of flagellated bacteria is an essential aspect of biofilm formation and develpoment, where inter-bacteria interactions, such as steric and hydrodynamic interactions, are fundamental. What is the importance of the observed morphological cell modifications? Specifically, in which way are the elongated and highly-flagellated cells propelled? What is the role of the interference of flagella of adjacent cells?

- Active turbulence is a fascinating phenomenon, which is observed experimentally in bacterial suspensions. Several theoretical models have been employed to explain this effect, from active Brownian rods (without hydrodynamics) to dipole microswimmers. So far, it is neither clear how active turbulence is best characterized, nor is the driving mechanism resolved for the various—dry versus wet— active systems. Other questions are: What is the role of the rotlet dipole? Is the phenomenon universal, like high-Reynolds-number hydrodynamic turbulence? What are the parameters to be tuned to reach the active-turbulent state?

- Flows, ubiquitous in microbial habitats, strongly modify the swimming behavior of microbes and lead, e.g., to shear-induced trapping and flow-driven accumulation at surfaces. Moreover, microswimmers modify the rheological properties of fluids. Here, studies are necessary to resolve the peculiarities of swimmer shape, coupling of propulsion and flow, and run-and-tumble motion of bacteria. This is paramount for transport of microbes in large-scale flows and the usage of microswimmers in applications as carriers or viscosity modifiers.

- Microswimmers usually accumulate at surfaces. The design of surfaces to enhance or avoid accumulation (biofouling), or to direct microswimmer motion in microfluidic devices, is therefore very important. In particular, geometrical structuring of surfaces is interesting. Can the combination of flow and surface design be employed for the sorting of microswimmers of different shape, propulsion strength, and circling motion?

- Flagellated propulsion is a universal mechanism, which does not depend on the type of embedding fluid, the density of swimmers, etc. Therefore, it is also an interesting construction principle for artificial microswimmers. This is mainly an engineering challenge, because miniature motors have to be constructed to drive the flagellar motion.

**Concluding Remarks**

Motion by flagellar propulsion is a universally employed concept by biological microorganisms. While the motion of individual swimmers is by now well understood, much remains to be discovered in the role of flagella in collective motion.

**Acknowledgements**

We want to thank J. Elgeti, T. Eisenstecken, J. Hu, K. Qi, S.-Y. Reigh, S. Rode, G. Saggiorato, and M. Theers for stimulating, fruitful, and enjoyable collaborations. Support of this work within the DFG priority program SPP 1726 on ''Microswimmers – From Single Particle Motion to Collective Behaviour'' is gratefully acknowledged.

**Sperm as microswimmers**
U. Benjamin Kaupp and Luis Alvarez
Molecular Sensory Systems
Center of Advanced European Studies and Research
Bonn, Germany

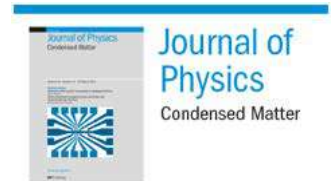

**Status**

Sperm have risen to one of the most useful model systems of biological microswimmers. Sperm carry a hair-like protrusion – called flagellum – that serves as antenna for chemical and physical cues in the environment, as motor that propels the cell, and as rudder that steers sperm in sensory landscapes. A sensory landscape or sensory field refers to a field of any chemical or physical stimulus that can be registered and processed by the cell. For steering in a sensory field, sperm modulate the asymmetry of the flagellar beat: When the flagellum beats symmetrical, sperm move on a straight path; when the beat becomes more asymmetrical, fluid is pushed aside, and the swimming path becomes curved. In all sperm species, the flagellar beat is modulated by the intracellular $Ca^{2+}$ concentration ($[Ca^{2+}]_i$) [1]. Sperm use various sensing mechanisms to gather physical or chemical cues to spot the egg [2]. In mammals, three mechanisms have been proposed that guide sperm through the narrow oviduct: chemotaxis, thermotaxis, and rheotaxis. In marine invertebrates, chemotaxis has been firmly established. During chemotaxis, sperm couple stimulation by chemoattractants to changes in $[Ca^{2+}]_i$ and, ultimately, flagellar beat asymmetry. In sea urchin, the best-understood model system, a cGMP-signaling pathway controls the opening of sperm-specific $Ca^{2+}$ channels in the flagellum (Fig. 1). A fascinating feature of sea urchin sperm is that a single chemoattractant molecule can elicit a $Ca^{2+}$ response. Such an elementary $Ca^{2+}$ response involves (1) the synthesis of about 10 cGMP molecules; (2) a change in the intracellular proton concentration ($\Delta pH_i$) of about 0.01 units; and (3) a voltage response of about 2-3 mV. Upon return to resting $V_m$ and at more alkaline $pH_i$, $Ca^{2+}$ channels open and sperm adjust their swimming path.

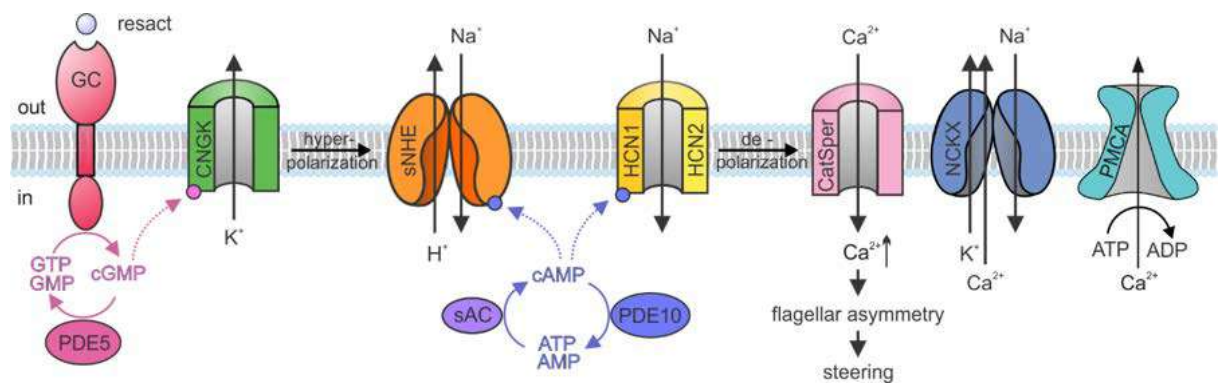

**Figure 1.** Signaling pathway in sea urchin sperm. The guanylate cyclase (GC) serves as receptor for the chemoattractant resact. Resact binding activates the GC, resulting in cGMP synthesis. cGMP activates a $K^+$ selective cyclic nucleotide-gated (CNGK) channel. Opening of CNGK hyperpolarizes the cell and activates a hyperpolarization-activated and cyclic nucleotide-gated (HCN) channel and a sperm-specific voltage-dependent $Na^+/H^+$ exchanger (sNHE). Opening of HCN channels restores the resting potential, whereas activation of sNHE increases the intracellular pH. Both events activate the $Ca^{2+}$ channel CatSper, leading to $Ca^{2+}$ influx. The $Ca^{2+}$ level is restored by $Ca^{2+}$ extrusion through a $Na^+/Ca^{2+}/K^+$ exchanger (NCKX) and a $Ca^{2+}$-ATPase PMCA; cGMP is hydrolyzed by a phosphodiesterase (PDE5). On resact binding, sea urchin sperm not only synthesize cGMP, but also cAMP, probably through activation of a soluble adenylate cyclase (sAC), which, at least in mouse sperm, forms a complex with the sNHE. Targets for cAMP are the HCN channel, whose voltage-dependent opening is modulated by cAMP, and the sNHE exchanger.

**Current and Future Challenges**

*Supramolecular organisation of the signalling pathway.* The sequence of signalling events shown in Fig. 1 is usually organised from left to right, from the receptor to the $Ca^{2+}$ channel. During forward signalling, several biochemical reactions are triggered that initiate the recovery from stimulation. Such feedback mechanisms include the hydrolysis of cGMP and extrusion of $Ca^{2+}$ from the cytosol. Thus, to depict signalling as a relay race, where the baton is conveyed from one stage to the other, is too simple. Probably, signalling components form molecular or functional complexes. To achieve full quantitative understanding, three pieces of information are required. First, the concentrations and stoichiometric ratios of signalling components must be known. In principle, this can be accomplished by quantitative mass spectrometry. Second, a great challenge will be to identify potential complexes among signalling molecules. Finally, future work needs to determine the kinetics and properties of signalling proteins in quantitative terms. Only then, robust quantitative network models can be built that reach beyond guesswork.

*Differences among species.* Sperm preferentially adopt unique, sperm-specific signalling molecules and pathways that differ from signalling in sensory cells and somatic cells in general [3]. Many of the signalling molecules are also species-specific; the differences among the molecules are not restricted to trivial details. Below the masquerade of seemingly similar faces (i.e. amino-acid sequences), signalling molecules adopt new structures, mechanisms, and functions. Thus, although $Ca^{2+}$ orchestrates motility in flagellated sperm, the molecules and signalling pathways that control $Ca^{2+}$ entry into sperm are diverse across phyla. Understanding the commonalities and differences of sperm signalling will be a daunting task, considering the diversity in sperm shapes, fertilisation cites, ionic milieus, hydrodynamics, and motility patterns. For example, sperm from marine external fertilisers operate in seawater with an osmolarity of about 1,100 mOsm, whereas fish living in lakes and rivers spawn into freshwater of only a few mOsm. Future work needs to explain the different sperm properties in the light of the respective habitat, reproductive organ, or reproduction strategy.

*How does the 3D flagellar beat control the swimming path?* Due to technical limitations, sperm chemotaxis has primarily been studied in two dimensions (2D). While swimming in shallow observation chambers, sperm move towards the glass/water interface, where they accumulate due to steric and hydrodynamic interactions with the walls [2]. Accumulation facilitates the observation of cell movement in the focal plane. While swimming in a chemical gradient of chemoattractant, sea urchin sperm alternate between episodes of higher and lower asymmetry of the flagellar beat. As a result, sperm move on drifting circles up the gradient [4]. Inspired by this periodic swimming pattern, a theory was developed that captures the essence of chemotactic navigation [5]. A cellular signalling system transforms the periodic stimulation $s(t)$ (chemoattractant binding) into a periodic intracellular signal $i(t)$ that in turn produces a periodic modulation of the swimming path curvature $\kappa(t)$. The resulting looping path guides sperm up or down the gradient, depending on the phase relation between $s(t)$ and $\kappa(t)$. The phase relation is determined by the latency of the $Ca^{2+}$ response and the ensuing motility response. The swimming path of unrestricted sperm was tracked in 3D using digital inline high-speed holographic microscopy [6]. Freely swimming sperm move along helical paths. The flagellum beats almost in a plane that slowly rotates around the helix axis; a full plane rotation is achieved after completion of a helix period. The principles of chemotactic steering on a helical path were studied in 3D chemoattractant gradients, which were established by photolysis of caged chemoattractants. The chemoattractant concentration field can be predicted at any point in space and time; thereby, swimming behaviours can be linked to the spatiotemporal pattern of the chemotactic stimulus. In a gradient, sperm adjust their swimming path by two different types of responses: gradual smooth alignment of the helical axis with the gradient and abrupt turns. Irrespective of the response strength, sperm align the helical axis with the chemical gradient. Thus, in contrast to bacteria that navigate using a stochastic strategy, sperm swim with high directional persistence and follow a navigation strategy that is deterministic [2,6].

Many challenges are still ahead for a full understanding of sperm navigation. An important task will be to record the 3D flagellar beat with high spatiotemporal precision and to reconstruct the 3D trajectory from the beat. Furthermore, it would be interesting to understand how molecular motors are synchronised or sequentially activated to produce flagellar beating waves, and which mechanisms produce different beat patterns for steering. Progress in this direction was provided by a recent electron microscopy study of the flagellar axoneme [7].

*Recursive interactions between sensory field, cellular signalling, and behavioural response.* Sperm, and cells in general are exposed to multiple chemical and physical environmental cues that initiate a cellular and ultimately, a behavioural response. It has been proposed that sperm rely on fields of fluid flow, chemicals, and temperature for navigation [2]. A grand aim of biology is to understand the transfer function between stimulus input $s(t)$ and response output (for sperm the path curvature $\kappa(t)$) in quantitative, molecular, and mechanistic terms. This endeavour is challenging because cells encounter a rich, spatially complex, and highly dynamic environment composed of puffs, plumes, ramps, periodic cycles, or voids of diverse stimuli. Naturalistic sensory field, if known at all, are difficult to emulate in the laboratory. The complexity is further enhanced in motile cells (microswimmers), because the trajectory $\mathbf{r}(t)$ shapes a cell's "perception", that is the stimulus function $s(t)$ acquired while cruising in a sensory field. This concept is known as information self-structuring [8]. The ultimate goal for sensory microswimmers is to map in real time the recursive interactions between a sensory landscape, the encoding biochemical circuits, and ultimately locomotion (Fig. 2).

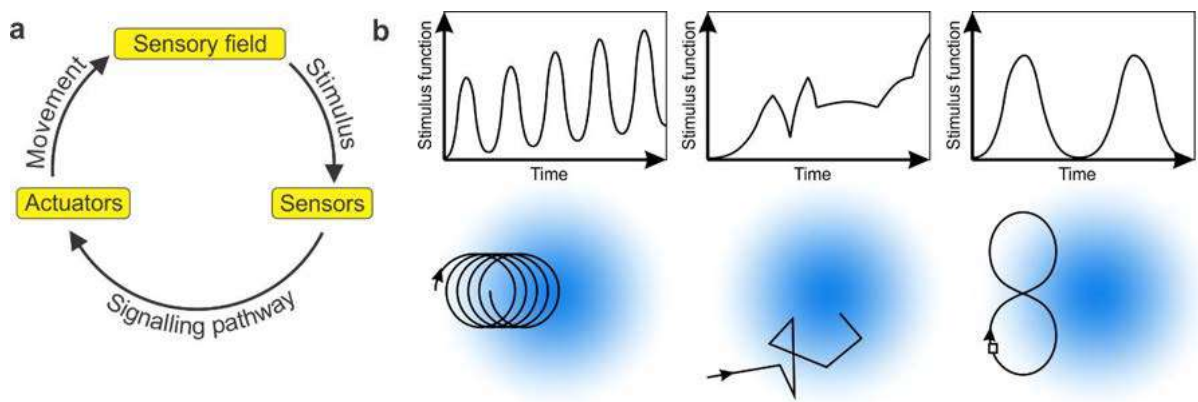

**Figure 2.** Recursive interactions between a sensory landscape, cellular signalling, and the resulting motion trajectory. **a)** The sensory-motor loop. Sensory cells capture information from their environment. The stimulus registered by the cell sensors is translated by a signalling pathway into a motor output. As the cell moves in the sensory field, the stimulus is also altered. Thus, the stimulus function is not only given by the sensory field, but also by the active field exploration executed by the cell. **b)** Three examples of variations in sensory stimulus (top) that result from alternative exploration paths of the same stimulus field by a cell (bottom). The stimulus field is shown in blue shades.

**Concluding Remarks**

A powerful approach for probing a cell's percept combines time-varying perturbations with signalling read-outs from live cells. Genetically encoded light-sensitive ion channels or enzymes (optogenetics) and chemical photo-triggers or photo-switches [9] are most popular for probing cell signalling. We envisage that, in future work, photonic techniques will allow deciphering rapid cellular computations and the underlying molecular mechanisms during signalling and locomotion in full quantitative terms. The computational principles gathered from sperm navigation might inspire the design of artificial microswimmer robots, remote control of motile cells and organisms, or artificial intelligence concepts in general.

**Acknowledgements**

This work was supported by the German Research Foundation DFG; SPP 1726 Microswimmers.

**How flagellated protists feed.**

*Thomas Kiørboe, Centre for Ocean Life, DTU-Aqua, Technical University of Denmark*

# Status

Many aquatic unicellular protists are equipped with flagella that facilitate not only the motility of the cells but – maybe more important – their resource acquisition. These flagellates can be auto- or heterotrophic, i.e., acquire resources through photosynthesis and uptake of dissolved mineral nutrients or by foraging on prey, respectively, but many species are indeed mixotrophic, i.e., can acquire resources in both ways. Their main prey is submicron sized bacteria and small photosynthetic cells. The action of the flagella create flows past the cell that, one way or another, enhances resource acquisition. These flows, however, also make the cells susceptible to rheotactic (flow sensing) predators, and there is, therefore, a conflict – a trade-off - between resource acquisition and survival. The number and arrangement of flagella as well as flagella beat patterns and kinematics in free-living protists must have evolved to optimize the trade-off between resource acquisition (near field flow) and survival (far field flow). Together with environmental conditions this fundamental trade-off is a main determinant of the structure and function of microbial communities. In the ocean, microbial communities play a pivotal role in marine pelagic food webs and for ocean biogeochemistry and global carbon budgets through the flagellates photosynthetic activity and consumption of bacteria and phytoplankton, and by flagellates being consumed by microzooplankton and thus making energy and carbon available to higher trophic levels. A mechanistic understanding of this fundamental trade-off is, therefore, essential to understand and model the role of microbes for ocean biogeochemistry.

The action of cilia and flagella in unicellular organisms are most often studied in the context of propulsion and steering [1]. However, these organelles play a key role in resource acquisition: the beating flagella create swimming and feeding currents that facilitate both the advective enhancement of diffusive uptake of dissolved nutrients and, in particular, the encounter of prey, and the flagella may also be directly involved in capture and handling of prey. The pioneering work of Fenchel [2] progressed our understanding of the functional ecology of free-living, single celled aquatic protists significantly, and until recently very little has be done to further explore the issue [3]. Fenchel described interception feeding protists that encounter prey as the flagellate swims through the water, or microbial filter feeders that force water through a filter where prey particles are strained, but his account of the fluid dynamics of feeding was incomplete. A fundamental problem in encountering prey is that viscosity impedes predator-prey contact at the low Reynolds number at which protists operate, yet heterotrophic flagellates are capable of clearing huge volumes of water for prey, typically corresponding to $10^6$ times their own body volume per day for microscopic prey. Commonly used simple fluid dynamical point-force models, where the flagellate is modelled as a sphere and the action of the flagella as point forces, typically underestimate observed clearance rates by orders of magnitude [4]. This is likely because such models do not describe near-cell flow fields with sufficient accuracy, and does not account for hydrodynamic interaction between multiple flagella and the cell body and the physics of prey encounter is consequently poorly understood. Therefore, the key process of resource acquisition remains unexplored for most forms.

The simple point-force type models often used to describe self-propelled micro swimmers may be better suited to describe the far field flow disturbance generated by these organisms and, hence assess the predation risk that they experience while feeding. Depending on the position of the force(s) generated by the beating flagella and the kinematics of power strokes, stokeslet, stresslet, impulsive stresslet, or quadropole models captures far field flows relatively well [5], and predict predation risks with surprising accuracy, at least in slightly larger forms[6].

## Current and future challenges

Recent advances in microscale particle tracking and particle image velocimetry (μ-PIV) have allowed the quantification of near-cell flow fields in swimming flagellates [7] and, most recently, the description of feeding flows generated by some flagellates [8]. The latter have demonstrated the existence of feeding flows that are sufficient to account for observed clearance rates and that are dedicated to enhance prey encounter and solute uptake rather than propulsion. In mixotrophic dinoflagellates, for example, this flow is generated by the combined action of two flagella, one trailing behind the swimming cell, and one placed in a grove around the equator of the cell. The latter flagellum is further embedded in a 'sock' and thus drives an undulating sheet, and is additionally equipped with hairs. The joint action of these very specialized flagella both propel and steer the cell though the water and pull streamlines close to the cell at the point where prey is contacted. The generated flow also enhances the diffusive uptake of nutrients by a factor of about 2 (over that of uptake by pure diffusion alone). The details of the fluid dynamics is, however, unresolved, and may be clarified only through the application of CFD.

Choanoflagellates offer another example of specialized structures required to acquire resources. Choanoflagellates have attracted considerable attention because they are the presumed ancestor of multicellular life, and the cell type still exists, in modified form, in the kidneys of mammals. The cell has a single flagellum that is surrounded by a 'collar' of closely spaced filter stands that extends from the cell. The flagellum drives a current through the collar filter, where bacterial prey is retained. The feeding flow can be accurately described using particle tracking. Both simple analytical models as well as detailed CFD suggest that the power produced by the flagellum can drive water through the collar filter, but only at a rate that is 1-2 orders of magnitude lower than what can be observed and measured [9]. This discrepancy between model and observation can only be reconciled if one assumes that the flagellum drives a vane, which, however, has never been observed. A vane is well known form the choanocytes of the closely related sponges.

The above is just two examples of the large diversity in flagellar arrangements that exists among unicellular protists (Fig. 1) and the consequent very different prey encounter mechanisms. The number of flagella varies among species between one, two, multiples of 2

(2, 4, 8, 16), and many (the latter are called ciliates), their beat patterns and kinematics vary, and the flagella can be naked or equipped with hairs or vanes. And many flagellates have multiple swimming gaits. In addition, in forms with multiple flagella, their kinematics may be coupled via hydrodynamic interaction or otherwise [10].This large diversity has evolved through natural selection to facilitate swimming and to optimize the trade-off between resource acquisition and survival. Obviously, the differences in flagellation and kinematics will produce different flow fields and, presumably, consequent differences in the efficiency of resource acquisition and in exposure to predators. For most of these forms, however, neither is known. Specifically, the actual prey encounter mechanism, i.e., how flows created by the beating flagella overcome the impeding effects of viscosity and bring the flagellate into contact with its prey, remains elusive. Also, the role of the flagella in handling encountered prey is unknown. Preliminary observations suggest that the flagella can maneuver prey to ensure phagocytosis, that different flagella on the same cell have different functions in this process, and that the flagella are involved in deciding whether a particular prey is selected or rejected [3].   Detailed microscopic observations and high speed cinematography together with particle tracking and μ-PIV are possible tools to achieve quantitative information, and CFD a possible way to model and understand prey encounter mechanisms (Fig. 2).

## Conclusions

The evolution of flagellation and kinematics and structural details of the flagella is governed not only by the need to swim, but also by the needs to eat and not be eaten. Ocean biogeochemistry is governed mainly by the smallest organisms in the ocean, and the combination and function of pelagic microbial communities are governed to a large extent by trade-offs, first and foremost the trade-off between resource acquisition and survival. Even though my overarching question and motivation is of an ecological nature, we need to get down into the details of the fluid physics in order to mechanistically quantify the magnitude and shape of this trade-off for the main functional forms in pelagic systems. Once we understand the mechanism we can forget about the nerdy details and simplify the trade-off functions in a meaningful way that is also suitable for inclusion in ecological models.

## Acknowledgements

Financial support was received from the Danish Council for Independent Research (grant # 7014-00033B). The Centre for Ocean Life is supported by the Villum Foundation.

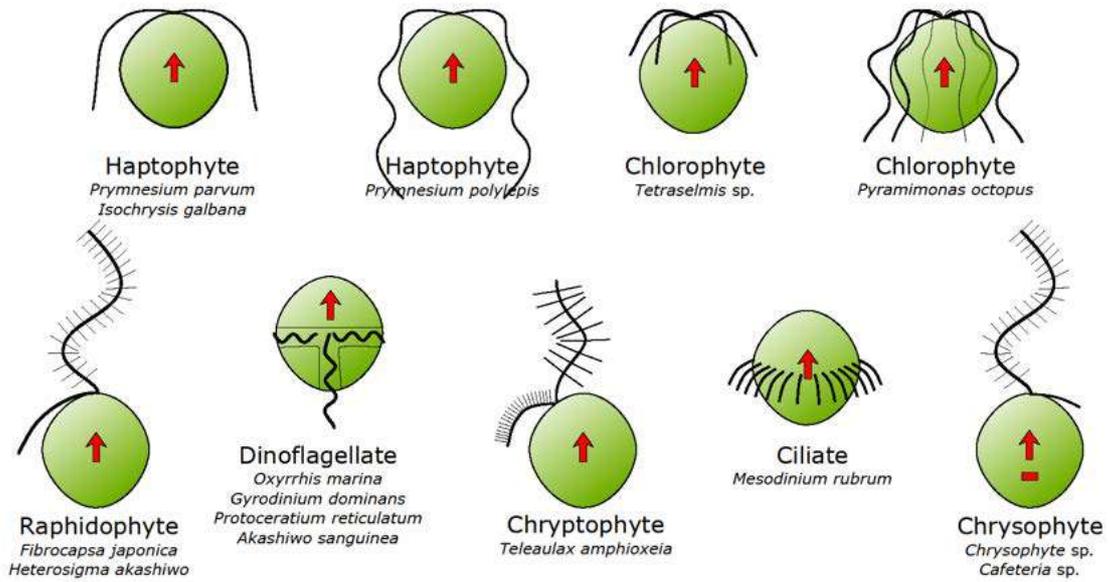

Fig. 1. Schematic illustrating some of the diversity in flagellar arrangement among marine protists. Artwork by Lasse Tor Nielsen.

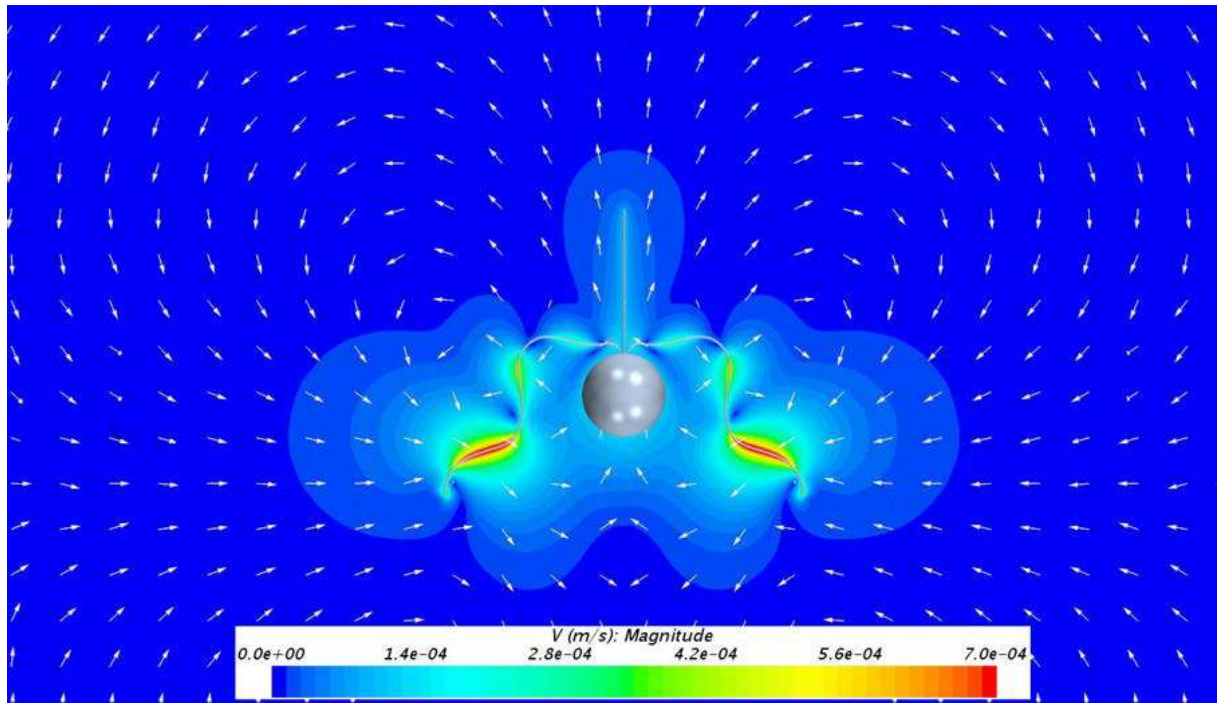

Fig. 2. Snapshot of flow field generated by swimming haptophyte, *Prymnesium polylepis*, computed by CFD. The simulation is based on the observed kinematics and beat pattern of the two flagella. The straight, slender structure in front of the cell is the haptonema (not a flagellum) that is involved in prey capture. Simulation by Seyed Saeed Asadzadeh (DTU).

# Swimming bacteria
## Eric Lauga, DAMTP, University of Cambridge

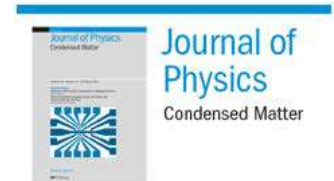

**Status**

The most primitive of all microscopic swimmers are bacteria, the largest group of prokaryotic organisms in biology. These single-celled organisms without a cell nucleus have diverged on the evolutionary tree from the eukaryotic domain a few billion years ago. Bacteria constitute the bulk of the biomass of our planet and not only are they responsible for many infectious diseases, they also play a critical role in the life of higher organisms by performing chemical reactions and providing nutrients. The behaviour of bacteria is strongly influenced by the physical constraints of their habitat and the most important of these constraints is the presence of a surrounding fluid.

Despite being far from higher organisms such as animals and plants in an evolutionary sense, prokaryotes swim in viscous fluids by exploiting similar physical principles to those used by eukaryotes: they actuate slender filaments in order to take advantage of the anisotropic drag they experience in viscous fluids. In the case of swimming bacteria, the slender filaments have the shapes of perfect helices. A few microns in length, the semi-rigid helices are rotated by motors, which are embedded in the bacterial cell wall. These rotary motors operate with approximately constant torque. Each motor transmit its rotation to that of the associated helical filament using a universal joint in the form of a short flexible polymeric beam called the hook, which is tens of nanometres long. The group made up of the motor with its hook and filament is called a flagellum (see Fig. 1, left). Some swimming bacteria are monotrichous and swim by actuating a single flagellum, while others are peritrichous and have multiple flagella distributed over their surface.

Most of what we know about the physics of swimming bacteria has been discovered via the peritrichous model organism *Escherichia coli* (*E. coli*, see Fig. 1, right) [1]. Equipped, on average, with four flagella distributed approximately randomly on its cell body, *E. coli* cells explore their environment using a run-and-tumble motion. During runs (approximately one second long), flagellar filaments gather in a helical bundle behind the cell and propel the cell forward. Runs are interrupted by quick (tenth of a second) tumbles where the counter-rotation of some rotary motors leads to a disruption of the bundle, and reorientation of the cell. By modulating the duration of runs in response to their environment, cells can perform chemotaxis and move up (or down) chemical gradients.

Microbiology is a field with a long history and, given the context of the worldwide consequences of infectious diseases, it remains an active area of research. But beyond biological aspects, swimming bacteria have also long been of interest to the physics community. Much effort from the biological physics side has focused on understanding and quantifying the behaviour of cells in fluid environments. The invention of the tracking microscope in the 1970s was a major breakthrough in the field, which has allowed to follow individual cells as they perform run-and-tumble motion and to understand stochastic cell dynamics as the basis for chemotaxis [2]. A second important development came in the 2000s with the discovery of a fluorescent staining method enabling the dynamics of flagella on individual cells to be observed in real time (Fig. 1, right) [3].

One particular aspect of the physics of bacteria that has received considerable attention is the dynamics of the fluid flow created by the swimming cells [4,5]. Indeed, stresses created by the fluid set in motion are at the heart of force generation by the rotating flagella. Furthermore, swimming bacteria are influenced by the flow fields created by neighbouring cells and by their immediate environment (such as the presence of boundaries). Therefore, and unsurprisingly, hydrodynamic interactions have been shown to affect the dynamics of cells in complex and crowded situations.

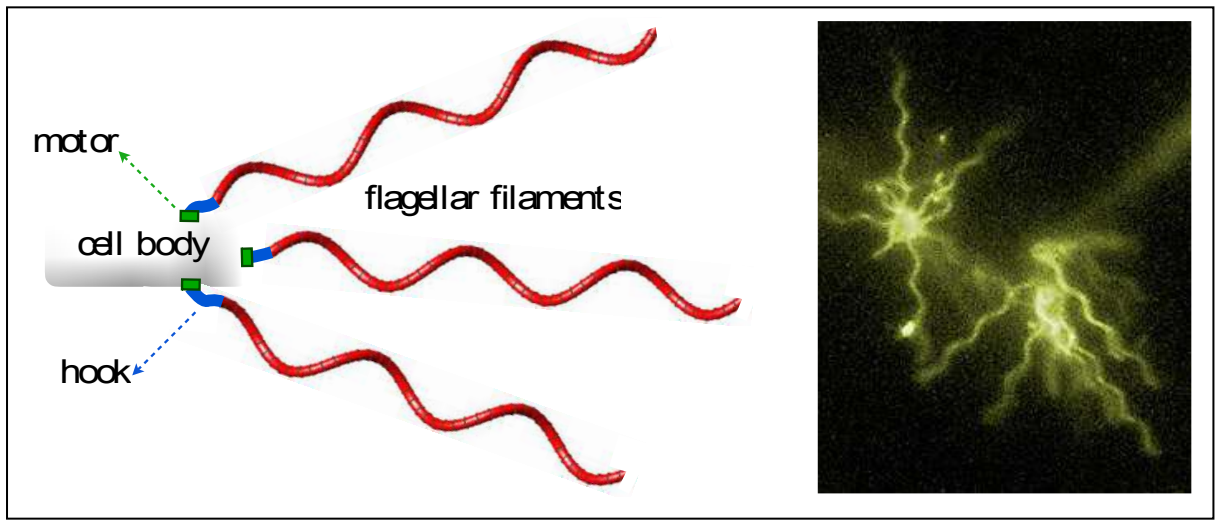

**Figure 1. Left:** Sketch of a swimming bacterium equipped with three flagella. Each motor rotates a hook, which in turn rotates a helical flagellar filament, leading to forward propulsion of the cell in fluid environments. **Right:** *Escherichia coli* as a model organism for swimming bacteria. Pictured are two cells, each equipped with multiple flagellar filaments visualised using fluorescent staining. Photo credit: Howard Berg, Rowland Institute, Harvard University, reproduced with permission.

**Current and Future Challenges**

The principal difficulty in deriving rigorous models for the swimming of flagellated bacteria is the nonlinear nature of the underlying external physics. Indeed, it involves nonlocal hydrodynamic interactions between flagella, short-range steric and electrostatic interactions, and elastic deformations of the flagellar filaments, which not only bend and twist but also undergo conformational changes [5]. Although the list of open questions in the domain is numerous, we highlight here three specific topics at the level of individual swimming cells for which a modelling approach combining analysis and computations will enable important progress.

Between the end of a tumble and the next run, flagellar filaments go from being oriented in random directions to all gathering on one side of the cell. An outstanding question remains regarding the exact role played by hydrodynamic interactions in this process. Similar to the dynamics of long hair behind the head of a swimmer at the pool, flagellar assembly and wrapping can take place passively without the need for these interactions. In this case, the filaments are dragged behind the cell body as the bacterium swims and they passively gather behind the cell since the hooks are flexible [7]. Furthermore, since the flagella rotate, the cell body counter-rotates, which also leads passively to a wrapping of the flagella around each other. However, long-range hydrodynamic interactions also lead to attraction. Indeed, the propulsive force created by each flagellum pushes fluid away from the cell body and thus create attractive flows driven along the cell wall. These flows then result in flagellum-flagellum attraction. Similarly, the swirling flows created by flagellar rotation lead to flagellar wrapping. Future mathematical modelling will be required in order to unravel the fundamental aspects of active vs. passive interactions at the heart of bacterial bundling.

When the bacteria are swimming, all flagellar filaments are orientated and gathered on one side of the cell. Since the flows induced by flagella have a slow, ~1/r spatial decay, we need to incorporate the impact of all nonlocal hydrodynamic interactions in order to understand the force generated by the rotation of the synchronized helical bundle. This has typically involved formidable computational efforts. Recent preliminary work has shown, however, that one can take advantage of the separation of length scales to mathematically evaluate these hydrodynamic interactions. Indeed, three different length scales characterize a flagellar filament: (i) its cross-sectional radius (tens of nanometres), (ii) the separation between the flagella, which is set geometrically by the typical helical amplitude (hundreds of nanometres) and (iii) the length of the filaments themselves (a few microns).

We thus see a clear separation between the three length scales, which in turn implies that we can represent the flows induced by the translation and rotation of the flagella using line distributions of flow singularities interacting locally as if they are present in infinite filaments [8]. The implication of this scale separation is far-reaching. It means that nonlocal hydrodynamic interactions can be integrated out and, therefore, in the future we will be able to capture the hydrodynamics of flagellar bundles mathematically.

A final topic of recent interest concerns the solid mechanics of the hook. During a run, the filament of each flagellum is pushing on the hook which itself pushes on the cell, and the hook is thus under compression. An elastic beam under compression is known to buckle if the compression exceeds a critical value (Euler buckling). In fact, some monotrichous marine bacteria equipped with a single flagellum exploit buckling of the hook to induce cell reorientation [9]. The hooks of peritrichous bacteria are known to be sufficiently flexible to enable an elasto-hydrodynamic instability from no swimming to locomotion [7] but no buckling has been reported for peritrichous cells under seemingly identical propulsive conditions to those of monotrichous bacteria. Is there something fundamentally different about spatially distributed motors in terms of structural stability? Do hydrodynamics forces in the bundle provide a source of stability?

**Concluding Remarks**

New high-fidelity capabilities in experimental methods give us the ability to gain ever-growing detail on the dynamics of swimming bacteria, thus necessitating the need for a new wave of interdisciplinary mathematical modelling. In this short overview, we focused on questions at the level of a single cell, but countless problems in the interactions of cells with their environments remain poorly understood: from the ability of bacteria to swim in complex, structured fluids and interacting with obstacles, to their collective modes of locomotion and their adaptation to chemical changes. Even in the absence of swimming, multiple microbiological systems present modelling challenges, in particular biofilms, which are communities of surface-bound bacteria embedded in an extracellular matrix (see Fig. 2) [10]. The life of bacteria in fluids still has much room to amaze us.

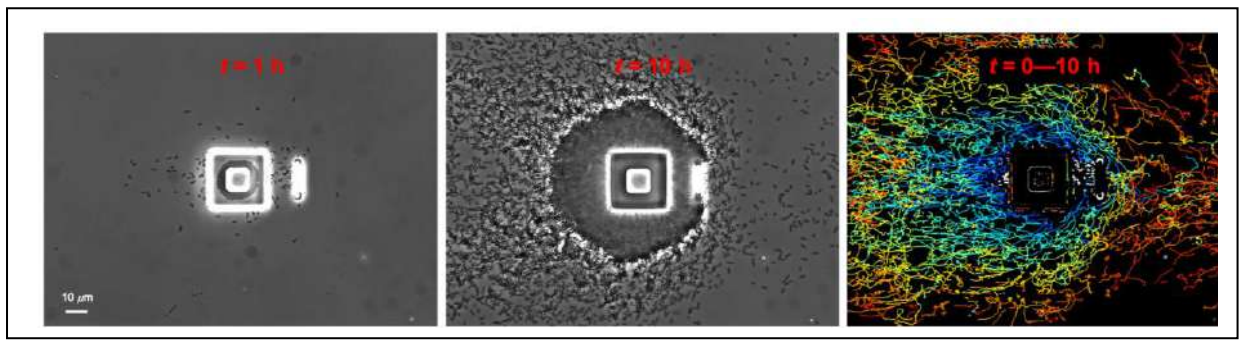

**Figure 2.** Submerged biofilm of the human pathogen *Pseudomonas aeruginosa* grown in the centre of a microfluidic setup. Left: cells at t=1 h; Middle: biofilm at t=10 h. Right: trajectories of biofilm cells where colour denotes cell speed (warm colours indicate higher mean cell speed) showing that cells speed up as the biofilm expands. Photo credit: Nuno Miguel Oliveira, DAMTP University of Cambridge, reproduced with permission.

**Acknowledgements**

*I thank Nuno Miguel Oliveira for help with Fig 2. Funding from the European Research Council (ERC) under the European Union's Horizon 2020 research and innovation programme (grant agreement 682754 to EL) is gratefully acknowledged. I thank Alex Chamolly, Debasish Das, Hélène de Maleprade, Christian Esparza López, Michael Gomez, Nuno Miguel Oliveira and Maria Tătulea-Codrean for feedback on an early draft of this paper.*

# Establishing model active colloids

**Wilson C. K. Poon**

SUPA and School of Physics & Astronomy, The University of Edinburgh, Peter Guthrie Tait Road, Edinburgh EH9 3FD, UK

E-mail: `w.poon@ed.ac.uk`

23 July 2019

**Abstract.** I argue for the importance of widely-adopted model systems for active matter research, assess how close motile *Escherichia coli* bacteria are to becoming a model active colloid, and point out some remaining issues over which caution needs to be exercised in assessing published data to date using these bacteria.

## 1. Status

Progress in any new field of 'small-science' physics requires good experimental model systems. The paucity of good model systems constitutes a bottleneck in current active matter research. Few, if any, experimental system have yet gained widespread acceptance. As a host of interesting phenomena are discovered, the lack of common models renders it is hard to distinguish generic from system-specific features. Confronting theory and simulations is hampered because many experimental parameters remain unknown.

Microtubules-kinesin mixtures [1] can be prepared reproducibly and are widely used as model extensile active gels, with many parameters already known. The situation is less satisfactory in active colloids. A model passive colloids requires reproducibility, known particle size/shape distributions, quantifiable and 'tuneable' inter-particle interactions and accurate particle concentrations. For an active colloid [2], we must additionally understand the propulsion mechanisms and motility-specific interactions (hydrodynamics, chemical fields, . . . ). A constant but tuneable swimming speed is also desirable.

At first sight, self-propelled Janus particles should make ideal model active colloids. Since the original suggestion and first demonstration, many related systems have been synthesised. The possibility of sphericity is a big attraction, as is the absence of any biology. A high-throughput method now exists for characterising their swimming [3].

Light-activation based on embedding hematite [4] or other physics [5] adds to the appeal. However, other features are troubling.

First, many aspects of the propulsion mechanism remain poorly understood [6]. Secondly, bulk experiments are difficult. The dense metal coating causes sedimentation; in other cases, gravitaxis causes particles to accumulate at the top. The proximity of surfaces further complicates swimming mechanisms and inter-particle interactions. Finally, the rapidity with which such particles consume fuel ($H_2O_2$, etc.) requires a 3D reservoir to sustain a finite-concentration 2D system. While small 3D systems can be achieved using special arrangements [7], this situation is not ideal.

To overcome some of these difficulties, one could use light-powered Janus swimmers [4, 5]. Indeed, such particles may yet become a good model as more groups learn to handle and use them. Perhaps surprisingly, another possibility is to use motile *Escherichia coli* bacteria.

The reader should consult a practical introduction to *E. coli* as active colloids [8] for details and references. Cells behave as hard spherocylinders.‡ Length polydispersity can be minimised reproducibly by rigorous control of growth and harvesting. Differential dynamic microscopy (DDM) gives the 3D speed distribution of $\sim 10^4$ cells in *circa* 2 min and the fraction of non-motile cells (minimisable by careful preparation).

Suspended in phosphate motility buffer (PMB), *E. coli* does not grow, and swims using internal resources, obviating the 'fuel problem'. There is little evidence of interaction via chemical fields due to nutrient depletion or active chemical communication, at least not over a few hours in PMB. Nutrients can be added to tune the speed to a limited extent. A more versatile method is to use light-activated *E. coli* whose speed increases with illumination intensity up to some saturation level [10]. Motile cells can be trapped in circular orbits at surfaces; but they do escape, and many remain in the bulk for 3D experiments.

The swimming mechanism is well understood, from individual rotary molecular motors to the hydrodynamics of flagellated propulsion. Indeed, to date, how a single *E. coli* swims is probably less controversial than how a Janus colloid self propels! Fitting the known flow field around a single motile cell gives the propulsive force dipole [11].

Mutants are widely available (and authors are obliged to share them). Wild-type cells change direction by rapid tumbles, but non-tumbling 'smooth swimmers' exist, mirroring two paradigmatic classes of micro-swimmers: run-and-tumblers and active Brownian particles. The original light-activated *E. coli* stops sluggishly at light off; deleting the $F_1F_0$-ATPase complex reduces the stop time to $< 0.1\,\mathrm{s}$ [10].

There is therefore 'a lot to like' about *E. coli*. A significant body of work now exists using it to study active matter physics. This literature is fast reaching the point of mutual correction, validation and reinforcement. One example is the ongoing attempt to understand how flagellated bacteria swim in high-molecular-weight polymer solutions, where many data sets for *E. coli* exist and can (more or less) be compared with each other [9]. However, such work also reveals that a 'dark side' of *E. coli* remains, which needs to be addressed in the community's quest for a good model system.

‡ Occasionally, a cell becomes heritably sticky by switching on production of the protein Ag43. Starting from fresh stock solves the problem.

## 2. Current challenges and future directions

An early finding from DDM was a time-dependent average speed, $\bar{v}(t)$. In PMB sealed inside a sample chamber, $d\bar{v}/dt < 0$ until $\bar{v} \to 0$ abruptly at oxygen exhaustion. At low cell densities ($n \lesssim 10^9$ cells/ml), a small amount of nutrient, e.g. glucose, gives $d\bar{v}/dt \approx 0$ between oxygen and nutrient exhaustion [8]. At higher $n$, accumulating carboxylic acids from fermentation lead to $\bar{v} \to 0$ abruptly before nutrient exhaustion. A stronger buffer may help up to $n \gtrsim 10^{10}$ cells/ml. Another solution is to use light-activated strains, but now $d\bar{v}/dt < 0$ possibly due to prolonged illumination [10]. In any case, cells age, and $\bar{v}$ drops with increasing time elapsed since preparing the initial culture. An unquantified $d\bar{v}/dt$ leads to uncertainties or even wrong conclusions. Consider, e.g., measuring $\bar{v}(n)$ to study motility-induced phase separation [12]. Since data points at different $n$ are taken at different times, any observed $\bar{v}(n)$ dependence is suspect until proven otherwise.

As with passive colloids, accurate concentrations are needed, and this leads us to one of the darkest corners in the 'dark side' of *E. coli*. 'Plating' [8] is the 'gold standard' for measuring $n$, but is lengthy and laborious. Typically, the turbidity, or optical density (OD), is calibrated to give $n$ by one-off plate counting. The OD depends on cell shape and refractive index, which vary with strain and growth conditions. Alas, spectrophotometers can differ by up to 100%. The dark art of OD measurements has received some recent attention [13], but issues remain. Polydispersity complicates converting $n$ into cell body volume fraction $\phi$, which does not account for excluded volume effects of flagella bundles, the whereabouts of which remains little probed, especially at high $\phi$. At present, all reported $n$ and $\phi$ should be treated with caution.

In extending statistical mechanics to elucidate the origins of collective behaviour in active colloids, it is important to deal with the issue that in passive colloids is known as polydispersity. All atoms of (say) argon are the same; but a distribution of at least the size is inevitable in synthetic colloids, where shape and interaction may also be polydisperse. In *E. coli*, as Kenny Breuer memorably describes it, individual cells have their own 'personalities' even if their genomes are identical, a phenomenon that biologists call 'phenotypic heterrogeneity'. A genetically identical population nevertheless contains cells that show different physical (and biological) properties. Thus, e.g., genetically identical *E. coli* show five different phenotypes in their adhesion to glass surfaces [14]. New work is increasingly revealing that individual cells show complex swimming behaviour not describable by some average 'run and tumble'. Figuring out whether such 'living polydispersity' matters or not is an important research question.

Finally, it is important for physicists to take note that different strains of *E. coli* can be substantively different in ways that affect the comparability between data sets. In particular, deleting gene X in strain A may give rise to a different ensemble of phenotypic effects, e.g., $\bar{v}(t)$ and growth characteristics (and hence OD calibration), than deleting the same gene in strain Y. This may be the source of some of the variability between labs reported in the literature.

## 3. Concluding remarks

The active colloids community will benefit from converging on a handful of synthetic and natural model systems. This should allow us to pin down generic features, and confront theory and simulations with parameter-free stringent experimental tests. I suggest that motile *E. coli* can be one of these models. Establishing this model will require care with time-dependent effects, reliable ways to determine $\phi$ to, say, $\pm 10\%$, an agreed set of common strains that are periodically re-sequenced to spot random mutations, and shared protocols. Then, we may find that 'whatever is true of *E. coli* is often true for all active colloids'.§ *Bacillus subtilis* is also widely used, and can become another model natural micro-swimmer with significantly more anisotropic cell bodies. It is also possible that one type of Janus particles, perhaps most probably one of the light-activated systems [5], can also be established as a model. However this turns out, if the experience of polymer and colloid physics is anything to go by, the emergence of widely-accepted models should be a milestone for active colloids research, and must be a key calling point on any roadmap.

## Acknowledgments

I thank the EPSRC (EP/D071070/1 and EP/E030173/1) and the ERC (AdG-PHYSAPS) for funding.

§ Jacques Monod on the role of *E. coli* as a model organism for biology: 'Tout ce qui est vrai pour la bactérie *Escherichia coli* est vrai pour l'éléphant.' [15]

## Motility by shape control in unicellular organisms

Antonio DeSimone

The BioRobotics Institute, Scuola Superiore Sant'Anna, Pisa, Italy

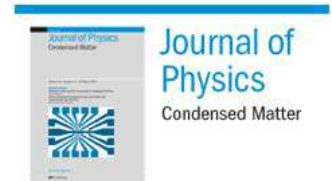

**Status**

The study of the propulsion mechanisms of individual micro-swimmers does not exhaust the field of micro-swimmers [1]. Collective effects are of paramount importance, but in this section we concentrate on understanding motility at the level of a single unicellular organism (more on this in the concluding remarks).

Understanding the propulsion mechanisms of unicellular organisms on the basis of detailed bio-physical models is important both from the point of view of biology (in order to shed light on the physics of cell motility phenomena which play a fundamental role in biology, from the spreading of metastatic tumour cells, see, e.g., [2], to the response of the immune system, see, e.g., [3]) and from the point of view of technology (as for example in Soft Robotics where new bio-inspired concepts for the design of more responsive, efficient, and autonomous devices is being sought, see, e.g., [4]).

It is foreseeable that this two-way interaction will lead in a not-so-distant future to significant advances in the design of new, bio-inspired engineering devices and to a more quantitative understanding of biological processes that are fundamental for life.

**Current and Future Challenges**

One of the key roles that quantitative bio-physical models of the motility of unicellular swimmers can play is to help identify general, over-arching principles for locomotion and shape-control.

Swimming at low Reynolds numbers is a fluid-structure interaction problem where, by action and reaction, one may equivalently focus on the equations of motion of the fluid subject to the forcing of a shape-changing micro-swimmer, or on the equations of motion of the swimmer, subject to the drag due to the presence of the fluid.

The first point of view has been fruitful, for example, in the analysis of the signature a swimmer leaves in the fluid flow it generates. By representing the latter as the result of a suitable combination of singular sources (multipole expansions), the widely used notion of pusher and puller swimmers has emerged [1].

We have chosen the second point of view, which is typical of the engineering problem of controlling the movement of vehicles (guidance, navigation, and control), and exploited the tools of nonlinear geometric control theory [5], [11]. This point of view has delivered the celebrated scallop theorem of Purcell [1], which can be rephrased as the statement that loops in the space of shapes are necessary to produce non-trivial displacements through periodic shape changes [5] (see, however, [9] for some caveats).

By focusing on the mechanics of individual swimmers, the problem of what are the mechanisms the organisms have at their disposal to execute shape changes becomes natural, and studying it can be rewarding in terms of the possible discovery of new morphing mechanisms [5-8]. These can be exploited in the design of innovative engineering devices such as deployable structures, dexterous robotic manipulators, and biomedical applications such as smart endoscopic capsules.

In spite of several exciting discoveries, and some success in replicating biological mechanisms and constructs with artificial machines, there is plenty of room for using bio-inspiration at a much deeper level. With the exception of the rotary motors powering bacterial flagella (which are *concentrated* at few locations on the cell wall, not unlike the engines of typical man-made vehicles) locomotion is mostly powered by molecular motors that are *distributed*. Indeed, molecular motors are distributed along eukaryotic cilia and flagella, or inside the body of the organisms (dyneins driving the sliding of microtubules in eukaryotic flagella, kinesins driving the sliding of microtubules and pellicle strips in *Euglena gracilis* [8], myosin motors driving the sliding of hierarchical assemblies of actin filaments in muscular contraction, etc.). Organised collective behaviour emerges in these systems spontaneously, from the body architecture, thanks to the constraining action exerted by the elastic resistance of the sub-structure of the body on which the motors exert their forces. Replicating these concepts in engineered systems capable of self-regulating their mechanical response, and possibly made with bio-hybrid materials mixing passive components and active molecular motors is an interesting (grand) challenge for the future.

The study of motility in unicellular organisms offers also opportunities to understand the response mechanisms in simple organisms, where it is not mediated by a nervous system and a brain. Recent developments include some preliminary understanding of the response to light (phototaxis) [12] and to confinement [8] in *Euglena gracilis*.

*E. gracilis* swims thanks to the asymmetric beating of a single anterior flagellum, which produces helical trajectories accompanied by body rotations around the helix axis [10]. These rotations cause the fact that the photosensitive organelle, positioned laterally in the body behind an opaque shutter (the eyespot), is periodically lit or sheltered from light coming from a lateral source, while it is continuously exposed to light if the source is aligned with the body axis. This suggests that the swimming style may provide the key mechanism for the phototactic response of *E. gracilis*, possibly exploiting a common biochemical "trick" to use periodic signals as a means to navigate, whereby existence of periodicity implies lack of proper alignment [13].

When confined (in crowded environments, between glass plates, or in a capillary), *E. gracilis* switches its motility behaviour from flagellar swimming [10] to amoeboid motion (metaboly), powered by peristaltic waves across the whole body [8]. The mechanisms by which the beating flagellum "senses" the presence of external obstacles and activity of the motors is "transferred" from the flagellum to the body of the organism is currently unknown. Moreover, the peristaltic waves enable a form of very efficient and versatile crawling motility in severely confined environments, which can be presumably exploited in robotic applications. Intriguingly, it is also unknown whether this mode of locomotion is used by the organism in natural conditions.

**Concluding Remarks**

The bio-physics of unicellular swimmers is an ideal playground to understand fundamental mechanisms of response in unicellular organism (to light, confinement, etc.). It provides opportunities for the design of artificial systems exploiting radically new bio-inspired principles including self-organised material response, bio-inspired control in robotic systems, and bio-hybrid materials. The demonstrated responsiveness of unicellular organisms (to light, confinement, etc.) poses new challenges also to the study of collective behaviour in active matter, whereby the individual particles making up the whole are not only active, but also responsive.

**Acknowledgements**

The research reported here stems from joint projects with F. Alouges, M. Arroyo, G. Cicconofri, G. Corsi, N. Giuliani, G. Noselli and is funded by the ERC Advanced Grant 340685-MicroMotility

# Emergent collective phenomena through active particle control by light

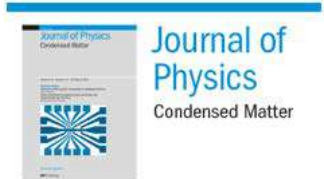

S. Muiños-Landin, A. Fischer, N. Söker and F. Cichos

Molecular Nanophotonics Group, Leipzig University, Linnéstr. 5, 04103 Leipzig

**Status**

Adaptive behavior is a key feature of biological systems. In response to environmental conditions living matter has evolved specific functions and behaviors out of intrinsically noisy molecular processes. In many cases, these adaptive processes require activity, i.e., a form of active transport or energy consuming motion together with a control signal that is processed to modify the systems response. Inspired by this biological adaptivity, for example, Machine Learning techniques have successfully coarse grained the complexity of biological adaption processes into mathematical strategies. These models are applied to imitate the adaptive behavior of natural systems, and to reach the autonomy of different robotic schemes in several fields. The almost omnipresent application of Machine Learning nowadays makes it a very multidisciplinary research field going from mathematics to neuroscience, passing through computer science, biology and of course physics. Yet, introducing such computational capabilities into artificial microscopic or molecular systems is still challenging. On one hand, the motion of microscopic objects such as active particles is strongly influenced by Brownian motion and thus actions are inherently noisy. On the other hand, adaptive systems need some type of sensing and feedback mechanism that allows the system to gain information about the environment and to feed this information back into an appropriate response.

Artificial microscopic active systems rely on different propulsion mechanisms [1]. The specific propulsion mechanism and the non-equilibrium nature of those systems already provide a wide range of new collective phenomena due to physical interactions. Self-assembly by hydrodynamic interactions [3] as well as motility-induced phase transitions belong to these phenomena [2]. However, biological systems go beyond such physical interactions by an exchange of signals, which control their actions even at interaction distances beyond the range of physical forces.

Such signals can be introduced into active particles with an appropriate control. The control of colloidal systems by light has been sparked with the development of Photon Nudging [4] of light-controlled self-thermophoretic active particles. It introduces a tool for a feedback control of Janus-type active particles [5]. Based on the real-time tracking of the position and orientation of such particles with respect to a target location the activity of the particle is controlled. Thus, the "vision" of a target location generates a mean drift towards a target location. The control of such active particles is nevertheless limited by the speed at which particles reorient their sight towards the target by rotational diffusion. Inspired by this, feedback control of the motility of Janus-like particles has opened up the study of self-organization principles following quorum sensing-like interaction rules [6].

Recently, it has been shown that the use of self-thermophoretic symmetric active particles removes the importance of active particle reorientational dynamics for the control (see Fig. 1A). The absence of rotational diffusion time scales as a limiting factor for the control of the particle activity in these symmetric particles provides a whole new experimental platform. Information-based interactions can be applied to control artificial active matter and its self-assembly. It has been shown that information that is extracted from a system of active particles allows the self-organization into dynamical structures, so called active particle molecules [7]. Such molecules show oscillations that are solely controlled by the speed of information flows (see Fig. 1B). The possibility of engineering interactions among the particles can lead to a new era of a bottom-up study of emergent structures and dynamics. The precise control of these interactions allows the optimization of the collective responses.

Optimization of actions according to environmental signals is the fundamental principle behind Reinforcement Learning [8]. In this field of Machine Learning optimal behavior is reinforced by rewards provided upon actions that are carried out in a specific state of the system. The technique of Reinforcement Learning has been theoretically studied for the optimal navigation of microscopic swimmers [9] and also large-scale active systems [10]. Recently it has been shown how microswimmers controlled by light can also optimize their navigation (see Fig. 2) through Reinforcement Learning [11]. This work shows how the influence of the stochastic noise of such a context plays a role in the strategies developed by these learning routines.

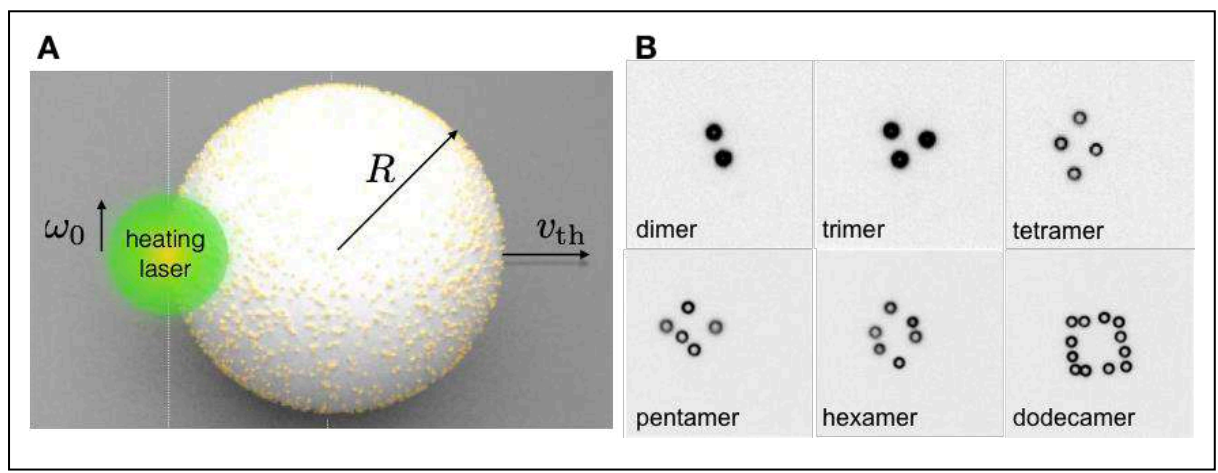

**Figure 1**. A symmetric particle as an active agent propelled through self-thermophoresis and self-emergent structures driven by information-based interactions. **A)** Illustration of a polymer microparticle (radius R) covered homogeneously with gold nanoparticles over its surface. The propulsion at a thermophoretic velocity ($v_{th}$) is reached through the asymmetric energy release using a focused laser beam (at a distance d from the center and with a beam waist $\omega_0$). **B)** Dark-field microscopy images (inverted color scale) of self-organizing structures formed by virtual interactions which are solely based on an information exchange in a feedback loop between different swimmers (2 μm diameter particles as described in A). The structures are dynamic and their characteristic dynamical properties are emergent from the time constants in the feedback loop and the noise in the system [8].

**Current and Future Challenges**

*3D challenge.* Current experiments study cases in two-dimensional confinement. The self-thermophoretic propulsion mechanism of the symmetric microswimmer and other controllable swimmers also allows for three-dimensional control. It is a challenge to maintain the feedback control of light-activated individual microswimmers in a structure or swarm migrating through a soft matter environment in three dimensions. The implementation of 3D tracking methods based on the use of convolutional neural networks might substantially improve the efficiency in terms of speed and precision.

*Activity waves challenge.* Recently a strategy was proposed to direct ensembles of active Brownian agents by imposing time and space periodic activity landscapes [12], where the migration is strongly dependent on the swimmer scales compared with the activity wave parameters. Even a change of the migration direction is predicted in a certain parameter range. This could be shown in an experiment with light-activated Janus-type swimmers.

*Optimal delay challenge.* The temporal delay intrinsic to the reaction of any real active system has been proposed to be a critical parameter in the emergence of single and collective behavior [7]. While in a biological context the dependency of collective phenomena can be just observed and modeled, experimental demonstrations of the influence of such delays remain still at the macroscopic scale [13]. The above described control mechanisms of symmetric microswimmers now enable the study of the temporal delay influence in emergent phenomena at the microscopic regime. The arbitrary implementation of artificial interactions between synthetic microswimmers also allows the artificial introduction of delayed responses to those interactions. The study of the behavioral responses of large ensembles of artificial microswimmers to dynamic or heterogeneous delays will shed light into the understanding of critical phenomena observed in nature. The level of control and the scalability of microswimmer-based experimental platforms in contrast with macroscopic robotics systems makes those systems a particularly powerful tool for such understanding and employing such phenomena for potential applications.

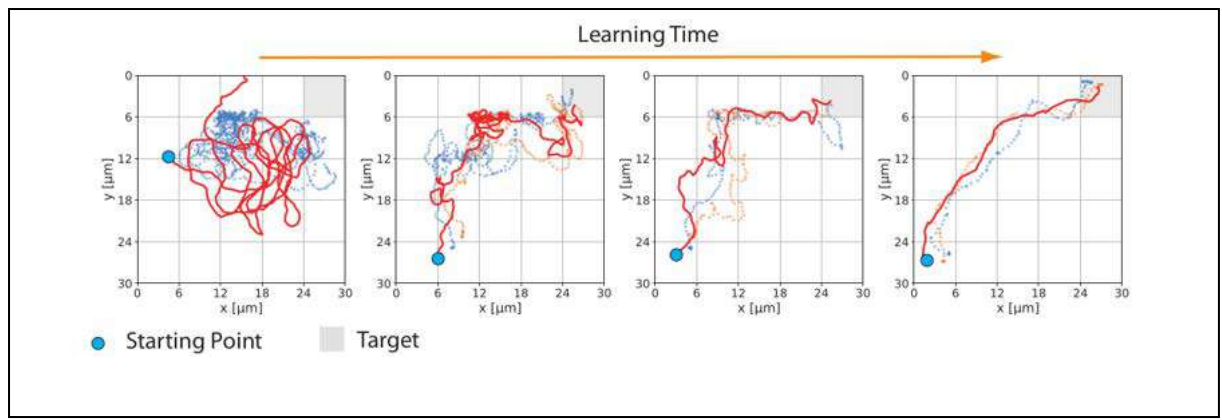

**Figure 2**. Machine Learning with artificial microscopic swimmers. The control schemes that Photon Nudging offers allowed the implementation of Reinforcement Learning navigation in a grid world for a self-thermophoretic microswimmer. The figure shows the evolution of the trajectories followed by the swimmer at different points during the learning process. An optimal chain of actions depending on the swimmer position is developed as a solution for the navigation problem towards a target point.

*Many particle challenge.* At this point, all control mechanisms rely on feedback loops with an intrinsic delay time that depends on the detection/reaction mechanisms. The timescale of such procedures depends essentially on the image processing routines and the calculations involved in either the information-based interactions or, for example, in a developed learning routine. There are two parameters that can increase substantially the complexity of the experimental schemes. On the one hand the reduction of the size of the agents (increasing the magnitude of diffusion processes) and on the other hand the increase of the number of active agents. While the first one remains as a hard boundary of the state-of-the-art experiments, the second one is a technical challenge to software and hardware optimization, including the accurate and simultaneous laser positioning. The success facing those issues would imply, for example, the implementation of Multiagent Learning procedures or the usage of a spatial light modulator (SLM) for particle control, or the real-time individual control of large amounts of active particles. This represents a significant experimental challenge with colloids controlled by light.

**Concluding Remarks**

The control of artificial microswimmers by light has considerably evolved during the last years. These experimental schemes are delivering the possibility of understanding the emergence of complex collective phenomena in biological active matter by a bottom-up approach designing the basic rules of interactions in colloidal active matter. Novel propulsion and steering mechanisms increased the versatility of active particle control, such that studies of microscopic information bound structures, the behavioural optimization by Machine Learning algorithms and out of equilibrium driven self-organization are now possible. This field of study is emerging as a very multidisciplinary one where soft matter physics, chemistry, biology or Machine Learning meet. Applications such as cargo transport or material properties modification, autonomous transport might be facilitated by the coordinated effort including all of those fields.

**Acknowledgements**

*We acknowledge the support of the Deutsche Forschungsgemeinschaft (Projects CI 33/16-1 and CI 33/16-2) within the Priority Program 1726 "Microswimmers".*

# The dynamics of diffusiophoretic motors

Raymond Kapral
Chemical Physics Theory Group, Department of Chemistry,
University of Toronto, Toronto, Ontario M5S 3H6, Canada

Pierre Gaspard
Center for Nonlinear Phenomena and Complex Systems, Université
Libre de Bruxelles (U.L.B.), Code Postal 231, Campus Plaine, B-1050
Brussels, Belgium

## Status

Studies of the motion of colloidal particles in concentration gradients have a long history and the continuum theory of the diffusiophoretic mechanism responsible for this motion is well understood [1,2]. Also, the self-propulsion of active colloidal particles (motors) has been studied extensively in experiments and through deterministic continuum theory [3,4]. In the self-diffusiophoretic mechanism propulsion occurs because concentration gradients on the surface of the particle arising from asymmetric chemical activity lead to a body force on the motor and, since the system is force-free, this force is compensated by fluid flows in order to maintain momentum conservation (Fig. 1 left). If the motors of interest have micron or sub-micron dimensions thermal fluctuations become important necessitating stochastic formulations of the dynamics. Although there has been a considerable amount of research in this area, some of the basic elements that are involved in the description of diffusiophoretic propulsion of small colloidal particles are often overlooked, treated in special limits or under specific conditions.

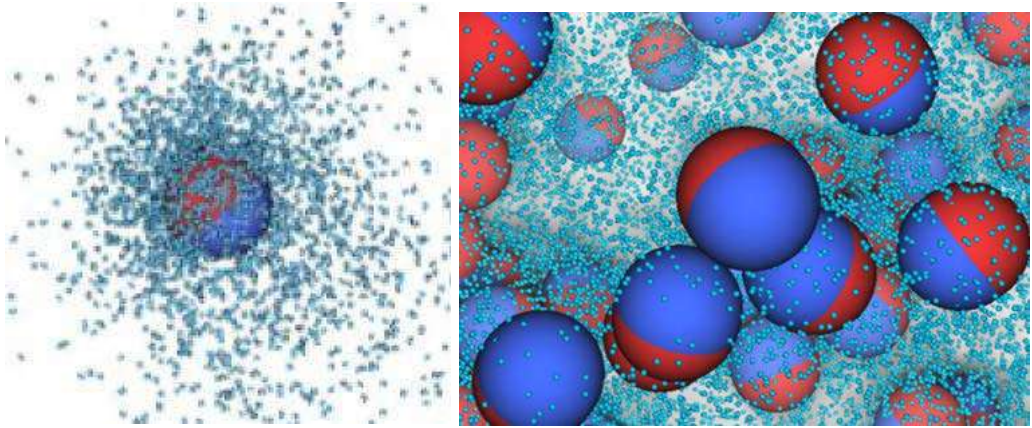

**Fig. 1:** *(Left) A Janus motor produces an inhomogeneous distribution of product molecules (small spheres) as a result of chemical reaction on the catalytic face (red) of the motor. (Right) A collection of Janus motors and the inhomogeneous product concentration fields they produce.*

Self-propulsion is a nonequilibrium phenomenon. A system at equilibrium satisfies detailed balance that follows from the microscopic reversibility of the laws of motion governing the time evolution of the system. These basic dynamical principles must be respected in theoretical formulations since they have implications for the forms that the macroscopic laws and dynamical properties take. In order for motor propulsion to occur the system must be in a nonequilibrium state where detailed balance is broken. Such nonequilibrium conditions are typically obtained by placing the system in contact with reservoirs that control the concentrations of chemical species so that the chemical affinity takes non-zero values. Thus, some of the key elements that are required for the construction of a dynamically consistent theory of self-propulsion of small motors are the respect for the fundamental properties that stem from microscopic reversibility, the implementation of constraints to drive the system out of equilibrium, and the incorporation of thermal fluctuations in the dynamical description.

A great deal of current research concerns the collective behavior of many-motor systems, often employing simple generic models to capture essential features of the collective dynamics [5]. A full theoretical description of the many-body dynamics of diffusiophoretic motors presents additional challenges since effects arising from chemical gradients are often the most important factor responsible for the collective phenomena. This requires the development of more complex models that account for chemical reactions and the many-body gradients they produce. Since diffusiophoretic motors generate fluid flow fields as part of the propulsion mechanism, these flow fields also contribute to the collective dynamics. In fact, with few exceptions, theoretical studies of the collective dynamics of active particles often focus on one or the other of fluid flow or chemical gradient effects to model collective behavior. The nature of the collective behavior of diffusiophoretic motors using models that incorporate all relevant effects remains to be fully explored. Challenging problems in theory, simulation and experiment remain, and potential real-world applications of these motors remain to be developed [6].

## Current and Future Challenges

**Mechanochemical coupling**: It is essential to understand the coupling between chemistry and mechanics for both biological molecular machines and chemically-powered synthetic motors. The nature of this coupling has been investigated in detail for molecular motors, but it is only recently that the importance of mechanochemical coupling has been recognized for our understanding of colloidal motors self-propelled by diffusiophoresis. This coupling is responsible for the generation of a velocity slip between the solid surface of the colloidal motor and the surrounding fluid arising from the reaction-generated concentration gradients of some molecular species. Since the motion of small colloidal motors is stochastic, the problem is to understand the nature of the coupling between their random spatial displacements and the fluctuations in the number of reactive events providing chemical free energy for their propulsion. Since the dynamics at the molecular scale is time-reversal symmetric, the mechanochemical coupling should obey Onsager reciprocal relations, as is the case for molecular motors [7,8]. The theoretical description of this coupling provides expressions for the diffusiophoretic velocity and the reciprocal effects of an external force and torque on the reactions catalyzed by the colloidal particle [9,10].

**Colloidal surface properties**: Given the central role played by surface forces in the diffusiophoretic mechanism, a main component in the experimental design of new motors is the engineering of their equilibrium and nonequilibrium surface properties. Theory is expected to play an important part in guiding the experimental work. Materials science has produced many theoretical tools to predict the bulk properties of materials; however, the large variety of surface properties remains uncharted territory, especially for nonequilibrium surface properties.

Therefore, the development of theoretical methods to predict these material properties is an important and challenging task for research. Using surface tension, an equilibrium surface property, the engineering of hydrophobicity has already been developed extensively. However, colloidal motors function out of equilibrium, consuming and dissipating chemical free energy by catalytic surface reactions, sliding friction, and the mechanochemical coupling provided by diffusiophoresis, in addition to bulk fluid viscosity. The surface processes involved in reaction, including the adsorption, dissociation, surface diffusion, and desorption of the different molecular species involved in the reaction, can be engineered using the methods of heterogeneous catalysis for the various surface materials and fuel species. In addition, the forms of the interaction potentials between the solute species and the solid surface on both catalytic and noncatalytic parts of the surface are important in order to optimize the diffusiophoretic effect and, thus, the propulsion of the colloidal particle. Very little information is available on the quantitative aspects of these interactions. Moreover, we also need to consider the sliding friction due to the interaction between the solvent molecules and the solid surface, which gives rise to the slip length. In this regard, the theoretical development is still in its infancy.

**From micro- to nanometric motors**: The problems associated with the surface properties of diffusiophoretic motors become even more challenging when very small synthetic motors with nanometer-scale dimensions, comparable to those of molecular machines, are considered. On these scales the application of even fluctuating continuum treatments are problematical, although microscopic analogues of the diffusiophoretic mechanism should operate. This motor domain is largely unexplored.

**On the conditions to achieve nonequilibrium steady states**: On the experimental side there is a need for controlled experiments to study fundamental aspects of motor motion. Most experiments on chemically-powered motors are carried out under quasi steady state conditions where fuel is initially supplied in excess and consumed as motor motion takes place. In such circumstances motor motion will continue with variable velocity until the fuel is exhausted. True steady state conditions can be implemented by coupling the system to constant fuel and product concentration reservoirs, similar to the continuously-fed-unstirred reactors used in investigations of pattern formation in nonequilibrium chemical systems. In this way the nonequilibrium system state could be controlled by the reservoir concentrations. Motors could be made to move forward or backward by varying reservoir concentrations, and the chemical affinity can also be fixed to determine the distance from equilibrium. Such experiments would allow various aspects of active motion to be studied, including the effects of external forces and torques on the reaction rates.

**Collective dynamics**: Many of the issues raised above have consequences for the collective behavior of diffusiophoretic motors (Fig.1 right). Experiments are often again performed by initially supplying fuel to the system and observing the quasi steady state dynamics. In fact, the "single-motor" experiments are almost invariably carried out on very dilute suspensions of many-motor systems. While such experiments provide a great deal of information on the collective behavior, they cannot describe all aspects of a system with prescribed boundary concentrations to drive it from equilibrium. For example, one of the common nonequilibrium motor states is dynamic cluster formation, where groups of motors form a cluster which may propagate or dissociate and reform. Since motors require chemical fuel for propulsion, and chemical gradients provide one of the driving forces for motor aggregation, the availability of fuel surely plays an essential part in the cluster formation and dynamics. If the nonequilibrium state is fixed by constant concentration reservoirs, fuel may be depleted in cluster interiors changing the character of the dynamics. If fuel is supplied in the interior of the system, say, by other nonequilibrium reactions involving fuel and product molecules, the collective behavior could be very different. Thus, studies of the effects of system constraints on many-body dynamics are needed. In addition, as high-density cluster or other dynamical states arise, effects related to direct intermolecular interaction must be taken into account. The motors exhibit highly correlated motions that are outside the often-used mean-field models that neglect such possibly important correlation effects. The development of a comprehensive theoretical theory for such collective behavior is at an early stage.

## Concluding remarks

The constituents of active matter, and in particular the synthetic diffusiophoretic colloidal motors that are the subject of this article, are fascinating objects that challenge theory since their behavior is an intrinsically nonequilibrium phenomenon, and their dimensions place them in the borderline region where the transition from microscopic to macroscopic takes place. They are potentially very useful but it remains to be seen when practical applications will come to fruition [6]. Nonetheless, Nature has learned how to exploit molecular machines for a myriad of tasks, and one can hope that we will also learn how to use synthetic motors with various length scales, operating singly or collectively, in simple or complex media, as effectively as Nature.

## Acknowledgements

Financial support from the International Solvay Institutes for Physics and Chemistry, the Université libre de Bruxelles (ULB), the Fonds de la Recherche Scientifique-FNRS under the Grant PDR T.0094.16 for the project "SYMSTATPHYS", and the Natural Sciences and Engineering Research Council of Canada is acknowledged.

## ORCID IDs
Raymond Kapral https://orcid.org/0000-0002-4652-645X
Pierre Gaspard https://orcid.org/0000-0003-3804-2110

**Synthetic swimmers with tunable behavior:
self-phoretic colloids**
Marisol Ripoll, Forschungszentrum Jülich, Germany

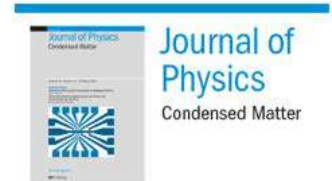

**Status**

Most examples of active matter are biological in origin, but current interest in the design of synthetic active systems is quickly raising [1, 2]. Besides the manufacture of structures mimicking biological microswimmers, artificial micrometer-sized colloidal swimmers can be synthesized nowadays. The shape and composition of these swimmers can be tailored aiming at systems with various properties, which will certainly find applications in lab-on-a-chip devices, or drug delivery, and which will also be of fundamental theoretical significance in non-equilibrium statistical physics and transport processes. A relatively simple and effective strategy to design artificial microswimmers are phoretic effects.

*Colloidal phoresis* refers to the mechanical force that arises from an inhomogeneous fluid environment interacting with a suspending colloid, which translates in the colloidal drift motion. Such inhomogeneities can be gradients of electric potential (electrophoresis), concentration (diffusiophoresis), or temperature (thermophoresis). These gradients are frequently a consequence of external constrains, but interestingly, in case one particle is able to produce a local gradient field, self-propulsion may occur. For diffusiophoretic microswimmers [3], the metal coated part catalyzes a chemical reaction inducing a local concentration gradient, while for thermophoretic microswimmers, the metal coated part is able to effectively absorb heat from e.g. an external laser, which creates a local temperature gradient [4]. Catalytic swimmers are experimentally more abundant, although they require that fuel and product are constantly added and removed to and from the solution, in order to keep stable gradients. Thermophoretic swimmers are activated by heat, such that they do not require any modification of the solvent, and the absence of surfactants or chemical fuels, makes them easily bio-compatible. Due to their intrinsic nature, these two swimmer types will have potentially different applications, and they can eventually even be combined, providing them with additional functionality. On the other hand, the two phoretic phenomena are, from a fundamental viewpoint, equivalent. This equivalence implies that most conclusions can be interchanged from catalytic to thermophoretic and vice versa. For practical reasons, here we will mostly focus on the thermophoretic effect. It is important to note that the drift of a colloid in a solvent gradient can in principle occur towards or against the gradient. In this way, thermophobic swimmers are those that propel against the heat source (the most frequent case), while thermophilic swimmers are those that drift towards heat source. The osmotic motion of the colloid surrounding fluid, occurs together with the colloid drift, the direction, range, and intensity of both displacements is determined by the surface solvent detailed interactions, as well as by the system boundary conditions [5].

Simulations performed with Janus and dimeric colloidal swimmers [6] have explored the precise solvent behavior and show a fundamental different hydrodynamic behavior. A very powerful approach to the study of phoretic swimmers is the use of *mesoscopic computer simulations*, and in particular the use of a particle based method known as Multiparticle Collision Dynamics (MPC), in which the solvent is explicitly considered. Hydrodynamic and phoretic interactions emerge as a natural consequence of the imposed boundaries. In the case of diffusiophoresis, these are the different interaction of the two components in the solvent, and in thermophoresis the temperature dependent interactions. Janus particles displayed the typical neutral behavior which means that their hydrodynamic interactions are short ranged, and axi-symmetric [6]. Hydrodynamic interactions between self-phoretic Janus particles are then expected to be negligible in comparison to the effects of concentration or temperature gradients. Other structures of relevance are dimeric Janus colloids [7]. The solvent velocity fields around dimeric microswimmers with slightly separated beads, show to behave as the well-known hydrodynamic character of force dipoles [8], being the flows originated by thermophilic and thermophobic simply symmetric and with reverse direction with respect to each other. If another dimer or particle is placed lateral and close to the dimer, the flow field will exert an attraction in the case of a thermophilic microdimer and a repulsion in the case of a thermophobic microdimer, which allows us to identify them respectively as *pushers* and *pullers*. In the case of the dimeric swimmers, the separation between the beads, and their size ratio is also very important, determining not only the overall swimmer velocity, but also the qualitative shape of the hydrodynamic interactions. The flow field of the asymmetric dimer in Fig. 1a shows a strong and long ranged lateral hydrodynamic attraction which is qualitatively different from the symmetric case, as the quantitative comparison in Fig. 1b shows. Such cross-over from attraction to repulsion in the lateral part is due to the complex geometry of the dimer, since simpler spherical geometries do not exhibit similar features.

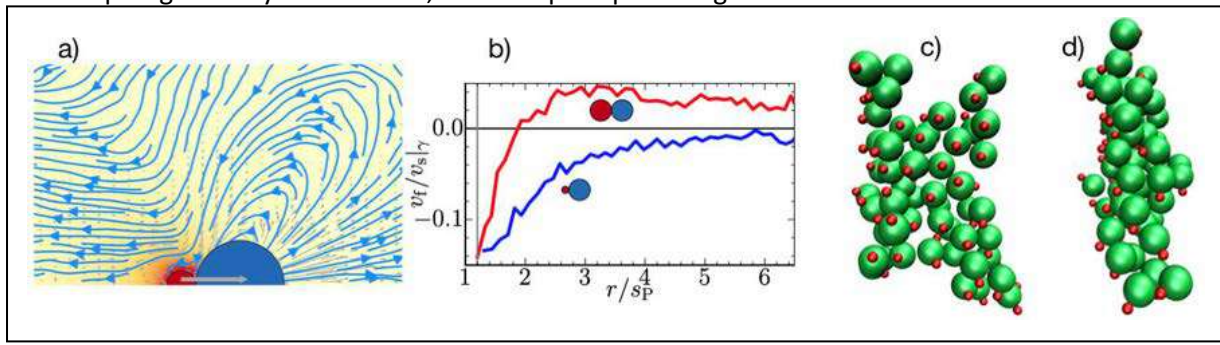

Figure 1. a) Flow field and stream lines in the area close to an asymmetric thermophobic dimer. b) Fluid velocity as a function of the distance to the center of the phoretic bead in the axis perpendicular to the swimmer orientation for a symmetric and an asymmetric thermophobic swimmers. Fluid velocity is normalized with the corresponding single swimmer velocity, and the distance with the size of the phoretic bead. The sign of the fluid velocity refers to hydrodynamic attraction (negative) or repulsion (positive). c), d) Simulation snapshots corresponding to the front and side view of one large oriented cluster of asymmetric thermophobic swimmers. Figure from [9].

The collective behavior of self-phoretic colloids is determined by various interactions of which phoretic and hydrodynamic are the most relevant ones. The heated bead of each dimer produces a temperature gradient in the surrounding solvent which affects, not only the attached phoretic bead, but also any other colloid nearby. Two thermophoretic dimers might therefore have different combinations of hydrodynamic and phoretic attraction and repulsion, with which the overall collective behavior can be tuned. In particular, two thermophobic asymmetric dimers will hydrodynamically attract each other laterally, and phoretically repel axially, making them to assemble in single-layered structures, with hexagonal and high orientational order (see Fig. 1c,d), and directed collective motion [9]. In contrast, thermophilic dimers form large static clusters, showing that in this case the phoretic attraction dominates any other hydrodynamic interactions, behaviour also observed for similar chemically active Janus colloidal particles [10].

**Current and Future Challenges**

The collective behaviour of phoretic swimmers has already shown to display reversible clustering, the formation of planar swarming structures, and several other phenomena can still expected to be proved. Furthermore, the mechanisms involved in the formation of these structures, the importance of the phoretic effect, and hydrodynamics, as well as the behaviour of phoretic swimmers as a function of their shape and intrinsic properties, are still open and relevant questions. The combination of experimental and theoretical research that mutually support the finding of novel behaviors is therefore one of the main actual challenges. Natural extension of this research is the investigation of quasi-two dimensional confinement, and other multimeric structures such as triplets, quadruplets, or swimmers of mixed phoretic character. These are expected to propel with much more persistent trajectories, or to behave as rotors which could eventually originate resonant states. Other features such as convection, shadowing, or the influence of solvents with various phase behaviors, like nematic states, different phase co-existences, or viscoelasticity, are all aspects of great interest. Eventually the behavior of mixtures of different swimmer types, and alternative phoretic swimmers like those based on mixed phoretic effects, or local phase transitions, are also of great fundamental and technological relevance. Of particular interest is for example the directional heating of the swimmers to investigate guided motion. This takes into account that a solution partially heated, for example by an illuminating laser, can heat only those dimeric colloids with a particular orientation such, that the colloids would tend to migrate towards or against the heated areas.

**Concluding Remarks**

Phoresis provides a powerful strategy to design synthetic microswimmers, micromotors, or micropumps, which has become a promising tool in the field of microfluidics and bio-compatible material development. These systems have the great advantage of behaving like passive unless they are chemically, electrically, or thermally activated. In particular, temperature gradients can be locally controlled at very small scales, allowing to engineer lab-on-chip devices with effects that could be switched on and off by the local application of temperature sources such as lasers. Furthermore, all applications of phoresis depend not only on the system geometry, the intensity and direction of the applied gradient, but also, and very importantly, on the material properties. This is characteristic of both the suspended material, and the employed solvent, as well as various other ambient conditions like the average temperature, pressure, concentration, or the solvent pH. In this way, the properties of the phoretic machines are very subtly dependent on all the different contributions, providing a very large versatility to these devices, and a fertile investigation field.

**Acknowledgements**

I am very thankful to current and past students and collaborators in this topic: Mingcheng Yang, Daniel Lüsebrink, Martin Wagner, Zihan Tan, Sergi Roca, and Alberto Medina. This work was supported by the Priority Program "Microswimmers" (SPP 1726) of the German Research Foundation (DFG). We also gratefully acknowledge the generous computing time granted on the supercomputer JURECA at Jülich Supercomputing Centre (JSC).

# Active vs. driven colloidal systems

F. Sagués

Departament de Ciència de Materials I Química Física, Universitat de Barcelona

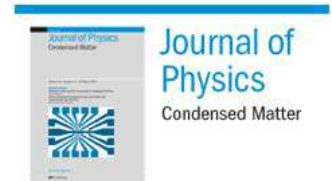

## Status

Interest for colloidal systems of motile constituents has exploded during the last couple of decades [1, 2]. Motility is here understood either originated from an intrinsic *activity* of the dispersed entities, or, contrarily, be *driven* externally, following a purposed (electric, magnetic, or optic) actuation. The difference between active and driven, although sometimes faint appears to us essentially rooted in the fact that autonomous swimmers, while consuming internal or ambient-available energy, mostly follow unguided trajectories. Conversely, driven colloids normally admit to being steered by the same forces that set them into motion.

Research in the field has progressively shifted from the description of individual behaviour towards understanding the complexities that emerge from large assemblies. Within the first perspective, experiments on Janus-like colloids and models such as those based on Active Brownian Particles or squirmer-based swimmers appear as highly celebrated achievements. [1, 2]. Equally popular, are collective phenomena such as anomalous density fluctuations related to swarming/flocking, motility-induced phase separations, or adapted versions of fundamental thermodynamic concepts (pressure and temperature), to mention a few [1, 2].

Apart from the basic interest in understanding new aspects of non-equilibrium physics, progress in the study of motile colloids has been spurred both from the perspective of practical applications, mainly in relation to their biomedical use (swimming micro/nano robots), as well as being often considered as simple realizations of much more complex biophysical materials (biofilms, cell tissues, cytoskeleton reconstitutions, etc).

## Current and Future Challenges

In this section I will briefly address two issues that I consider among the biggest challenges for future research in the field of active and driven colloids. To balance the discussion, one example refers to an active colloidal system, while the second considers a scenario of driven electric actuation for colloids dispersed in anisotropic fluids.

**Active nematics: Control from interfaces.** A rewarding example of active material is the aqueous assembly of bundled microtubules powered by kinesin motors under ATP consumption [3]. Most typically, the material is prepared at the interface with an isotropic oil and constitutes a paradigmatic realization of an *Active Nematic* (AN) [4]. Experiments and dedicated numerical simulations have unveiled rich dynamical behaviour of defect-driven active flows under non-confinement, and, more specially, much effort has been directed towards understanding the characteristics of an apparent turbulent state yet endowed with a characteristic length scale [5, 6] (Fig. 1a). Very recently we proposed a versatile and robust strategy towards the control of such chaotic-like state. The simple idea consists in replacing the isotropic interfacing oil with a *Liquid Crystal* (LC) in its smectic phase. Without any further intervention, the active turbulent state gets confined into circular domains (focal conics) (Fig. 1b) [6]. Moreover, by applying a moderate magnetic field, we observe active flow regularization organized in stripes of a well-defined and activity-dependent periodicity [7] (Fig. 1c).

Both situations are explained in terms of the arrangement of the passive LC molecules, in absence and under magnetic forcing, respectively.

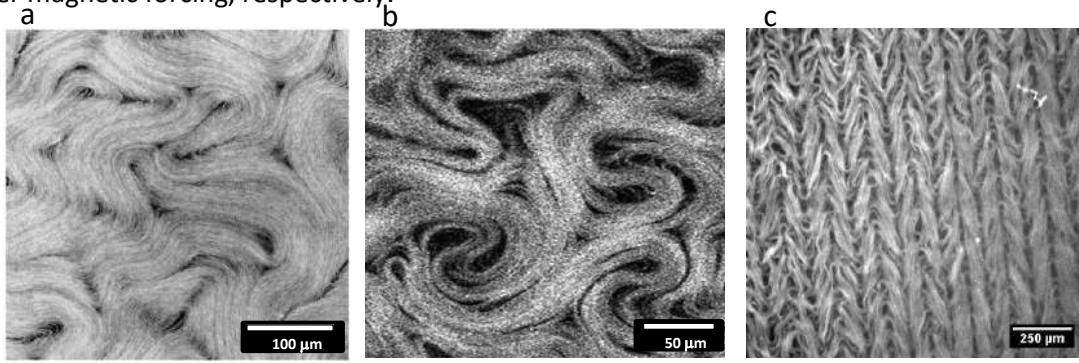

Figure 1. The active nematic system based on the microtubule-kinesin preparation. In panel (a) the system is interfaced conventionally with an isotropic oil and develops a turbulent-like regime. Panels (b) and (c) correspond to interfaces prepared with a Liquid Crystal, without (b) and with an imposed magnetic field (c). In both cases active flows are rheologically conditioned following the textures of the passive LC: conical in panel (b), and lamellar (bookshelf) in panel (c) (the magnetic field is directed perpendicular to the stripes).

**Phoretic nematic colloids: Individual and collective effects**. Phoresis of colloids is a standard technique, commonly employed in aqueous environments under the application of a DC electric field. On the other hand, the use of a LC dispersing medium permits to explore non-conventional electrokinetic phenomena of nonlinear nature (*Liquid Crystal Enabled Electrokinetics LCEK)* [8]. Specifically, electrophoresis of colloids is quadratic in the electric field, permitting to apply an AC forcing. Moreover, electrophoresis in this case has a tensorial nature, permitting displacements perpendicular to the applied electric field. We have recently explored different situations in this context, addressing aspects of the individual motion, as well as features of collective assembling [9, 10].

In the first context, we have considered LCEK, as well as gravity-induced sedimentation as benchmark, of colloids in a cell of homogeneous director orientation enforced to be parallel to the enclosing plates. Spherical silica particles with different functionalization protocols ensure situations of homeotropic (perpendicular) (Fig. 2a), as well as planar (tangential) anchoring of the LC at the particle surface. A broad temperature range has been investigated and the particle size has been similarly varied. Both under gravity- and electrically-induced driving nematic colloids follow ballistic motion along the far-field direction imposed by the nematic matrix. However, when looking at diffusion characteristics transversal to the nematic director we found striking differences depending on the particle/LC contact. More specifically, we find normal diffusive behaviour for planar anchoring, while super-diffusion is evidenced for hometropic interaction (Fig. 2b, c). The latter exponents turn out to largely depend on particle size and temperature, with a saturating trend at a value close to 1.8 for the highest temperatures explored, still well-below the nematic/isotropic transition [9]. We tend to interpret this marked difference in terms of the backflow currents originated in the LC host by the moving colloid.

In relation to collective effects, a simple strategy to assemble LCEK-driven colloids consists in locally creating defect-based accumulation sites. We implemented this technique by using a photosensitive (azo-based) surfactant, that change locally the LC anchoring from homeotropic to planar on one of the enclosing plates [10]. In doing so, nematic colloids are not only driven through the cell but collectively reunited by thousands around a pure splay or mixed splay/bend singularities, depending on the illumination protocol. In the first case, a revesible aster-like assembly is observed, while a rotating cluster of colloids is composed in the second scenario. The spatial distribution of the colloids is in both cases far from trivial (Fig. 2d). In the simpler case of an arrested cluster, the colloidal density displays a marked dependence on the radial coordinate, with a striking phase coexistence: an interior core region of close packed colloids that grows with time, coexists with a liquid-like corona of fixed width that finishes at a gas-like phase of very diluted colloids. According to a recently formulated model this phase organization is the result of a delicate interplay between repulsive forces of different nature (elastic, hydrodynamic and electrostatic) acting on top of the direct LCEK driving. This collective situation, as well as the apparently simpler individual dynamics mentioned in the precedent paragraph are remarkable clues of the rich promises open for the study of colloidal motion if one attempts to go beyond the isotropic and non-structured fluids considered conventionally.

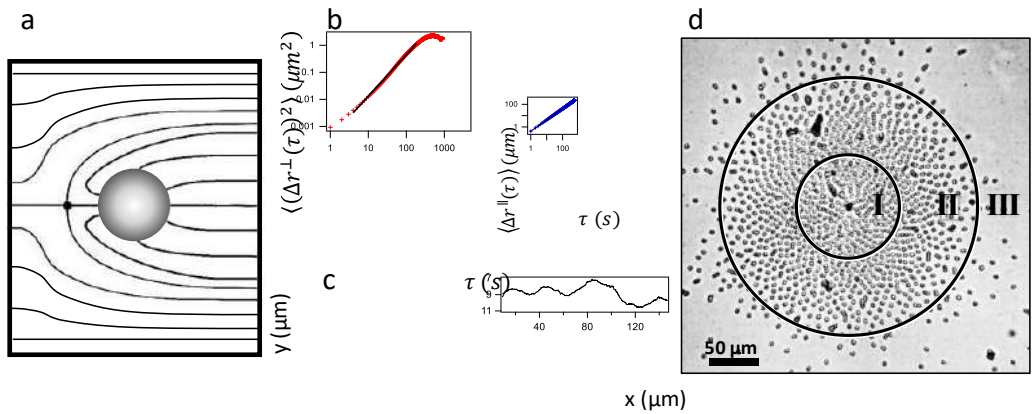

Figure 2. *Nematic colloids driven by nonlinear electrophoresis. Panel (a) represents schematically a colloid dispersed in a nematic Liquic Crystal (LC) featuring a hedgehog defect from the homeotropic contact of the LC at the surface. Panels (b) and (c) reproduce the super-diffusive behaviour of the elementary displacements of the phoretic colloid transversal to the far-field LC director. Panel (d) shows an assembly, with coexisting phases of colloids aggregated around a topological defect in the LC matrix.*

**Concluding Remarks**

We conclude by emphasizing the need to assume novel challenges, both from experiments and theory, to better understand the behaviour of single and collective colloidal motion. Active and driven materials as well, both animate or inanimate, provide us with remarkable possibilities towards the control and harnessing of their motility in different physicochemical or biophysical contexts.

**Acknowledgements**

I thank the past and present collaborators who have participated at different levels in the two commented projects. J. Ignés-Mullol as general experimental supervisor, P. Guillamat, B. Martinez and J. Hardoüin in the active nematic subject, and J. M. Pagès, A. Straube and P. Tierno in the phoretic colloids study. Financial support from MINECO (AEI/FEDER) under Projects FIS 2013-41144P and FIS2016-78507-C2-1-P is acknowledged.

# Confined Active Matter

Amin Doostmohammadi and Julia M Yeomans
The Rudolf Peierls Centre for Theoretical Physics
Clarendon Laboratory
Parks Road
Oxford, OX1 3PU, UK

When dense active matter is confined, the interplay of the confinement size and the length scale of patterns that are formed by activity, together with the possible creation of motile topological defects, can lead to surprisingly complex behaviour. Investigating this is relevant to understanding active systems that thrive under confinement, such as biofilms, tissues or organoids, and to developing novel means of controlling active matter for possible uses to power microscale devices.

**Active flow in a channel:**

In the bulk, dense active nematic or polar systems with hydrodynamic interactions show active turbulence characterised by chaotic flows, engendered by fluid jets and vortices, and motile topological defects [1]. The scale of the vortices is the active length, $l_a$, which is set by a competition between the elastic restoring force and the strength of the activity. When the system is confined in a channel of dimension $l < l_a$ the hydrodynamics is screened and active turbulence cannot develop. Instead more regular flows are possible: spontaneous shear flows, unidirectional flows, and vortex lattices. However, these depend sensitively not only on the fluid parameters and the channel width but also on the details of the boundary conditions and the strength of intrinsic fluctuations.

So far, almost all experimental and numerical results have been restricted to two dimensions. Here, as the width of the channel is increased at fixed activity, the evolution in flow configurations can usually be characterised as no flow => laminar flow (shear or unidirectional) => a one-dimensional line of flow vortices => active turbulence (Fig. 1) [2]. The system starts to flow when the active stresses can overcome the pinning effect of the boundaries, velocity vortices can form once the channel becomes wide enough to accommodate them, and then a further increase in channel width allows relative motion of the vortices, corresponding to active turbulence.

Motile, active topological defects add complexity to this sequence. Channel walls are preferential sites for defect formation, particularly if any boundary alignment is weak and if active particles can freely slip along the walls. The stationary -1/2 defects remain close to the walls due to elastic interactions, whereas the self-propelled +1/2 defects move towards the centre of the channel. If the flow is laminar +1/2 defects can traverse the channel to be annihilated by the -1/2 defects at the opposite wall. In the vortex regime, however, they can be captured by the flow vortices and perform a "ceilidh dance", with right and left moving defects moving past each other on sinusoidal trajectories (see Fig. 1d) in a way reminiscent of the great chain figure in country dancing [2].

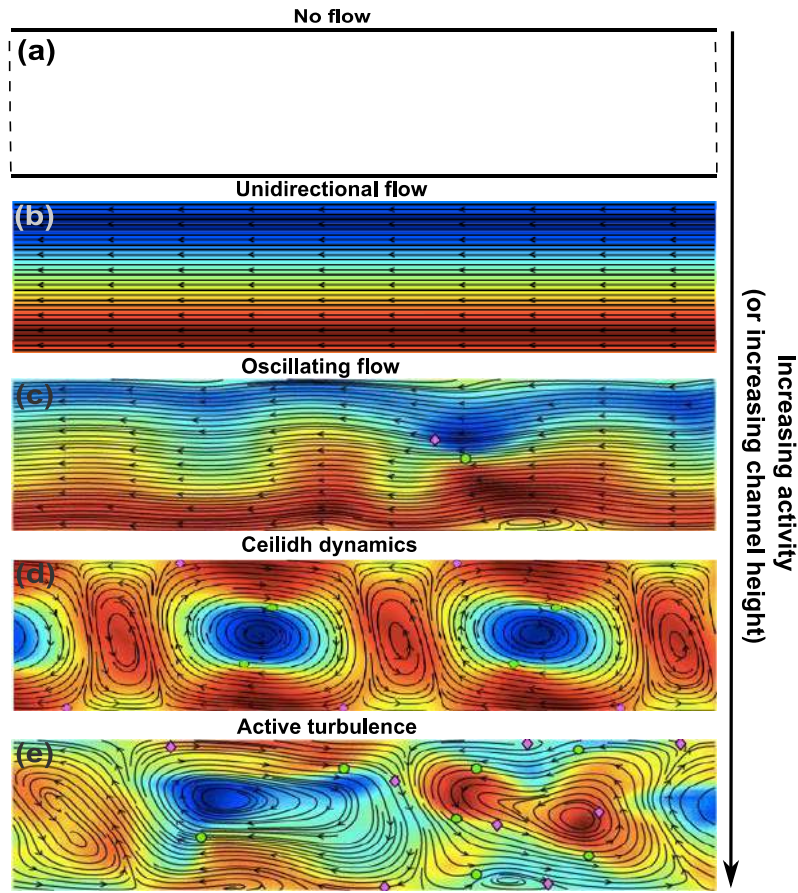

Figure 1: Flow states of an active nematic confined in a channel. The black lines represent the streamlines of the velocity field and the black arrows indicate the direction of the flow. The colourmap represents the vorticity field. Circles (green) and diamonds (magenta) mark +1/2 and -1/2 topological defects, respectively. (Taken from [4]).

Simulations have been used to study the full range of confined behaviours. Experiments on various two-dimensional active systems have so far been able to address more restricted, but diverse, regions of the available parameter space. A transition from a quiescent state to spontaneous shear flow has been observed for **cells** confined to an adherent stripe [3], and there is evidence of velocity vortices in other experiments which track cells dividing and expanding along a channel, but the vortices do not form a regular pattern. Dense suspensions of **bacteria** moving around a racetrack show a crossover from laminar flow to active turbulence with increasing track width. Near the crossover vortices form, but again these do not order into a regular lattice. Experiments which confine **microtubule-kinesin** mixtures to move at an oil-water interface are not able to access the quiescent state, but transitions from shear flow to a one-dimensional vortex lattice, and then to active turbulence are observed as the channel width is increased. This is a system where defects form easily at the walls because the microtubules have weak, planar anchoring and are able to freely slide along the channel walls. In the shear state the defects form in periodic bursts and then move across the channel to annihilate with the -1/2 defects at the opposite wall. In the flow vortex lattice state +1/2 defects become entrained by the vortices to perform a recognisable, albeit noisy, ceilidh dance. See [1], [4], [5] for more complete references.

Very little is currently known about the behaviour of confined active material in three dimensions. As a first important step in this direction, recent experiments [6] have shown that confining a microtubule-kinesin mixture in a three-dimensional micro-channel can produce long-range coherent flows up to metre scales. For channels with a cross section of aspect ratio ~ 1 the flow was coherent for all the channel sizes (μm to mm) considered. Once the channel aspect ratio increased beyond ~ 3 the coherent flow was replaced by active turbulence. This scale-independent behaviour is surprising and not yet understood. The authors of [6] suggest that it may be related to layers of aligned active nematic on the walls of the micro-channel that power the whole fluid.

**Circular confinement and spherical shells:**

Similar interplays between active flow and the dynamics of motile topological defects govern the behaviour of active materials confined to circles [7] (Fig. 2). Flows that rotate around the container can be established for small circles, crossing over to active turbulence at higher activities or in larger traps. Planar or homeotropic boundary conditions introduce a +1 topological defect which splits into two +1/2 defects to minimise the elastic energy of the system. Simulations show that these motile defects move towards opposing walls where their behaviour depends on the details of the system. For example, defects can remain pinned and stabilise arrays of flow vortices, continue to move along periodic closed orbits, or continuously move in and out of the walls.

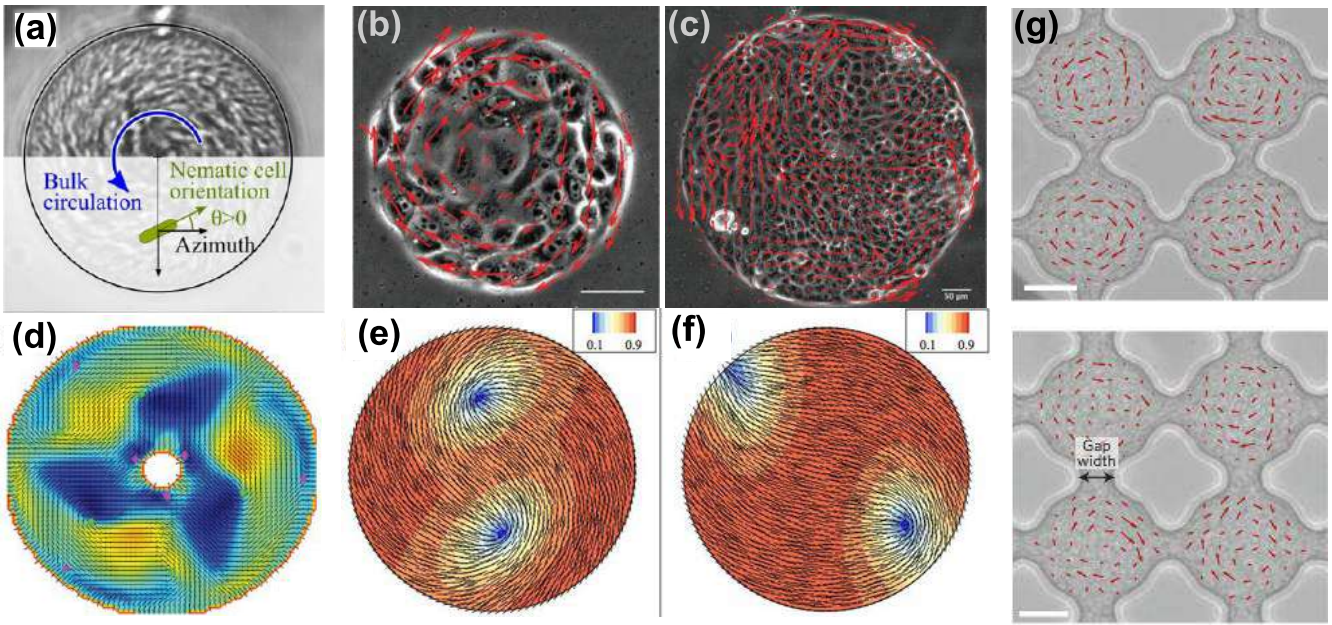

Figure 2: Flow and topological defect patterns in circular confinement. (a) Circulation of confined *B. subtilis* bacterial suspension. (b) Uniform circulation and (c) active turbulence of (Madin-Darby Canine Kidney) MDCK cells within different circle sizes. (d) Active nematic simulations of flow within an annulus. The colourmap represents the vorticity field. Circles (green) and diamonds (magenta) mark +1/2 and -1/2 topological defects, respectively. (e),(f) Time evolution of director patterns and defect trajectories in simulations of active nematics confined in a circle (g) Counter-rotating and co-rotating circulations of *B. subtilis* bacteria within interconnected circular confinements. (Taken from [4]).

Experiments on **cells** correspond to low activities, and persistent circular currents have been observed in confluent epithelial cell sheets. The coherent motion was destroyed by knocking out cell-cell junctions that enable force transmission between the cells, emphasising the importance of collective interactions in determining the dynamics. An innovative way of confining a higher-activity **microtubule-kinesin mixture** to circles is to interface it with a passive liquid crystal which is in the smectic A phase [8]. The passive smectic spontaneously self-assembles into polydisperse domains, known as toroidal focal conics, which have circular footprints formed by concentric SmA planes, at the interface with the active layer. The boundary shear stress experienced by the active filaments is much lower along than perpendicular to the SmA planes and hence the filaments preferentially move along circular trajectories, centred on the focal conics. Each conic must trap a total topological charge of +1. This is achieved by the microtubules folding to form at least two +1/2 defects which drive rotating swirls. **Bacteria** also drive rotating flows in circular cavities and this has been exploited to create patterned flow over long distances by using short channels to connect lattices of circular cavities each of which contains a bacterial vortex. Coupling between the cavities led to a long-lived checkerboard pattern of clockwise and anti-clockwise vortices [9].

Turning now to geometries embedded in three dimensions, experiments which confined microtubule-kinesin mixtures to the surface of a lipid vesicle were key to the development of the field [10]. The geometry of the resulting active shell implies a total topological charge of +2, which corresponds to four +1/2 defects. For moderate activities, the defects follow closed trajectories that quasi-periodically oscillate between tetrahedral and coplanar arrangements. Changing the shape of the shell can localise the defects: for prolate spheroids two of the defects are localised at the poles and the other two oscillate around the waist of the shell, while on oblate spheroids defects can undergo relative rotational motion, reminiscent of ceilidh dance dynamics. The effects of confining cytoskeletal material within a droplet have also recently been studied using Xenopus egg extracts. The extracts comprise extensile active bundles which pushed against the droplet boundary resulting in the bundles aligning into a rotating vortex structure. Cells are confined active systems and this research highlights questions about whether or how active flows are relevant to their functionality. See [1], [4], [5] for more complete references.

**Research Challenges:**

We suggest that there are two particularly timely and exciting research questions to be addressed in the near future. **The first** is to design machines that are powered by active matter. Motor proteins have evolved to be the engines driving cellular processes and bacteria are natural swimming microrobots. Can we usefully exploit the principles designed by evolution, and will dense active matter have a part to play in this development? If so, its dynamics will need to be controlled, and confinement is one promising way to ensure this.

**Secondly**, it is increasingly apparent that the concepts of active matter can be usefully applied to biological systems, both on sub-cellular and on tissue length scales. For example, topological defects have been identified and shown to have biological functionality in both bacterial and cellular monolayers. Given that in nature many of these biological systems interact with geometrical constraints, it is important to understand the interplay between active matter, topological defects, confinement and biological functionality.

**Liquid Crystalline Active Matter**

Igor S Aranson, Dept of Biomedical Engineering, Pennsylvania is State University

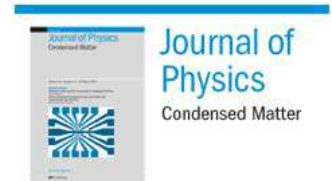

**Status**

Active matter encompasses a wide class of systems out of equilibrium. These systems are composed of interacting self-propelled "agents" or particles. The agents, such as bacteria or artificial microswimmers, transduce the energy stored in the environment (e.g. in the form of some nutrient or chemical fuel) into persistent mechanical motion. The examples range from biological cytoskeletal filaments [1], swimming organisms (bacteria and unicellular algae) [2], to synthetic catalytic nanomotors [3] and colloidal self-propelled Janus particles [4]. Active materials demonstrate a remarkable tendency towards self-organization and exhibit properties that are not present in traditional materials and composites.  Namely, active materials may self-repair, change shape on signal, or even adapt to new conditions.

Possibly the simplest and the most studied realization of active matter is a suspension of microscopic swimmers, such as motile bacteria or self-propelled colloids [5].  The studies over the last ten years revealed a plethora of new phenomena, from the reduction of the effective viscosity to rectification of chaotic motion.  Artificial rheotaxis, i.e. drift against the imposed flow, was also observed. Most of the contemporary studies of active matter are performed in an isotropic medium like water. Typical suspending liquid also contains some sort of a "fuel" in the form of a nutrient for bacteria or hydrogen peroxide for catalytic bimetallic microswimmers (often called nanomotors). It is desirable developing a realization of active matter based on an anisotropic and highly structured suspending medium such as a liquid crystal. Such a system would have the potential to control and manipulate active matter by the external electric or magnetic field, chemical gradients, light, or by surface treatment.

A

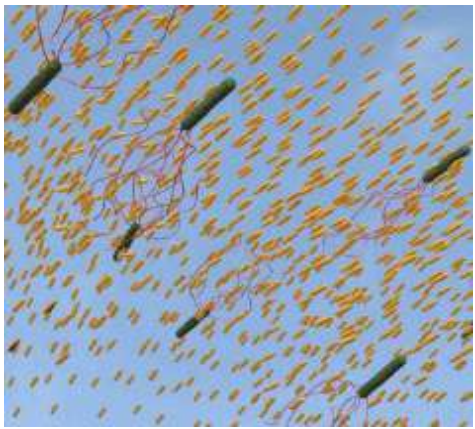

B

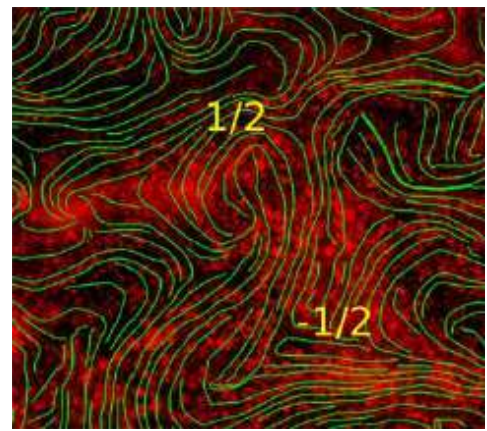

**Figure 1.**  (A) Living liquid crystal: lyotropic liquid crystal infused with swimming bacteria. (B) A fluorescent image showing bacterial concentration and nematic director orientation in a film of living liquid crystal.  Bacteria predominantly accumulate in the cores of positive topological defects and expelled from the cores of negative defects. Two select half-integer topological defects are marked [7]

Liquid crystals occupy an important niche between solids and liquids.  Due to a crystal-like ordering of molecules, liquid crystals may flow like a liquid and respond to deformations as a solid. A new class of composite materials is formed by mixing a water-soluble (lyotropic) liquid crystal with a small amount of active fraction, such as a suspension of swimming bacteria [6]. This active composite termed living liquid crystal exhibits a variety of highly-organized dynamic collective states, the spontaneous formation of dynamic textures of topological defects (i.e. singular configurations of the molecular orientation field), trapping and transport of bacteria in the cores of defects [7], as well as controlled and reconfigurable transport of cargo particles, and manipulation of individual trajectories of microswimmers [8], see Fig. 1. A predictive model of living liquid crystals is developed by coupling the well-established and validated model for nematic liquid crystals (so-called Beris-Edwards approach to liquid crystals) with the corresponding equations describing bacterial transport [7].

**Current and Future Challenges**

The studies of living liquid crystals provide fundamental insights into physical mechanisms governing a variety of active systems, such as tissues of migrating cells, cytoskeletal extracts, bacterial biofilms. Furthermore, living liquid crystals may stimulate the development of new classes of soft adaptive composites capable of responding to light, magnetic and electric fields, mechanical shear, airborne pollutants, or bacterial toxins [9].

For many technological applications, it is highly desirable to design a synthetic version of living liquid crystals, e.g., based on artificial microswimmers such as catalytic AuPt nanorods [3] or other autophoretic Janus particles [4]. The preliminary experiments show that the chemical composition of lyotropic liquid crystals is generally not compatible with the propulsion mechanisms of AuPt nanorods [3]. However, the anisotropy of liquid crystal can be utilized for guiding the motion of active Janus particles confined at the interface between the aqueous solution of hydrogen peroxide and nematic liquid crystals.

Microscopic rod-shaped particles can be also propelled by ultrasound [10]. Acoustic propulsion of nanorods in liquid crystals is relatively easier to achieve. Unfortunately, it is a challenge to create homogeneous distributions of the ultrasound intensity. Overall, it makes the control of nanorod motion and distribution in liquid crystal quite challenging. Colloidal particles in liquid crystals can be also propelled by electric or magnetic fields. Liquid crystal anisotropy can be potentially used to guide magnetic microswimmers. However, these systems were not mastered yet. The main bottleneck here is relatively high viscosity of liquid crystals. Therefore, practical use of the liquid crystal anisotropy for the control and guidance of artificial microswimmers remains an open challenge.

Up to date, only nematic liquid crystals were explored in active systems. The use of more ordered liquid crystalline phases, e.g., smectic (layered) or cholesteric (chiral) for the design of novel active systems is an intriguing opportunity. Preliminary experiments demonstrated that bacteria have a hard time swimming in smectic or cholesteric liquid crystals, mostly due to increased viscosity. However, here also comes an opportunity. For example, smectic liquid crystals often exhibit highly ordered textures of more complex defects, such as focal conical domains. Correspondingly, focal conical domains can be used as templates for the spatiotemporal organization of active matter. These domains can be controlled by imprinting surface patterns or applying a magnetic field.

Ultimately, there are many similarities between living and artificial active systems. For instance, bacteria and chemically powered nanomotors often have similar sizes, shapes, and propulsion speeds. However, this similarity turns out to be quite superficial. The main differences in this context stem from the inter-particle and boundary interactions. Bacteria typically bounce from the boundaries and scatter from other bacteria. In contrary, due to electric charging [3], chemically propelled microswimmers may hover very close to charged surfaces. Despite the apparent resemblance, living and artificial active matter systems do not demonstrate similar collective states or respond the same way to external stimuli. Bacteria often self-organize into large-scale vortices and coherent jets at the concentrations above the critical one [2]. In contrast, chemically propelled microswimmers form permanent static clusters [3].

From the biomedical perspective, a liquid crystalline order is often exhibited by common biological fluids, such as mucus, DNA solutions, or suspensions of viruses. In this list, mucus plays an exceptionally important role in higher organisms. For instance, the human body produces about one litre of mucus every day. Mucus is responsible for multiple biological functions, from aiding fertilization to protecting the reproductive tract. Mucus organization and microrheology affect motility and spatial organization of bacteria in a nontrivial way, see Fig. 2. Therefore, understanding the interactions between bacteria and biological liquid crystal such as mucus may have important implications for human health.

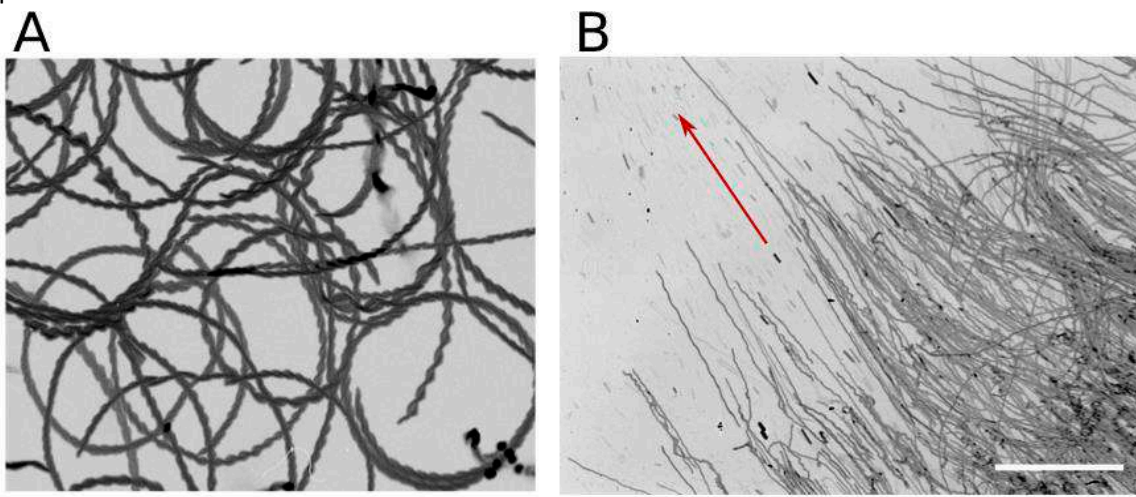

Figure 2. (A) Trajectories of bacteria *Bacillus subtilis* obtained by consecutive snapshot accumulation in Newtonian liquid. The bacteria display the well-known circles close to a wall. (B) Trajectories of bacteria over 8.5*s* near mucus/water interface. The bacteria penetrate mucus in the direction of the red arrow. Scale bar 100 µm. Courtesy of N Figueroa Morales, Aranson's lab.

**Concluding Remarks**

We surveyed opportunities and challenges in applications of living liquid crystals. Topological defects play a very important role in the spatiotemporal organization of this class of out-of-equilibrium materials. Topological defects, acting as elementary excitations or "atoms", can be used for control and manipulation of these living composites. Akin to vortices in superconductors, topological defects in living liquid crystals can be used in the simplified, coarse-grained description of active matter. Thus, studies of active topological defects and their spatiotemporal behaviour can provide deep insights into fundamental mechanisms governing the organization of active systems.

**Acknowledgements**

The work was supported by the National Science Foundation under Grant NSF PHY1707900

# Active Suspensions with Programmable Interactions

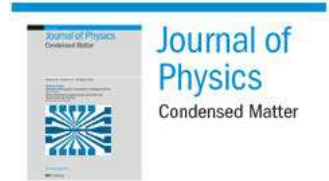

## Clemens Bechinger
Physics Department, University of Konstanz, Germany

## Status

Active colloidal particles (APs) are currently considered as model systems to mimic collective behavior under controlled conditions but may also find technical use as microrobots to perform complex tasks in liquid environments [1]. In the meantime, a plethora of propulsion schemes has been developed which allows APs to harvest energy from their surrounding and convert it into translational motion. A key question in this field is to understand, how active particles are able to organize from random disperse into dense and highly organized collective structures (swarms, flocks) as known e.g. from birds, fish, insects or even bacteria [2]. Clearly, such behavior requires robust communication mechanisms between individual group members. Conversely to living systems where collective behavior is achieved by complex internal processes which are not always fully understood, the response of APs is much simpler. When user-defined interaction-rules were be applied to synthetic systems, this would provide a novel and versatile platform to gain a better understanding of the relationship between communication and self-organization.

A possible experimental route to impose user-controlled interactions to a suspension of APs is the use of a feed-back mechanism by which the propulsion of APs can be controlled on an individual level [3]. Recently, this idea has been experimentally demonstrated using Janus particles which are propelled by a light-induced mechanism [4, 5]. When such particles are illuminated with a focused laser beam, their propulsion motion can be controlled on a single particle level. Using a rapidly steered laser beam and a feedback mechanism, this allows to adjust the AP's motion depending on the surrounding particle configuration, the latter tracked in quasi real-time employing digital optical microscopy.

## Current and Future Challenges

As a specific example for such user-defined interaction rule, we discuss how quorum sensing [6, 7] can be implemented in a suspension of synthetic ABPs. In bacterial systems, quorum sensing is achieved by the release and detection of diffusing signaling molecules which are characterized by a diffusion coefficient $D_c$, their production rate $\gamma$ and their finite lifetime $\tau$ [8]. The concentration sensed by an individual $i$ is then given by

$$c_i(t) = \tilde{c} \sum_{j \neq i} \frac{\sigma}{r_{ij}(t)} \exp(-r_{ij}(t)/\lambda)$$

<div align="right">Eq.(1)</div>

where $r_{ij} = \left| \vec{r}_i - \vec{r}_j \right|$ is the distance to its neighbour $j$, $\sigma$ the size of an individual, and $\lambda$ the decay length $\lambda = \sqrt{D_c \tau}$ (being a measure of the range of communication). The prefactor $\tilde{c}$ is given by $\tilde{c} = \gamma / 4\pi D_c \sigma$. When the concentration sensed by an individual exceeds a sharp threshold value $c_{th}$, this triggers a sudden change in its behavior, e.g. in its motility leading to the efficient formation of bacterial colonies. To mimic such quorum sensing behavior, the motility of particle i is set by the following rule: when the concentration $c_i$ (as determined from the experimentally measured particle configuration and the use of Eq.1) exceeds a user-defined threshold concentration $c_i > c_{th}$ the particle is non-motile, i.e. its propulsion velocity is zero. Otherwise it becomes motile and propels with velocity $v_i = v_0$, i.e. it is locally illuminated with the laser beam. Fig.1 shows how the normalized density distribution of a suspension of APs changes upon increasing the threshold $c_{th}$.

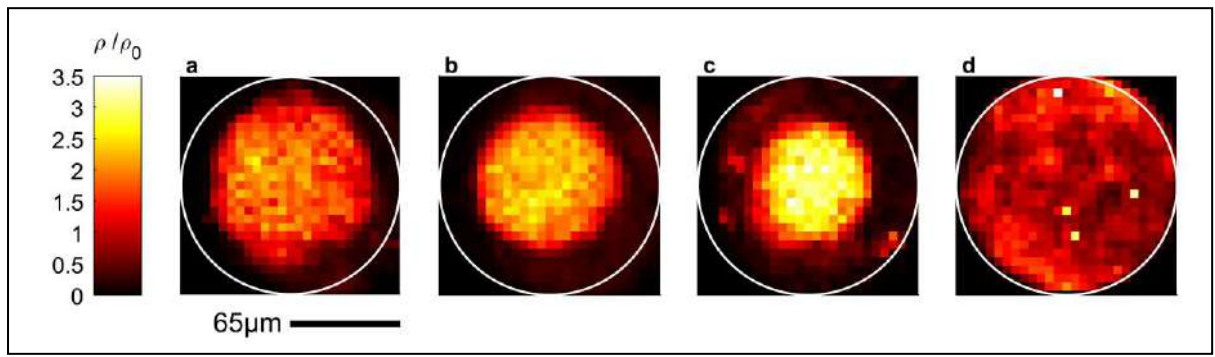

**Figure 1.** Density distribution of ABPs induced by quorum sensing with threshold values $c_{th}$ =5.9,7.3,8.6,c (a-d) and decay length =10. With increasing cth the clusters become denser and smaller until cluster formation breaks down for a too large threshold (figure adapted from reference [5].

To understand, why cluster formation is supported by quorum sensing, it is important to recall that the propulsion of APs becomes zero in regions with high local particle density. Accordingly, particles can only diffusively escape dense regions which then facilitates their growth. On the other hand, particles in dilute regions are motile, which largely increases their probability leaving such areas. Accordingly, particles will permanently switch from a motile to a non-motile state. Indeed, it can be shown, that cluster formation becomes most enhanced when the switching rate (motile to non-motile and vice versa) becomes large and thus supports the idea, that quorum-sensing rules lead to enhanced clustering similar to living systems.

The above approach can be easily changed to other types of rules, e.g. including memory-effects or non-reciprocal interactions where the mutuals response of two individuals is not symmetric (opposed to the above example of quorum sensing). Non-reciprocal rules are relevant e.g. when the environmental sensing of individuals is based on vision. For finite vision cones, an individual A may see individual B but not otherwise which will largely affect their collective behavior.

It is important to realize, that the above experimental approach does not aim towards modelling a specific collective state but imposes an interaction rule. Contrary to numerical simulations, all relevant particle interactions (phoretic, hydrodynamic) are fully contained within this approach and thus allows to observe collective behavior under realistic conditions. This apporach can be also extended to e.g. viscoelastic (opposed to Newtonian) swimming media which comprise the natural habitat of many living microorganisms. Such fluids are characterized by rather long stress-relaxation times which leads to a number of significant changes in their swimming motion [9-10].

**Concluding Remarks**

The possibility to impose user-defined interaction rules in systems of APs largely expands their use as model systems resemling the collective behavior of living systems. In addition to quorum-sensing rules, almost any other type of communication can be established, e.g. vision cones, time-delays, etc. Because mutual interactions in living systems are often not clear, comparison of the collective behavior with synthetic systems provides a promising route to uncover what type and over what distances information is exchanged between individuals to achieve a collective state.

**Acknowledgements**

I thank T. Bäuerle, A. Fischer and T. Speck for stimulating discussions. This work was supported by the microswimmers SPP 1726 of the German Research Foundation (DFG) and the ERC Adv. Grant ASCIR (693683).

# Microswimmers interacting by hydrodynamic fields

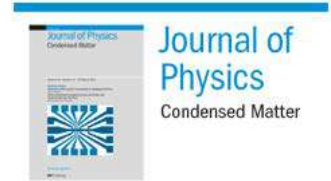

Holger Stark, Institute of Theoretical Physics,
   Technische Universität Berlin, Hardenbergstr. 36,
      10623 Berlin, Germany

## Status

Microswimmers use friction with the surrounding fluid to swim forward and thereby induce flow fields [1]. The details of hydrodynamic flow depend on the specific type of swimming mechanism involved. An intriguing example is the African trypanosome, the causative agent of the sleeping sickness [2]. It moves with a bending wave running along a flagellum, which is attached to the cell body. However, microswimmers can be categorized by their hydrodynamic far fields [1,3]. Flow of the leading force dipole decays as $1/r^2$ and flow of the source dipole as the next contribution decays as $1/r^3$, where $r$ is the distance from the swimmer. The force dipoles are divided into pushers and pullers also called extensile or contractile swimmers, respectively, that either push fluid out along their swimming axis or they pull it in. These swimmer types are often used to explore how the self-generated flow fields of microswimmers determine their behavior.

Microswimmers interact with bounding surfaces through their flow fields. For example, Ref. [4] shows how the detention time of a swimmer between reaching and leaving a wall strongly depends on its basic type and thus on its hydrodynamic interactions with the wall. Most prominently, microswimmers interact with each other through their self-generated flow fields and thereby induce appealing emergent collective motion, which is reviewed in Ref. [3]. In the following, we only mention a few illustrative examples including the author's own work.

First, a very prominent example is turbulent motion in a bacterial bath identified by a power-law decay of the spatial power spectrum of velocity fluctuations [5]. While a model with self-propelled rods and continuum theory reproduce the turbulent motional patterns [5], a recent study using lattice-Boltzmann simulations identifies the turbulent state only for pushers through large-scale fluid vortices and jets [6]. Second, colloidal rollers, which perform Quincke rotation in an external field, form propagating bands and at higher densities a polar liquid [7]. The rollers interact hydrodynamically through their hydrodynamic rotlet singularities and thereby form the observed polar order.

Third, in an own work we studied active Brownian particles in a harmonic trap potential. They interact with each other by the leading stokeslets that result from the trapping forces [8]. With increasing Peclet number or swimming velocity, the active particles show a relatively sharp transition between an isotropic configuration and the pump state, where they are aligned in a tightly packed region close to the trap center (Fig. 1, left). They form a regularized stokeslet, which we used to formulate a mean-field theory for the observed polarization (Fig. 1, right).

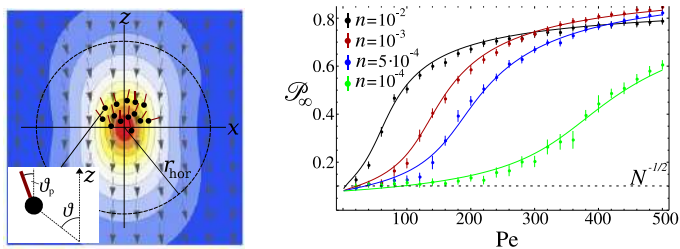

**Figure 1.** Left: Sketch of the fluid pump formed by aligned active particles. The flow field of the regularized Stokeslet is shown. Right: Polarization versus Pe for different densities. Solid lines are calculated from a mean-field theory, where single active particles align in the regularized Stokeslet. [Reprinted from Ref. [8] with kind permission of the American Physical Society; copyright (2014).]

Fourth, magnetotactic bacteria align in an external magnetic field. When swimming in a microchannel against a Poiseuille flow, they become more and more focused with increasing field strength and ultimately show an instability to a pearling state, where the bacteria circulate in droplets [9]. A recent theoretical treatment traces the instability to the interplay between activity, external alignment, and magnetic dipole-dipole interaction [10]. The authors of [10] surmise that in the experiments hydrodynamic interactions are most likely negligible. However, theoretical work without flow reports instabilities of aligned magnetic microswimmers due to hydrodynamic interactions [11] or stresses the generation of flow [12]. Thus, the role of hydrodynamics in this interesting problem should be further explored.

The squirmer is a model swimmer, which was introduced by Lighthill and later Blake to desribe ciliated microorganisms [13,14]. It has extensively been used by Ishikawa, Pedley, and coworkers to study their hydrodynamic interactions and collective motion (for a review see [13]). For the same purpose we have implemented it in the particle-based method of multi-particle collision dynamics (MPCD), which solves the Navier-Stokes equations including thermal noise, in order to perform large-scale simulations [14]. The squirmer is a spherical particle with a prescribed velocity field at its surface, which propels the particle. The swimmer type is tuned by the parameter β between a pusher (β < 0), neutral squirmer (β = 0), and puller (β > 0).

We have used squirmer model swimmers to analyze how they behave in a gravitational field. An important quantity is the ratio α of the swimming velocity to the sedimentation velocity in bulk. Even a single squirmer with α < 1 and at large Peclet number shows interesting bound states [Fig. 2a)]. They mainly correspond to stable fixed points in squirmer height above the bottom wall and orientation [15]. For example, in the floating state (β = 0,2) the squirmer points upwards. Then, the floating height is determined by a balance between the upwards swimming and the sedimentation velocity. The latter decreases when approaching the bottom surface due to increasing friction. We also simulated thousands of squirmers under gravity [16] and found a sedimentation profile with layering at the bottom and after a transitional region a very dynamic profile, which on average is exponential as for single microswimmers [Fig. 2b)]. The vertical squirmer current plotted in Fig. 2c) reveals a convection cell.

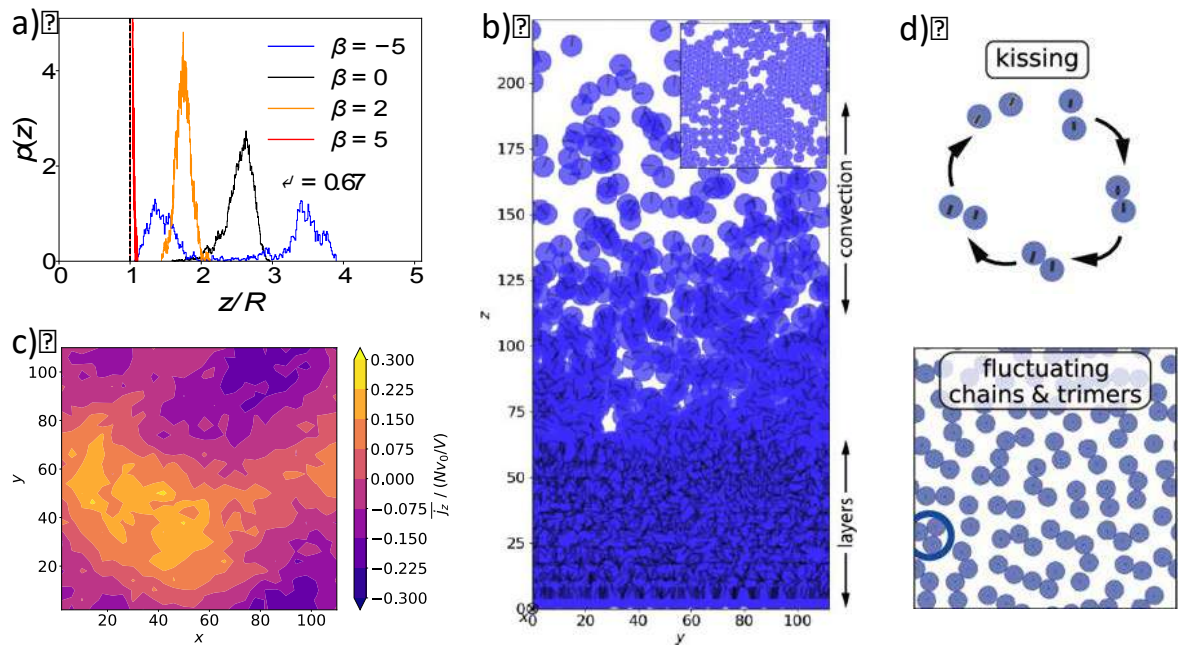

**Figure 2.** a) Distributions p(z) of heights z during motion of a single squirmer with α = 0.67. They represent stable floating (β = 0,2), wall pinning (β = 5), and the bimodal state of recurrent floating and stable sliding (β = -5). b) Snaphot of 2560 neutral squirmers with α = 1.5 and volume fraction of 0.244 in steady state. Inset: bottom layer of squirmers. c) The vertical current density of squirmers, averaged over the exponential region and time, is color-coded in the vertical plane. d) Monolayer of squirmers confined to the bottom surface by strong gravity. They form transient states such as kissing and flucutating chains and trimers, which form and break up. [Figure a) reprinted from Ref. [15] with kind permission of IOPscience; copyright (2018); Figs. b) and c) reprinted from Ref. [16] with kind permission of the Royal Society of Chemistry; copyright (2017); Fig. d) reprinted from Ref. [21] with kind permission of the Royal Society of Chemistry; copyright (2019).]

A prominent feature of active particles is the motility-induced phase transition (MIPS), which describes a phase separation into a dilute and dense phase of active particles. The question came up how hydrodynamic flow fields influence MIPS. A pure two-dimensional treatment of squirmer suspensions observes hydrodynamic suppression of MIPS due to the long-range nature of hydrodynamic interactions [17], while in a quasi-2D geometry of a monolayer of squirmers confined between two plates MIPS clearly occurs [18]. We have also mapped out the full binodal in the phase diagram of Peclet number versus mean density and found that it depends on the mean density of squirmers [19]. This is a clear feature of hydrodynamics and the non-equilibrium, which we could rationalize by adding hydrodynamic pressure to the active pressure balance between dilute and dense squirmer phases. A recent work [20] notices that the density of MPCD particles used in Ref. [18,19] are reduced in the dense squirmer phase. Thus the MPCD fluid is compressible. To reduce this effect, the authors of Ref. [20] strongly enhanced the density, which enhances the fluid viscosity. To keep the bare Peclet number (defined for a single squirmer) equal to the one used in [18,19], they reduced the squirmer veocity and MIPS could not be observed. However, the bare Peclet number is not the most suitable quantity to describe MIPS since hydrodynamic flow fields in multi-squirmer systems strongly enhance rotational diffusion [17,18]. Furthermore, larger system sizes are necessary to further assess the findings.

**Current and Future Challenges**

We start with two challenges, we are currently working on. Extending our work on squirmers under gravity, we study a monolayer of *squirmers* at the bottom surface that forms *under strong gravity* and observe different collective dynamics when varying area fraction and squirmer type [21]. This includes swarming and transient cluster/chain formation [see Fig. 2d)]. Most remarkably, for smaller densities neutral squirmers and pullers repel each other due to their self-generated flow fields and form what we call a hydrodynamic Wigner fluid, which shows glassy features.

We also investigate the collective dynamics of *bottom-heavy squirmers*, where a gravitational torque aligns them along the vertical to swim preferentially upwards reminiscent of algae. Already for the neutral squirmer the state diagram of ratio α (of swimming to sedimentation velocity) versus torque reveals interesting emerging collective dynamics. Between sedimentation (α < 1) and inverted sedimentation at large α, we identify stable convective plumes and hovering clusters that spawn single squirmers. Both, plumes and hovering clusters, also appear as transient structures, which initially form from a uniform state and then dissolve with time. We are currently analyzing them quantitatively.

Studying the collective motion of microswimmers in *external fields* certainly is a promising direction to further discover new phenomena in combination with their hydrodynamic interactions. This includes magnetotactic bacteria in an external magnetic field in various geometries (for example, group of Clement, Paris) and microswimmers in externally imposed flow fields such as Poiseuille flow in microchannels.

In a large class of active particles self-diffusiophoresis generates a velocity field at their surfaces through which they swim. Chemical interactions with nearby particles influence the surface flow field and vice versa. Studies of a *full coupled treatment of chemical and hydrodynamic fields* has just begun, for example, in the groups of Lauga (Cambridge) and Popescu (Stuttgart) and certainly needs further attention. This also includes the special case of active emulsions investigated in the group of Maas (Göttingen), where micelles formed by surfactants are important.

A further direction to pursue in connection with hydrodynamic interactions is the collective motion of microswimmers with more complex shapes starting from simple rods and proceeding to real dynamic swimming shapes observed, for example, for the African trypanosome. Finally, also the solvent can become more complex by making it viscoelastic.

**Concluding Remarks**

Research in some of the just proposed directions has already begun. They will contribute to identifying novel emerging collective patterns due to hydrodynamic interactions between microswimmers.

**Acknowledgements**

The author thanks A. Zöttl, J. Blaschke, M. Hennes, J.-T. Kuhr, F. Rühle, K. Schaar, and K. Wolff for collaborating on this inspiring topic and acknowledges funding from the DFG within the research training group GRK 1558 and the priority program SPP 1726, project number STA 352/11.

# Patterns of collective motion and escape in huge flocks of starlings

CK Hemelrijk, University of Groningen, the Netherlands, c.k.hemelrijk@rug.nl

## Status

Patterns of collective motion displayed by schools of fish and flocks of birds have attracted great interest in many fields of science. It is now widely acknowledged that the beautiful coordination among group members is achieved without a leader and largely by self-organisation [1]. Individuals that are moving are supposed to coordinate with others by (some of) the three A-rules: they Avoid collisions with others nearby, Align their heading to others at medium distance and are Attracted to those further away. The absence of a leader is particularly remarkable in the case of the starling flocks because they are exceptional in both their large number of flock members (up to many thousands) and their low number of individuals with whom they are supposed to coordinate their motion, namely their 6 to 7 closest neighbours only [2]. Models of individuals that move and follow the three A-rules, however, do not generate the flocking patterns typical of starlings, because they lack the great variability in shape of these flocks. The most frequent shape of schools of fish, instead, is oblong, also when turning. Their consistency of shape is captured by the models of moving and coordinating by the 3 A-rules whereby individuals avoid collisions by slowing down. The shape of flocks of starlings, however, changes with each turn [3]. In a new computational model, StarDisplay, by adding to the three A-rules a simplified model of aerodynamics of flying behaviour with rolling during turning, like in real birds and airplanes, the shape of the flock changes with each turn, like in empirical data of flocks of starlings (and also pigeons). Other detailed empirical traits of flocks described by physicists from Rome, are also found in StarDisplay, namely: the repositioning of individuals during turns, aspects of internal structure, i.e. the scale free correlation between the absolute length of the flock (in m), the correlation length of the deviation of the velocity and speed of individuals from that of the centre of gravity also in relation to speed control [4], and the degree of disorder or diffusion in the group [5].

Little attention, both empirically and theoretically, has been given to patterns of collective escape from an attack by a predator, even though predation is considered as the major evolutionary drive for group living [6]. Empirical patterns of collective escape are most impressive in huge flocks, such as those of starlings. Recently these patterns have been characterised, quantified, and classified, in relation to the hunting behaviour of the predator at different degrees of predation risk. The main patterns observed are blackening (where suddenly part of the flock darkens), wave event (where a dark band travels over the flock in the direction away from the predator [7,8]), split (the flock splits in sub flocks), and flash expansion (part of the flock 'explodes' with individuals flying in all directions away from the point of attack) (Figure 1)[9]. In computational models of fish schools under attack of predator, patterns similar to flash expansion and split have been shown to emerge [10]. In these models, individuals are moving and coordinating following the three A-rules. A predator was added that was attracted from a large distance to the members of the school moving towards them. Next to their rules of coordination, fish were supplied with rules to move away from the approaching predator in order to flee ahead of it. The collective patterns of flash expansion and split emerged in this model when the tendency of the individuals to coordinate with group members was low (weighted as 0.3) and that to escape the predator was high (weighted as 0.7)[11]. This is in line with the finding on starling flocks that flash expansion happens particularly when threat is intense, namely when the predator attacks at high speed by stooping from above rather than at low speed from the side or below [9]. This seems adaptive, since in case of intense threat behavioural motivation will be mainly oriented to escape rather than towards coordinating in a flock.

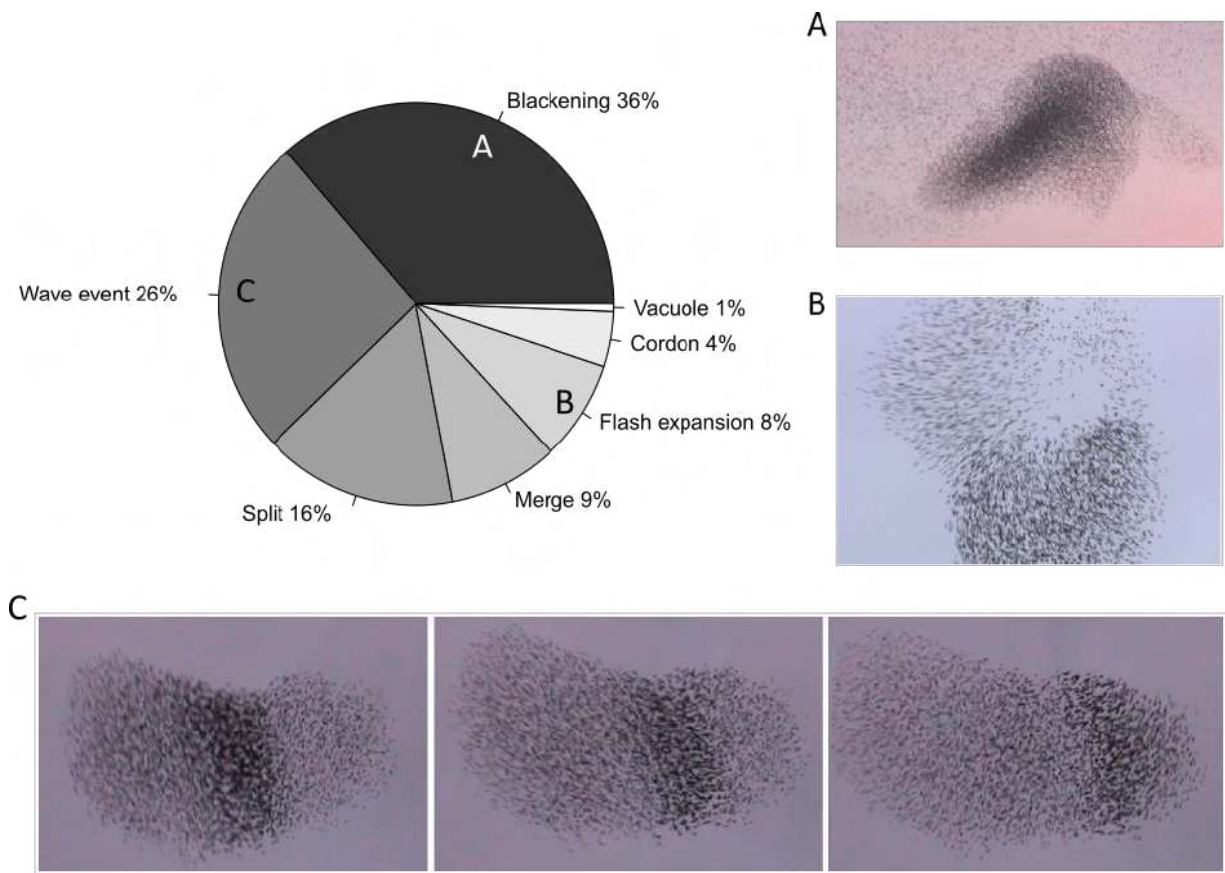

Figure 1. Patterns of collective escape in starling flocks above Rome. Pie chart shows relative frequencies of different patterns. A) Blackening, B) flash expansion, C) agitation wave with three subsequent time units showing a dark band moving over the flock to the right *(with permission from [9]*.

During waves of agitation individual birds are supposed to copy the escape manoeuvre displayed by neighbours. What kind of escape manoeuvre they copy, cannot be detected due to the large distance (hundreds of meters) from which humans observe the flock. Therefore, we have investigated for two escape strategies at the individual level whether dark bands emerge that are travelling over the flock in the model StarDisplay. The escape strategies were a) a manoeuvre to flee away fast from the predator inwards into the flock (resulting in a density wave) and b) a skitter motion in the form of a zigzag (Figure 2DE), which happened by banking sideward and back again (resulting in an orientation wave) [8]. It appears that in the model dark bands are observed only when the wave is an orientation wave based on the zigzag escape manoeuvre (Figure 2 A-C) and not when it is a density wave. In an orientation wave dark bands are visible due to the larger area of the wing that we temporarily observe when birds are banking while zigzagging (Figure 2)

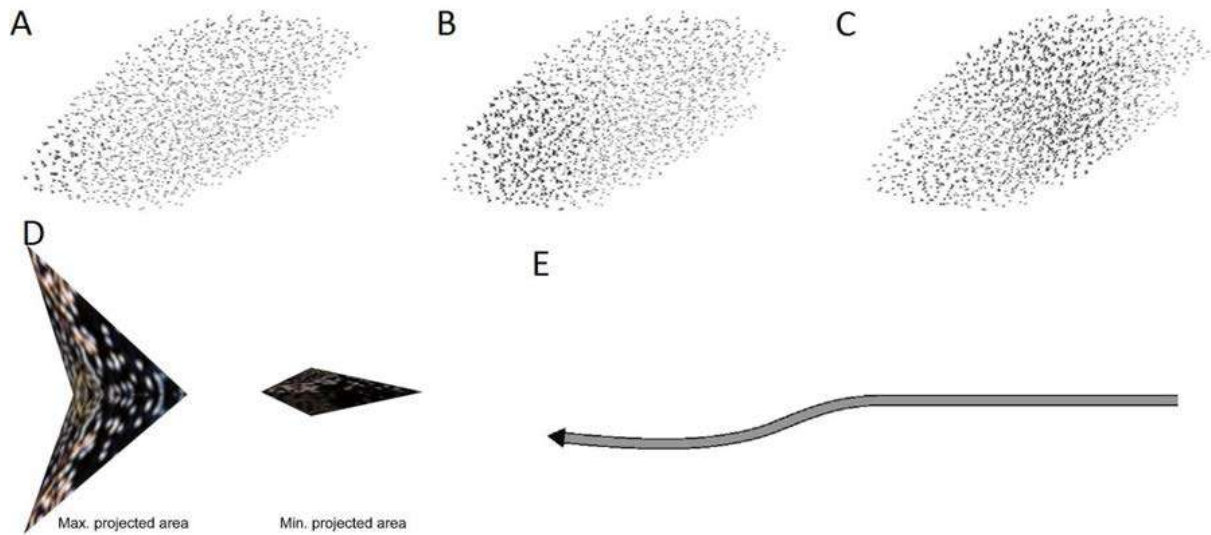

Figure 2. The agitation wave in the model, StarDisplay. A, B, C) three subsequent time units showing a dark band moving over the flock to the right. D) wing surface from above and the side (as shown when banking during turning). E) the zigzag motion of the bird where it rolls to turn *(permission form* [8]

## Current and Future Challenges

**Damping of a wave**

As to waves of agitation it is of interest to discover whether they are damping with time, and if so what causes this, because during an undisturbed turn of a flock all individuals copy the turning motion completely, without attenuation[11]. However, in starlings waves of agitation occur in response to a predator attack [7]. Thus, we may imagine that further away from the location of attack, birds are less frightened and, thus, skitter less, causing us to observe a smaller variation of surface of the wing. This process needs to be studied in detail in both models and empirical data, by measuring dark bands moving over the flock, and testing whether they become more greyish with time nearing the end location, thus, whether the amplitude of illumination decrease while travelling over the flock.

**Blackening of flocks**

Sometimes parts of flocks show sudden blackening. This may happen both when a predator is present and when it is not. What behaviour underlies blackening cannot be detected due to the large distance of the observer from the flocks. Therefore, we studied it in the computational model, StarDisplay. In the context of a predator being absent and the flock swirling above the roosting site (site for sleeping), we showed that blackening is due to changes in orientation of birds towards the observer (leading to changes in wing surface exposed to the observer), but is neither due to a change in density in the flock, nor to a change in distance towards it, nor due to a change of orientation of the complete flock towards the observer Costanzo et al (submitted). It still needs to be studied, however, what causes blackening in flocks of starlings when they are under attack of a predator. In this case, next to changing the surface of the wing relative to the observer, also folding of flocks and variation in density may be a cause.

**Strikes between airplanes and bird flocks**

The number of strikes between bird flocks and airplanes increases steadily, due to the larger population of birds, increased air traffic, the faster speed of airplanes and their flights being more silent[12]. In order to prevent bird strikes, at present we are working on developing a method for effectively chasing flocks away with an artificial RobotFalcon. Here, knowledge on patterns of flock escape is important, first, because flocks of different species may differ in their internal structure [13] and patterns of escape, and second, because the behaviour of the RobotFalcon should be customised to elicit different kinds of flock responses. Our method involves filming the patterns of collective escape from nearby using a camera on the head of the RobotFalcon and from a distance using a ground-based camera while we are attacking flocks with the RobotFalcon. We focus on flocks of several species, such as gulls, crows, starlings, and lapwings.

## Concluding Remarks

In order to gain understanding what processes underlie the different ways of collective escape in bird flocks, we need to learn more on both the trajectories of motion of individuals and of the group. This involves the integration of studies using computational modelling and empirical data of collective motion during both normal traveling and collective escape. Such empirical studies should involve both observations on flocks under natural attack and when being attacked by RobotPredators.

## Acknowledgements

I thank all collaborators who contributed to the work on collective motion and escape in starling flocks that inspired this paper. I am grateful for financial support by the 6th European framework under the STREP-project ''Star-Flag'' (n12682) in the NEST-programme of ''Tackling Complexity in Science,'' Cognitie Pilot project of NWO (051.07.006), the Nederlandse Organisatie voor Wetenschappelijk Onderzoek (NWO; file 823.01.017), the Netherlands Organisation for Scientific Research (NWO), the Open Technology Programme (OTP), project number 14723, two visitor grants from NWO (040.11.468 and 040.11.573/ 1798), a startup grant of the University of Groningen and a grant from the Gratama foundation.

# Simulating Active Networks of Cytoskeletal Filaments

Francois J. Nedelec
Sainsbury Laboratory, Cambridge University
Bateman Street, Cambridge, CB2 1LR, UK

## Status

Filamentous systems are ubiquitous in nature and offer a rich playground to experimental and theoretical investigators. Many of the most fundamental events of living cells, including division or migration, rely on intracellular processes involing cytoskeletal filaments such as actin or microtubules. These processes also contribute to morphogenesis at the level of the tissue and higher. Importantly, the movements of molecular motors and cytoskeletal filament dynamics often bring these systems far from thermodynamic equilibrium. Consequently, cytoskeletal processes are best viewed as microscopic "machines", with original behaviour that lies between inert and living matter. New theoretical tools have had to be developed to study them.

We focus here on mathematical models in which the filaments are explicitly represented. Each filament in the system will typically have a position, an orientation and a length, and may be treated as rigid or flexible. This level of description lies between more microscopic models in which some structural aspects of the constitutive proteins are included, and coarser models in which, for example, only the average direction of multiple filaments is considered. A serious disadvantage of more microscopic descriptions is their additional computational costs, which often preclude simulating many filaments on a relevant time scale. On the other hand, the coarse-grained approach has seen many applications and variations [1], but it often starts by assuming some sort of local order, thereby losing the ability to describe all possible filament configurations. The focus of this article, filament-based simulations, typically excel at describing small systems by allowing the necessary details to be incorporated, as described below. While adding more features may preclude thorough mathematical analysis, valuable insights can be gained from numerical results obtained through computer simulations, in a way that promotes our understanding of the biological system and guides the development of theory. These models are limited by computer performance and the algorithms used to perform the simulation. Thanks to the spectacular development of computer technology in the past decades, their range of applications has tremendously increased. Our focus has shifted from writing efficient code to exploring which combinations of assumptions best represent biological reality. Hardware limitations will remain, but nature offers many systems of a size that can be simulated rapidly enough that many conditions can be analysed on a high-performance cluster. Important progress is still to be done, as we realize that current models lack some essential biological features. Yet, precise models of processes such as mitosis, cytokinesis, endocytosis or cell migration should become available in the coming years, helping us to understand how these cytoskeletal machines operate more globally.

Filament-based simulations find their roots in polymer physics. The forces produced by immobilized molecular motors could readily be added to a semiflexible chain model to simulate gliding assays [2]. It took efforts to extend this approach to multiple filaments, initially using rigid filaments and later with flexible filaments. Filament dynamics is a topic in itself that is still under active investigation, but many models use a lattice to store the chemical state of the filament building blocks (GTP/GDP for tubulin). With this capacity built into each filament, one can study the crosstalk that naturally exists between mechanical force and filament dynamics. For example, it has been possible to study how an aster of dynamic microtubules would move within a fixed volume. Since then many other situations have been considered in which an organelle, the nucleus or multiple nuclei, move within the boundaries of the cell. Microtubules can push on the boundaries, and may be pulled by molecular motors (dynein) anchored at the boundary or within the cytoplasm. Due to the configurations generated by the anchoring of microtubules at their minus-ends, direct microtubule-microtubule interactions are rare in such systems and may be ignored for simplicity.

Some important problems were tractable without considering the movement of the filaments: the trafficking of motors and vesicles along an immobile cytoskeletal network or chromosome capture by microtubules. In plant cells, microtubules are specifically anchored to the extracellular cell wall, and although they may be treadmilling, their organization is determined by the assembly dynamics at their tips and filament severing. Important computational gains are obtained by freezing the irrelevant degrees of freedom of the filaments, leaving the collisions detection algorithm as the bottleneck of the calculation. On the other hand, it may be desirable to simulate the Brownian motion of filaments, including their bending, when this qualitatively changes the chance for a filament to find a suitable target.

In general, it is essential to model the movements of filaments to study the evolution of the system. This constitutes the bulk of the calculation on which algorithmic choices will have a major impact [3]. Filaments may contact, but the best approach for implementing steric interaction has not yet been determined. Multivalent motors may bridge adjacent filaments promoting their active sliding relative to each other. A crosslinking motor or a passive crosslinker can be represented by an elastic link between the filaments, that may be Hookean or non-linear, defining a tension $f$. The position where a motor is attached on a filament evolves according to the force-velocity relationship of the motor $v(f)$ that is linear in the simplest case. To simulate the self-organization of microtubules [4], or the contraction of actin networks [5], a continuous description of motor movements was sufficient, where their displacement is $\delta = h\,v$, during a time interval $h$. This simplification was however insufficient to model the elongation of the spindle during anaphase, because molecules compete for the binding sites on the microtubule lattice. In that case, the movement of motors is modelled as instantaneous jumps on a lattice of binding sites, reflecting better the reality of protein interactions. This stochastic representation makes it possible to simulate traffic jams between motors or crosslinkers, and is readily integrated with the mechanics of the filaments. Because molecules of different type may or may not interfere on the microtubules, the modeler may enable or disable this feature on a case-by-case basis. Figure 1 illustrates simulations done recently in my group, where ~10000 filaments were simulated for ~1000s in ~2 days using a single processor core[4,5]. Figure 2 illustrates a realistic actin network with ~300 filaments that is found in yeast cells.

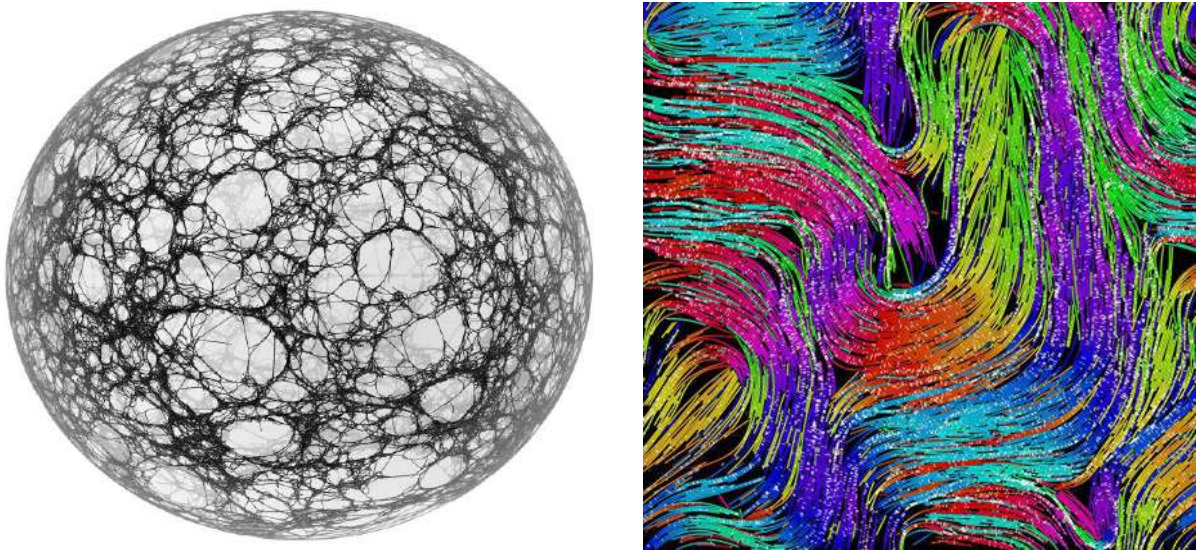

**Fig. 1:**   *Left: an network with 16k actin filaments of length 3μm on the surface of an ellipse of radii 10 and 11μm, contracts under the action of 240k motors and 240k crosslinkers* **[5]**. *Right: an active nematic pattern with ~5k dynamic microtubules of mean length 6μm driven by 16k crosslinking motors, within periodic boundaries 60×60μm* **[4]**.

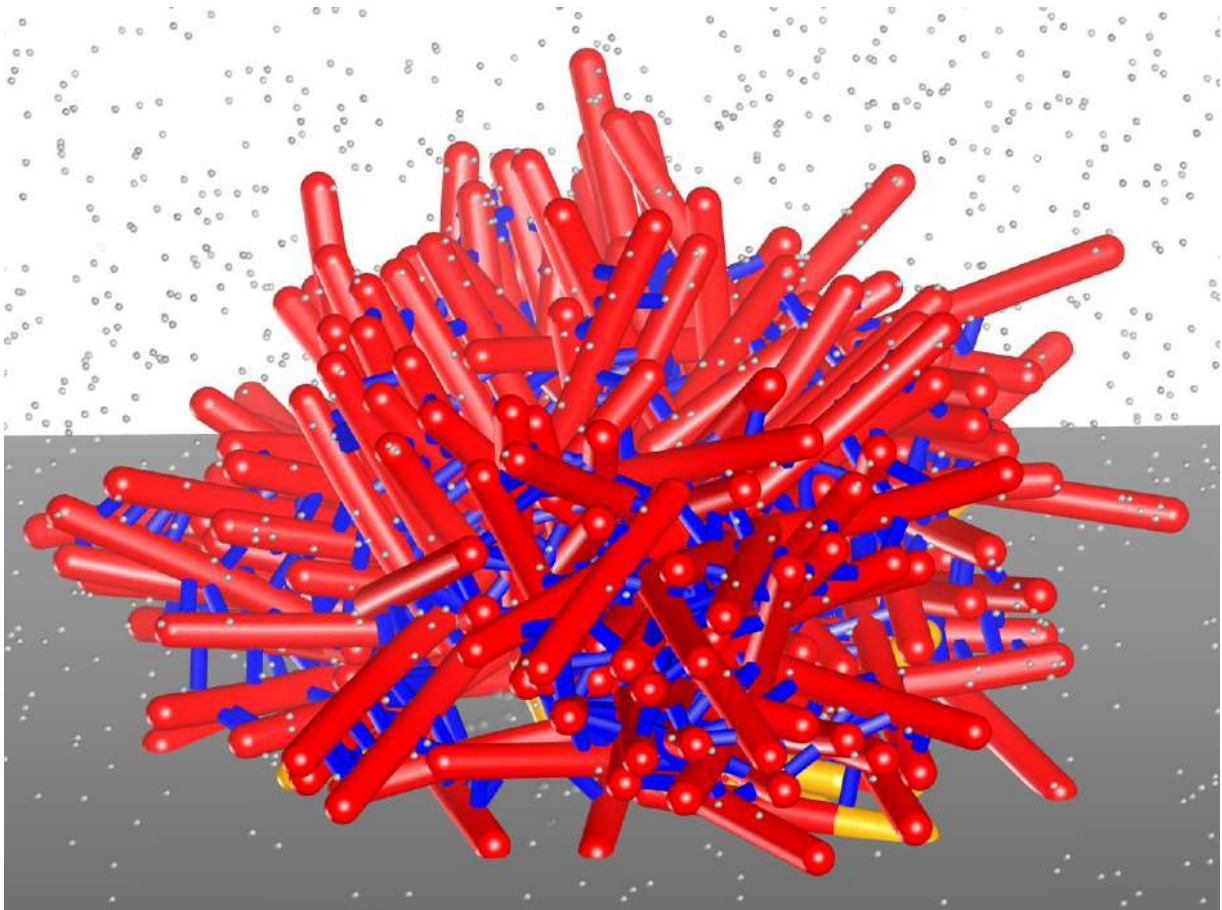

**Fig. 2:** *Predicted 3D organization of a branched network with ~300 actin filaments of length 55nm driving endocytosis in yeast. The filaments are represented with a realistic diameter of 7nm, and the whole actin burgeon is ~160nm across* **[6]**.

## Current and Future Challenges

Current development is aimed at problems involving the cytoskeleton and cell shape. At first one must introduce a general description of a volume, for example using a triangulation, to be able to study cytoskeletal systems within arbitrary shapes [7]. One can focus initially on situations where the shape evolves only slowly compared to the lifetime of cytoskeletal filaments within. This can be sufficient to study morphogenesis of multiple cells [8]. For a certain class of problems, it will be necessary to include hydrodynamic interactions between the filaments, as demonstrated with cytoplasmic streaming [9]. For increased realism, we may need to move beyond our current representation of filaments as having a backbone of vanishing thickness. This will be required for modelling systems where the motors follow helical paths on the microtubule surface [10]. Similarly, it is desirable to represent the natural twist of actin filaments, a source of chirality in actin networks that is of biological importance. While some of these developments have been done already by other groups, they cannot be readily combined until the underlying software is merged or the feature is reimplemented on a common platform. This illustrates that the inter-operability of the computational code developed by our community remains a challenge to be addressed.

While building a simulation one must face a difficult trade-off between simplicity and realism, bounded by computer power and the struggle of finding good algorithms. Simulations should start simply, and in view of the complexity of the underlying code, one should only parsimoniously add features and test every step. Yet, we have reached problems where the simplest description is not suitable anymore, convincing us to add more elements into the model to keep up with the latest experimental discoveries. The value of simulations is multifaceted. Building a rigorous model makes one think in unexpected ways, in my experience always leading to deeper understanding of the system. Potentially summarizing years of discovery by a whole community of researchers, simulations offer vastly under-exploited teaching and outreach opportunities. The pleasure of playing with an animated mock-up of the invisible reality is often limited to the lucky few who have tried to run the software, which is not conceived for the general public.

## Acknowledgements

I thank all my collaborators who over the years have contributed to the development of simulations through the work discussed here, and A. Gurzadyan, R. Carter and J. Rickman for critical reading. Current research in the group is supported by the Gatsby Charitable fund.

Revised version

# Active Cell Nematics: Multicellular Architectures and Flows

Trinish Sarkar, Thibault Aryaksama, Mathilde Lacroix, Guillaume Duclos, Victor Yashunsky, and Pascal Silberzan

Laboratoire PhysicoChimie Curie, Institut Curie, PSL Research University - Sorbonne Université - CNRS - Equipe labellisée Ligue Contre le Cancer, 75005 Paris, France

## *Status*

Spindle-shaped elongated cells in bidimensional cultures tend to align together and form large, perfectly aligned domains [1,2]. These domains are separated by nematic defects of charge ± 1/2 that prevent them to fuse (figure 1a). Interestingly, such a nematic organization is observed in several biological tissues such as muscles or liver [3]. As a result of cell activity, -1/2 defects generate only limited flows that remain balanced (no net direction) for symmetry reasons, while +1/2 defects that are not symmetric generate intense directed flows (figure 1b) that result in the net displacement of the defects themselves. Defects have been correlated with cell extrusion [4], the formation of tridimensional cell mounds [5] or the transition to low Reynolds number turbulence [6].

As time goes by, cell proliferation and subsequent increase in density [7,8] has two consequences on the time evolution of such systems: on the one hand, the nematic order is improved by excluded volume interactions whereas, on the other hand, movements are considerably slowed down by a process derived from Contact Inhibition of Locomotion that progressively leads to a jammed state [7,8]. The steady state at long times results from a balance between these two effects giving rise to finite well-oriented domains as large as 1 mm [1,8].

Confinement has a strong influence on the architecture and on the dynamics of these cell assemblies. When plated in stripes, NIH-3T3 fibroblasts tend to align with the stripe's edge. As cells proliferate, this order propagates toward the center of the stripe leading to cells perfectly aligned with the stripe direction [8]. This behavior holds as long as the stripe width remains smaller than the typical size of the above-mentioned domains in an unconfined culture.

Such a perfect alignment is the expected situation for passive nematic systems. Passive systems confined in disks are expected to organize with two facing +1/2 defects positioned on a diameter, at a well-defined distance from the origin that is set by minimizing the nematic distortion energy [9]. This is quantitatively what is obtained with several cells lines including NIH-3T3 fibroblasts, C2C12 muscles cells or RPE1 retina cells [10] (figure 1c,d). In this case, playing on the activity of the cells doesn't affect the defects' positioning. This passive-like behavior for cells that otherwise act as active entities can be attributed to a decreased activity induced by the jamming of the system or by an enhanced friction between the cells and their substrate (see discussions on the impact of friction in [7,10,11]).

Revised version

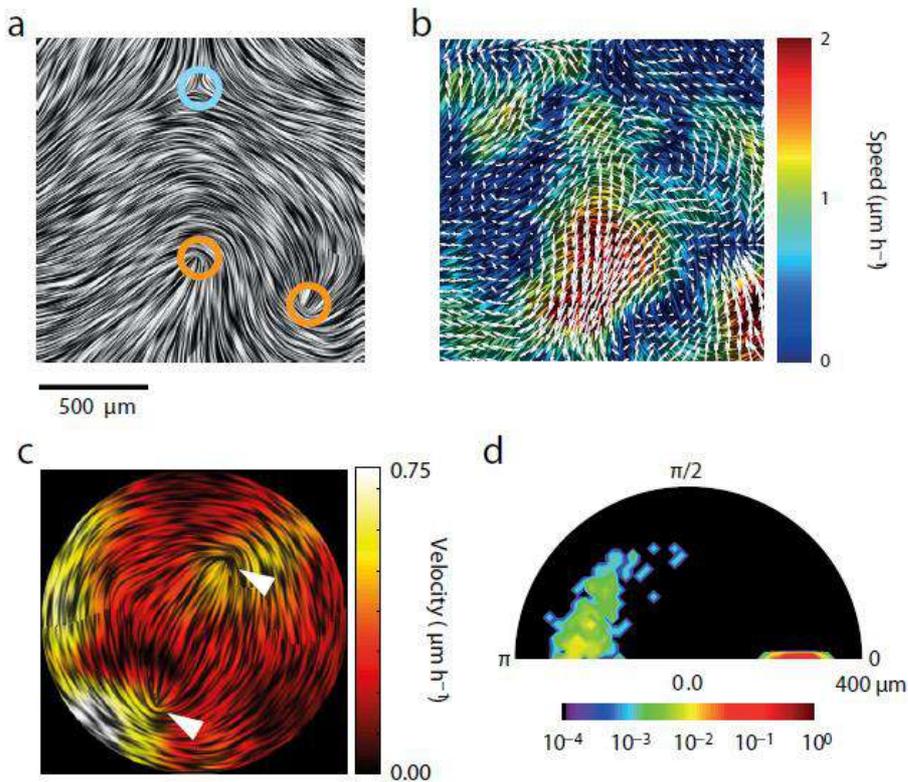

*Figure 1 : Topological nematic defects. a- Presence of +1/2 (orange) and -1/2 (blue) defects within a NIH-3T3 monolayer. The phase contrast image has been processed with line integral convolution to obtain the nematic field plotted here. b-Corresponding velocity field. The large flows correspond to +1/2 defects. c- When confined in disks, the system evolves toward a pair of +1/2 defects (white triangles) facing each other on a diameter (here, radius=350 μm). d- Probability density of the defects (N=486). Their position is well defined and corresponds to the configuration that minimizes distortions in a passive nematic. Adapted from [10].*

However, returning to the stripe geometry, C2C12 and RPE1 cells' self-organization markedly differ from the one adopted by the NIH-3T3 cells. Indeed, they make a finite angle with the stripe's direction (figure 2a). This tilted organization goes with a shear flow along the edges of the stripe (figure 2b). Similar antiparallel displacements have been observed in vivo in the collective migration of cancer cells [12]. For stripes narrower than a critical width, these two features disappear: cells align exactly in the stripe's direction and don't exhibit a net shear flow. In the framework of the active gels theory, this transition has been interpreted as a Fréedericksz transition controlled by the activity of the cells [11].

## Current and future challenges

### Nematic order and aspect ratio

If the nematic self-organization is relatively intuitive for cells that are intrinsically elongated, these architectures have also been reported for cells that are basically isotropic when isolated such as many epithelial cells [4,6]. In monolayers, although the cells' aspect ratio remains close to one, there is a slight bias sufficient to characterize a nematic order. A question therefore arises on the impact of the displacements of these deformable cells on their nematic order. In contrast, the analysis of the previous examples assumed that it was the elongated shape of the cells that triggered the nematic order, which in turn affected the flows and forces. A feedback mechanism between activity and cell deformation may be an important component to incorporate in the analysis [13].

Revised version

## Nematic order and polarity

Based on the typology of the defects (+1/2 and -1/2 only), there is no ambiguity on the nematic nature of the order. Yet, the nature of the motion of the corresponding isolated cells can be bipolar with no head or tail, or polar. Bipolar character amounts to apolar when integrated in time and is compatible with the nematic order while displacements of polar symmetry with a certain persistence are not. This is for instance the case with epithelial cells [4]. The reason why activity would foster nematic ordering of front-rear polar migrating cells remains an open question.

## Active or not active?

As mentioned above, depending on the experiment, the same cells have been observed to display either behaviors characteristic of a passive or an active system. For instance, C2C12 cells plated on disks position their +1/2 defects as passive systems do, while they exhibit a pronounced active shear flow when confined in stripes. One tentative explanation is that these observations reflect different surface densities. In stripes, mostly because of the convergent flow described below, cell density is maintained at a low level near the edges and shear flows can develop. In contrast, in circular confinements cell proliferation progressively jams the cell layer (that then amounts to an effectively passive system) [7,14], before it develops in 3D.

## Microscale vs. mesoscale contact guidance

As noted, cells in the low activity regime perfectly align in stripes whose width is up to a millimeter. Only the cells at the edges "feel" an external guidance cue (the edge itself): they align with this edge and this order propagates toward the center by self-organization. There are several bioengineering applications for which such an alignment is desirable [15]. Such alignments are classically obtained by plating cells on parallel arrays of lines or ridges whose width and separation are smaller than a cell size (a process called Contact Guidance). This way, every single cell is oriented in the prescribed direction and the whole assembly does too by juxtaposition of these individually aligned cells. The alignment in mesoscale stripes differs from this simple view by having the cells self-organizing within stripes that are very wide compared to a cell size. Both strategies yield perfectly oriented monolayers but no functional analysis of the resulting tissues has been reported so far.

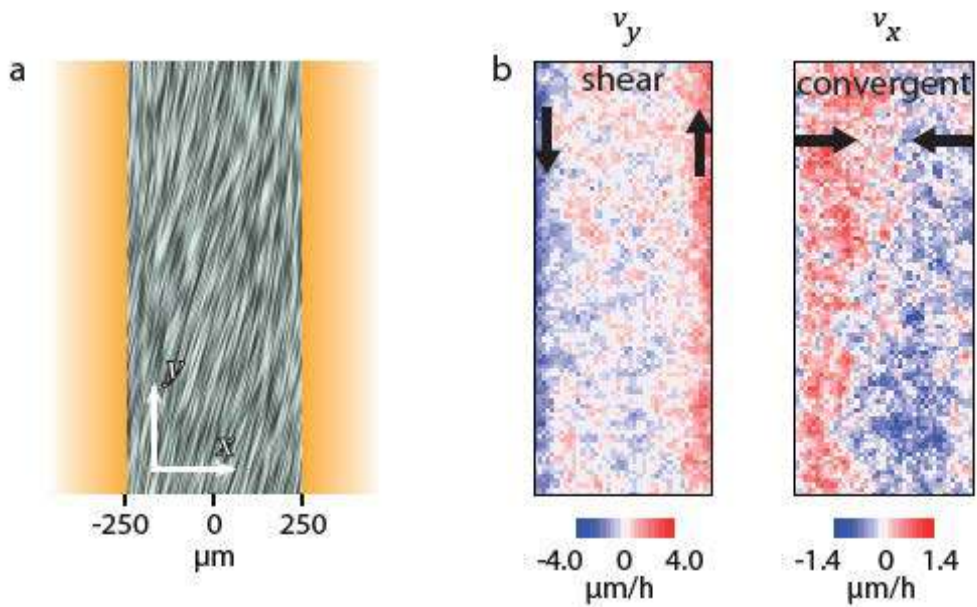

*Figure 2 : Orientation and flows in monolayers confined in stripes. a- Line integral convolution image of C2C12 cells confined within an adhesive stripe (light green: cell adhesive; orange: cell repellant) showing the spontaneous angle adopted by these cells. b- Velocity field of the cells. The velocities have been decomposed along their two components: along the stripe (shear)*

Revised version

*and across the stripe (convergent). The shear flow is confined close to the edges of the stripe because of the cell-surface friction. The peak amplitude of the convergent flows is smaller than the shear's (see scales) but they extend further within the stripe. Adapted from [11].*

## Shear flows and convergent flows

If the shear flow of active cells confined on stripes and their tilt with respect to the stripe's direction can be explained by an active gel theory, several other features that have been experimentally observed on these systems cannot [11]. Among those, the presence of convergent flows directed from both edges toward the center of the stripe is particularly interesting. These convergent flows that are dominant in the middle of the stripe where the shear flows are entirely screened by friction (figure 2b), need a source and a sink. The source is most likely cell proliferation. As for the sink, the cells converging to the center of the stripe tend to exit the monolayer and eventually build a tissue-scale tridimensional structure. Because of their different origins, convergent flow and shear flow are also subjected differently to surface friction. Their respective contributions to mesoscale Contact Guidance are in no way obvious and should be investigated further to develop efficient strategies of cell alignment.

## Concluding remarks

The present interest in active systems has spread to biology at different scales. Beyond the seminal active nematic systems based on microtubules or actin microfilaments [16], living cells have shown their potential as interesting systems to study these properties from a physics point of view. Yet, it should be kept in mind that living cells and tissues perform biological functions. However, the feedbacks between the above-described architectures and dynamics, and these biological functions remain largely to be studied. First results in this line have been obtained by the correlation between defects and cell extrusion, formation of cell mounds or displacements of cancer cells. There is little doubt that more of those correlations will be addressed in other systems.

## Acknowledgements

We thank the past and present members of the Biology-inspired Physics at MesoScales group as well as our collaborators from the Physical Approach of Biological Problems group. The BiPMS group is a member of IPGG and a member of the CelTisPhyBio Labex that we acknowledge for funding. VY acknowledges partial funding from the EU PRESTIGE program. TS is funded by the IC3i Phd program which has received funding from the European Union's Horizon 2020 research and innovation program under the Marie Skłodowska-Curie grant agreement No 666003.

Revised version

**Epithelial Monolayers as Active Materials**
Marino Arroyo and Sohan Kale, Universitat Politècnica de
Catalunya – BarcelonaTech, Barcelona, Spain

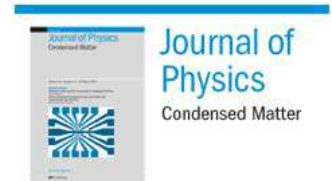

## Status

Cells are able to sense and react to their mechanical environment. Their various organelles form a mechanosensitive system that connects external mechanical cues to the gene expression. Furthermore, several functions of animal cells are mechanical in nature. Examples include controlling their shape during cytokinesis [1], migrating individually or collectively [2], or cohesively adhering to each other in epithelial tissues [3]. All these functions ultimately depend on the architecture and dynamics of the structural, force-bearing, elements in cells, which include the cytoskeleton, adhesive structures, the plasma membrane and the cytosol. The cytoskeleton is a composite structure of rheological components including microtubules (elastic), intermediate filaments (strain-stiffening coiled-coils), and actin networks (active viscoelastic gels). Cells and cohesive tissues can thus be seen as materials or structures. However, unlike inert engineering materials or structures, cells and tissues are constantly renewing the materials from which they are built and are actively generating forces, all of which requires chemical energy input [1].

Here, we focus on epithelial monolayers, which are very simple yet very important animal tissues. They consist of cohesive and highly organized cells that adhere to each other and often to an underlying matrix to form a barrier that lines internal and external organ surfaces, controls transport of gas or nutrients, and protects us from pathogens or desiccation. Moreover, epithelial mechanics is also important to shape embryos during development [4] and in diseases including cancer [5]. Epithelial tissues are subjected to mechanical deformations of various magnitudes and at different time-scales. For instance, as we breathe, cell monolayers in our lungs undergo areal stretches of more than 10% over seconds, whereas during blastocyst expansion in developing mammals, the trophectoderm undergoes a several-fold increase in area. Despite the fact that epithelial monolayers are active materials performing crucial mechanical tasks during adult life and development, the quantification of their mechanical response to stretch remains very limited, and how these tissues control tension, which if excessive could lead to tissue fracture and loss of function as barriers, remains poorly understood. Next, we review recent work on the characterization of the mechanical behaviour of epithelial monolayers under stretch and highlight initial steps and challenges in connecting tissue-scale mechanics with the architecture and dynamics of sub-cellular cytoskeletal networks (Fig. 1).

## Current and Future Challenges

Epithelial mechanics has been studied in various contexts, identifying a variety of mechanisms, cellular and supra-cellular, by which they respond to stretch. These include changes in the topology of the junctional network such as cell division, cell extrusion, and other types of network rearrangements, notably T1 transitions [4]. The fluid or solid rheology of epithelial monolayers has been linked to unjamming/jamming transitions dictated by the ease of undergoing a T1 transition leading to cell intercalation and flow. However, at shorter time-scales of minutes up to hours, epithelial tissues accommodate stretch by cellular deformation [4]. Focusing on this regime, only a handful of quantitative *in vitro* experiments have been able to measure the stress-strain relationship of epithelial monolayers. Charras and co-workers [6,7] have probed free-standing monolayers through creep and stress relaxation experiments at various time-scales and stretch/stress magnitudes. Casares *et al* [3] stretched adherent cultured cell monolayers on a flexible substrate while measuring tissue tension using traction force microscopy during stretch-unstretch manoeuvers. Latorre et al [8] developed a microscopic bulge test in which epithelial monolayers were stretched up to areal strains of 300% by transmural pressure forming 3D domes. These tests have unveiled, even in the seemingly simple case of frozen junctional network, a wealth of dynamical rheological behaviours, which we summarize below.

Under sudden stretch application, epithelial tissues initially build-up tension, which is then partially relaxed over time-scales of 10s of seconds [7] to a few minutes [3]. Upon unstretch or compression, tissue tension rapidly drops, and then partially recovers within a comparable time-scale [3]. In free-standing monolayers, sudden compression leads to formation of folds that can transiently flatten or stabilize depending on the unstretch magnitude [7]. These mechanisms have been interpreted as supra-cellular analogues of "reinforcement" and "fluidization" responses previously reported at the single-cell level [3], and have been understood with minimal rheological models. When subjected to a sudden tension in a creep testing setup, suspended epithelial tissues exhibit complex rheology behaving as viscoelastic solids at lower tensions and as complex fluids at higher tensions [6]. The equibiaxial stretching of suspended epithelial monolayers in bulging epithelial domes [8] has shown the ability of monolayers to reversibly undergo extreme deformations (up to 300% areal strains) at nearly constant tissue tensions. The measured tensional plateau was accompanied by heterogeneous cellular strain distribution, with coexistence of superstretched (~1000% areal strain) and barely stretched cells in a tissue with nearly uniform tension. These are hallmark features of superelasticity, a phenomenology associated prominently with superelastic Nickel-Titanium alloys, which in epithelial tissues were shown to be of an active origin.

As discussed earlier, all these mechanical behaviours should depend on the structural components of cells and their dynamics. The turnover time-scales of adhesion molecules (~10s of minutes) are significantly longer than those of cortical components including actin filaments (~min), myosin motors and crosslinkers (~sec) [7]. Consequently, at intermediate time-scales of a few minutes to hours, the junctional network of the tissue is conserved and epithelial mechanics is dictated by mechanical behaviour of the cell cytoskeleton, and more specifically the actomyosin cortex. The cortex is a 50-200 nm thin meshwork of crosslinked actin filaments anchored beneath the cell membrane [1]. The cortex is highly dynamic due to constant turnover of crosslinkers and actin filaments. Moreover, the pulling action of myosin motors powered by ATP hydrolysis collectively leads to effective contractile tensions. On a mesoscopic level, the active viscoelastic phenomenology of the cortex is captured by active gel theories grounded on non-equilibrium thermodynamics [9]. In principle, the tissue-scale phenomenology reviewed in the previous paragraph should be a consequence of the mechanics of the actin cortex. However, this connection between sub-cellular dynamics as described by active gel theories and tissue rheology has been lacking.

The predominant approach to theoretically understand epithelial monolayers is through vertex models, which represent the monolayer as a polygonal tessellation. The motion of vertices is governed by the mechanical forces arising from a work functional or pseudo-energy [10]. The mechanical forces have contributions from effective contractile tensions, cytoplasmic pressure, and cell-substrate interactions. The out-of-equilibrium nature of tissue mechanics is hidden in the cortical tensions, which can be assumed to be constant in a steady state. However, going beyond quasi-steady states and looking ahead, tissue-level dynamical vertex models based on out-of-equilibrium active gel theories of the actin cortex could help us establish a link between various molecular mechanisms, possibly tuned by cells, and the emergent epithelial phenomenologies. For instance, such a framework would help us understand how stress relaxation time-scales, tensional homeostasis, creep behaviour or transient buckling depend on the turnover of cross-linkers, the actin architecture and polymerization dynamics, or myosin activity. Such a theory would allow us to make specific predictions about how molecular mechanisms affect tissue rheology, which can be tested by biochemically perturbing cytoskeletal dynamics while measuring forces and deformation on tissue scales.

As an initial step in this direction, Latorre *et al* [8] explained active-superelasticity of epithelial tissues by explicitly accounting for cytoskeletal mechanics in vertex models. The classical vertex model was augmented with cortical turnover to account for stretch-induced cortical dilution due to limited amount of actin monomers ready for polymerization and re-stiffening from Intermediate filaments (IFs) at large stretches. The enhanced vertex model captured active-superelastic phenomenologies, including the tensional plateau, the sudden increase in strain variance amongst cells, and the extreme strain heterogeneity at later stages.

At the centre of this conceptual framework, active gel models should reflect the physics of the actin cytoskeleton. Various studies have shown that active gel models employing a simple description of cortical network architecture through actin density fields and nematic order parameter can capture the hydrodynamics of actomyosin gels. Yet recent work on living cells and reconstituted actomyosin gels have highlighted our lack of understanding of how actin network architecture, connectivity, and dynamics is controlled molecularly, and how this affects the effective material properties of the active gel [11,12,13]. To this end, increasingly realistic discrete models of the actomyosin cortex combined with super-resolution microscopy may help establish this connection between scales.

**Concluding Remarks**

Up to now, we have focused on the actin cortex as the major determinant of epithelial mechanics. However, at longer time-scales, tissue rheology depends on the ability of the junctional network to remodel, which ultimately depends on the dynamics of adhesion molecular complexes and adaptor molecules linking them to the cytoskeleton. How these ingredients integrate to control tissue mechanics remains a largely open question. Besides connecting sub-cellular mechanisms to in-plane tissue rheology, a similar connection could help us understand the collective migration of cohesive tissues or their three-dimensional morphogenesis. In coming years, an integration of experimental examination and theoretical modelling of tissues at various scales—from the dynamics of molecules to tissue mechanics—has the potential to provide a quantitative framework to understand epithelial mechanobiology during physiology, disease and development, and control it in epithelial technologies such as organs-on-a-chip.

**Acknowledgements**

We acknowledge the support of the European Research Council (CoG-681434), the European Commission (project H2020-FETPROACT-01-2016-731957), the Spanish Ministry of Economy and Competitiveness/FEDER (DPI2015-71789-R), and the Generalitat de Catalunya (SGR-1471, ICREA Academia award).

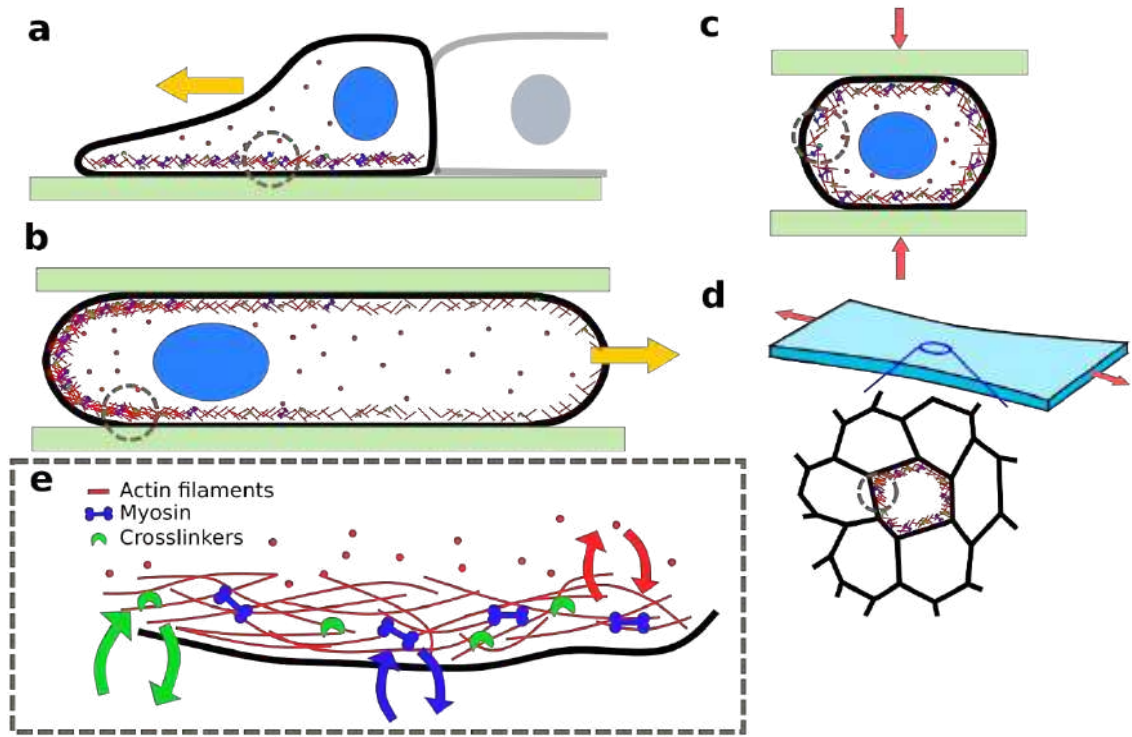

**Figure 1.** The active viscoelastic rheology of the actomyosin cortex, controlled by architecture, power input and turnover dynamics on the subcellular level (e) plays a dominant role in (a) mesenchymal and (b) amoeboidal cell migration, (c) cellular mechanics and (d) epithelial tissue rheology.